\documentclass[letterpaper, sigconf]{acmart}

\usepackage{natbib}
\usepackage{amssymb,amsmath,latexsym}
\usepackage{amsthm}
\usepackage{hyperref}

\usepackage{bm}
\usepackage{bbm} 
\usepackage{microtype}
\usepackage{url}
\usepackage{graphicx}
\usepackage{siunitx}
\usepackage{algpseudocode} 
\usepackage{xfrac} 
\usepackage{physics} 
\usepackage{bbm} 
\usepackage{wrapfig} 
\usepackage{enumitem} 
\usepackage{bytefield}	
\usepackage[underline=false]{pgf-umlsd}
\usetikzlibrary{shadows,positioning}
\usepackage{tabularx}
\usepackage{color}
\usepackage{subcaption}
\usepackage{amsmath}
\usepackage[T1]{fontenc}
\usepackage{float} 
\usepackage[english]{babel}

\usepackage[lined,ruled,vlined]{algorithm2e}

\usepackage{xcolor}
\theoremstyle{plain}

\newcommand{\X}{\mathcal{X}}
\newcommand{\Y}{\mathcal{Y}}
\newcommand{\Z}{\mathcal{Z}}

\newcommand{\Lab}{\textproc{Lab}} 
\newcommand{\Qlink}{\textproc{QL2020}}

\newcounter{Lprotocol}
\newenvironment{Lprotocol}[1]
  {\par\addvspace{\topsep}
   \noindent
   \tabularx{\linewidth}{@{} X @{}}
    \hline
    \refstepcounter{Lprotocol}\textbf{Protocol \theLprotocol} #1 \\
    \noindent\hrulefill}
  { \\
    \noindent\hrulefill
   \endtabularx
   \par\addvspace{\topsep}}

\newcommand{\sbline}{\\[.5\normalbaselineskip]}% small blank line

\usepackage{xargs}
\usepackage[colorinlistoftodos,prependcaption]{todonotes}
\newcommandx{\unsure}[2][1=]{\todo[linecolor=red,backgroundcolor=red!25,bordercolor=red,#1]{#2}}
\newcommandx{\change}[2][1=]{\todo[linecolor=blue,backgroundcolor=blue!25,bordercolor=blue,#1]{#2}}
\newcommandx{\info}[2][1=]{\todo[linecolor=green,backgroundcolor=green!25,bordercolor=green,#1]{#2}}
\newcommandx{\improvement}[2][1=]{\todo[linecolor=purple,backgroundcolor=purple!25,bordercolor=purple,#1]{#2}}
\newcommandx{\thiswillnotshow}[2][1=]{\todo[disable,#1]{#2}}

\usepackage{listings} 

\definecolor{lightgray}{gray}{0.9}
\DeclareSIUnit\dBm{dBm}

\renewcommand\footnotetextcopyrightpermission[1]{} % removes footnote with conference info
\setcopyright{none}
\acmDOI{}

\acmISBN{}

\acmPrice{}

\begin{document}
\title{A Link Layer Protocol for Quantum Networks}
\author{Axel Dahlberg$^{1,2}$, Matthew Skrzypczyk$^{1,2}$, Tim Coopmans$^{1,2}$, Leon Wubben$^{1,2}$, 
Filip Rozp\k{e}dek$^{1,2}$, Matteo Pompili$^{1,2}$, Arian Stolk$^{1,2}$, Przemys\l{}aw Pawe\l{}czak$^{1}$, 
Robert Knegjens$^{1}$, Julio de Oliveira Filho$^{1}$, Ronald Hanson$^{1,2}$, Stephanie Wehner$^{1,2}$}
\affiliation{%
\institution{$^1$QuTech, Delft University of Technology and TNO}
%\streetaddress{Lorentzweg 1}
%\city{Delft}
%\state{Netherlands}
\postcode{2628 CJ}
}
\affiliation{
\institution{$^2$Kavli Institute of Nanoscience, Delft University of Technology}
\postcode{2600 CJ}
}
\email{s.d.c.wehner@tudelft.nl}
\renewcommand{\shortauthors}{Dahlberg, Skrzypczyk, et al.}

\begin{abstract}
Quantum communication brings radically new capabilities
that are provably impossible to attain in any classical network.
Here, we take the first step from a physics experiment
to a fully fledged quantum internet system. We propose a
functional allocation of a quantum network stack and construct
the first physical and link layer protocols that turn
ad-hoc physics experiments producing heralded entanglement
between quantum processors into a well-defined and
robust service. This lays the groundwork for designing and
implementing scalable control and application protocols in
platform-independent software. To design our protocol, we
identify use cases, as well as fundamental and technological
design considerations of quantum network hardware, illustrated
by considering the state-of-the-art quantum processor
platform available to us (Nitrogen-Vacancy (NV) centers in
diamond). Using a purpose built discrete-event simulator
for quantum networks, we examine the robustness and performance
of our protocol using extensive simulations on a
supercomputing cluster. We perform a full implementation
of our protocol, where we successfully validate the physical
simulation model against data gathered from the NV hardware.
We first observe that our protocol is robust even in a
regime of exaggerated losses of classical control messages
with only little impact on the performance of the system.We
proceed to study the performance of our protocols for 169
distinct simulation scenarios, including tradeoffs between
traditional performance metrics such as throughput and the
quality of entanglement. Finally, we initiate the study of
quantum network scheduling strategies to optimize protocol
performance for different use cases.
\end{abstract}

\maketitle

\section{Introduction}

Quantum communication enables the transmission of quantum bits (qubits) in order to achieve novel capabilities that are provably impossible using classical communication.
As with any radically new technology, it is hard to predict all uses of a future Quantum Internet~\cite{Wehner2018,kimble2008quantum}, but several major applications have already been identified depending on

\begin{figure}[h!]
		    % \vspace{-12pt}
		\begin{center}
				\includegraphics[width=0.4\textwidth]{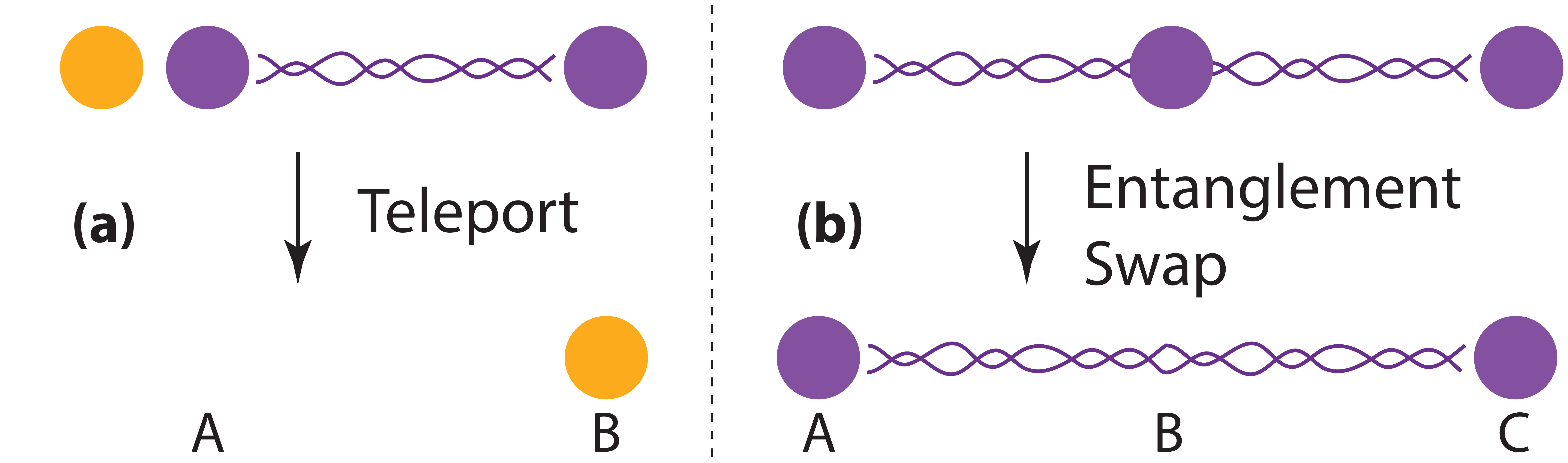}
					\end{center}
						    % \vspace{-10pt}
						    	\caption{Entanglement enables long-distance quantum communication: (a) once two qubits (purple/dark) are confirmed to be entangled (threaded links between qubits), a data qubit (yellow/light) can be sent deterministically using teleportation~\cite{bennett1993teleporting}, consuming the entangled pair; (b) long-distance entanglement can be built from shorter segments: If node $A$ is entangled with $B$ (repeater), and $B$ with $C$, then $B$ can perform \emph{entanglement swapping}~\cite{zukowski1993event} to create long-distance entanglement between the qubits at $A$ and $C$.}
								\label{fig:useEntanglement}
													        \vspace{-10pt}
\end{figure}

\noindent the stage of quantum network development~\cite{Wehner2018}, ranging from cryptography~\cite{bb84,e91}, sensing and metrology~\cite{gottesman2012longer, komar2014quantum}, distributed systems~\cite{ben2005fast, denchev2008distributed}, to secure quantum cloud computing~\cite{broadbent2009universal,fitzsimons2017unconditionally}.

%Possibly the most famous application is quantum key distribution~\cite{bb84} (QKD), which allows two remote network nodes to generate an encryption key, enabling the exchange of secret information.
%The security of an ideal QKD implementation follows from the laws of quantum mechanics, and \edit{is} thus fully future proof even against any attacker possessing a large-scale quantum computer.
%Quantum networks do, however, have many other applications already including clock synchronization~\cite{Jozsa2000}, extending the baseline of telescopes~\cite{telescope}, secure identification~\cite{secureID}, achieving efficient agreement on distributed data~\cite{Denchev2008}, exponential savings in communication~\cite{Buhrman2010}, quantum 
%sensor networks, as well as secure access to remote quantum computers in the cloud~\cite{blindQC}.

Qubits are fundamentally different from classical bits, which brings significant challenges both to the physical implementation of quantum networks, as well as the design of quantum network architectures. Qubits cannot be copied, ruling out signal amplification or repetition to overcome transmission losses to bridge great distances.
Two qubits can share a special relation known as \emph{entanglement}, even if these two qubits are stored at distant network nodes.
Such entanglement is central not only to enable novel applications, but also provides a means to realize a quantum repeater, which enables quantum communication over long-distances
(Figure~\ref{fig:useEntanglement}).

At present, short-lived entanglement has been produced probabilistically over short distances ($\approx100\,$km) on the ground by 
sending photons over standard telecom fiber (see e.g.~\cite{dynes2009efficient, inagaki2013entanglement}), 
%including commercial QKD systems~\cite{citations}, 
as well as from space over 1203km from a satellite~\cite{Yin2017}.
%Short-lived refers to the fact that the qubits were not stored at the end points.
%What's more, entanglement is produced probabilistically and there is no signal indicating the generation has been a success.
Such systems can allow the realization of applications in the prepare-and-measure stage~\cite{Wehner2018} of quantum networks
on point-to-point links, 
but cannot by themselves be concatenated to allow the transmission of qubits over longer distances.

In order to enable long-distance quantum communication and the execution of complex quantum applications, 
we would like to produce long-lived entanglement between two quantum nodes that are capable of storing and manipulating qubits.
To do so efficiently (Section~\ref{sec:qubits}), we need to confirm entanglement generation by performing \emph{heralded} 
entanglement generation. This means that there is a \emph{heralding signal} that tells us if we have been successful in an attempt
to generate entanglement. 
%I.e., generation is deterministic conditioned on a successful heralding signal. 

The current world distance record for producing such entanglement is 1.3km, which has been achieved using a solid state platform 
known as Nitrogen-Vacancy (NV) centres in diamond~\cite{Hensen2015}.
Intuitively, this platform is a few qubit (as of now maximum $8$~\cite{bradley2019solidstate}) quantum computer capable of arbitrary quantum gates, with an optical interface for initialization, measurement and entanglement generation.
Key capabilities of the NV platform have already been demonstrated, including qubit lifetimes of $1.46$ s ~\cite{Abobeih2018},
entanglement production faster than it is lost~\cite{Humphreys2018},
and using entanglement to teleport qubits between separated NV centres~\cite{Pfaff2014}.
%The QLE is the rate of entanglement generation in relation to the rate at which entanglement is lost due to limited lifetimes of quantum memories, demanding $QLE > 1$.
Other hardware platforms exist that are identical on an abstract level (quantum computer with an optical interface), and on which heralded long-lived entanglement generation has been 
demonstrated (e.g. Ion Traps~\cite{moehring2007entanglement}, and Neutral Atoms~\cite{hofmann2012heralded}).
Theoretical proposals and early stage demonstrations of individual components also exists for other physical platforms (e.g. quantum dots~\cite{Delteil2016}, rare earth ion-doped 
crystals~\cite{Valivarthi2016}, atomic gases~\cite{julsgaard2001experimental, chou2005measurement}, and superconducting qubits~\cite{narla2016robust}), but their performance is not yet good enough to generate entanglement faster than it is lost. 
%, as well as other color centers in diamond such as Silicon Vacancies~\cite{XX}.\axel{What to ref for atomic gases, supercon and SiV (Dirk Englund?)}

Up to now, the generation of long-lived entanglement has been the domain of highly sophisticated, but arguably ad-hoc physics experiments. 
We are now on the verge of seeing early stage quantum networks becoming a reality, entering a new phase of development which will require a 
joint effort across physics, computer science and engineering to overcome the many challenges in scaling such networks.
In this paper, we take the first step from a physics experiment to a fully-fledged quantum communication \emph{system}.

{\bf Design considerations and use cases:} 
We identify general design considerations for quantum networks based on fundamental properties of entanglement, and technological 
limitations of near-term quantum hardware, illustrated with the example of our NV platform.
For the first time, we identify systematic use cases, and employ them to guide the design of our stack and protocols.

{\bf Functional allocation quantum network stack:}
We propose a functional allocation of a quantum network stack, and define the service desired from its link layer to satisfy use case requirements and design considerations. In analogy to classical networking, the link layer is responsible for producing entanglement between two nodes that share a direct physical connection (e.g. optical fiber).

{\bf First physical and link layer entanglement generation protocols:}
We proceed to construct the world's first physical and link layer protocols that turn ad-hoc physics experiments producing heralded entanglement into a well defined service. This lays the groundwork for designing and implementing control and application protocols in platform independent software in order to build and scale quantum networks. 
%At the physical layer, we develop an abstraction of heralded entanglement generation which we name Midpoint Heralding Protocol (MHP), and propose a specific instance of such a protocol that depends on the platform used.
At the physical layer, we focus primarily on the quantum hardware available to us (NV platform) but the same protocol 
could be realized directly using Ion Traps or Neutral Atoms, as well as ---with small changes--- other means of producing physical entanglement~\cite{sangouard2011quantum}.
Our link layer protocol takes into account the intricacies of the NV platform, but is in itself already platform independent.
%The EGP turns entanglement generation into a robust service, and also creates a separation between the platform dependent part of the communication system and high layer control that can be the same independent of the underlying physical system.
%Such a stack is not only tailored towards enabling future control of quantum networks, but also to allow the execution of arbitrary quantum network applications in platform independent software.

%To realize the services desired from the link layer we propose the Entanglement Generation Protocol (EGP). The EGP 

\begin{figure}
\centering
\includegraphics[width=\linewidth]{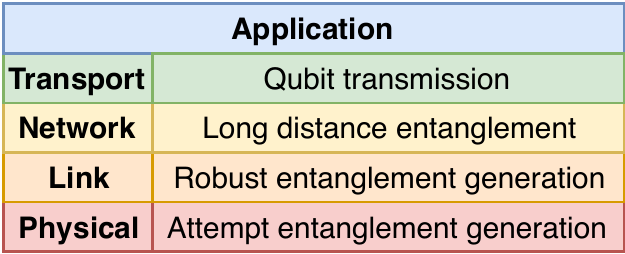}
\caption{Functional allocation in a quantum network stack. 
%Protocols differ significantly in that 
Entanglement is an inherent connection between quantum bits, 
%, and we wish to confirm (i.e. herald) entanglement generation already at the link layer to ensure the creation of long distance entanglement. 
which contrasts with classical networking where shared state is established at higher layers.}\label{fig:stack}
    % \vspace{-12pt}
\end{figure}

{\bf Simulation validated against quantum hardware:}
Using a purpose built discrete-event simulator for quantum networks, we examine the robustness and performance of our protocol 
using more than 169 scenarios totaling 94244h wall time and 707h simulated time on a supercomputing cluster.
To this end, we perform a complete implementation of our protocols and let them use simulated quantum hardware and communication links. 
To illustrate their performance, we consider two concrete short and long-distance scenarios based on the NV platform:
(1) \Lab\, where the nodes $A$ and $B$ are 2m apart.
Since this setup has already been realized, we can use it to compare the performance of the entanglement generation implemented on real quantum hardware against the simulation to validate its physical model, and
(2) a planned implementation of \Qlink\, where $A$ and $B$ are in two European cities separated by $\approx 25\,$km over 
telecom fiber.
Next, to investigate trade-offs between traditional performance metrics (e.g. throughput or latency) and genuinely quantum ones (fidelity, Section~\ref{sec:perf_metrics}), we take a first step in examining different quantum network scheduling strategies to optimize performance for different use cases.

\section{Related Work}
At present there is no quantum network stack connected to quantum hardware, no link layer protocols have been defined to produce entanglement, and no quantum networks capable of end-to-end qubit transmission or entanglement production have been realized (see~\cite{Wehner2018} and references therein).

A functional allocation of a stack for quantum repeaters and protocols controlling entanglement distillation (a procedure to increase the quality of entanglement) has been outlined in~\cite{vM:designing, vM:network, VanMeter2009, vM:protoDesign}, which is complementary to this work. This is very useful to ultimately realize entanglement distillation, even though no concrete control protocols or connection to a hardware system were yet given. 
We remark that here we do not draw layers from 
specific protocols like entanglement distillation, but focus on the service that these layers should provide (a layer protocol may of course choose distillation as a mean to realize requirements).
An outline of a quantum network stack was also put forward in~\cite{Pirker2018}, including an appealing high level quantum information theory protocol transforming multi-partite
entanglement. However, this high level protocol does not yet consider failure modes, hardware imperfections, nor the requirements on entanglement generation protocols and the impact of classical control. 
Plans to realize the physical layer of a quantum network from a systems view were put 
forward in~\cite{QI_Lloyd_2004}, however development has taken a different route.

In the domain of single-use point-to-point links for quantum key distribution (QKD), software has been developed for trusted repeater networks~\cite{Wehner2018} to make use of such key in e.g. VoIP~\cite{qphone_sigcomm_2013}.
However, these do not allow end-to-end transmission of qubits or generation of entanglement, and rely on trust in the intermediary nodes who can eavesdrop on the communication.
Control using software defined networks (SDN) to assist trusted repeater nodes has been proposed, e.g.~\cite{Yu2018,sdn1}.
These QKD-centric protocols however do not address control problems in true quantum networks aimed at end-to-end delivery of qubits, and the generation of 
long-lived entanglement.

In contrast, classical networking knows a vast literature on designing and analyzing network protocols.
Some ideas can indeed be borrowed from classical networking such as scheduling methods, but fundamental properties of quantum entanglement (Section~\ref{app:entanglement}), as well as technological considerations of quantum hardware capabilities (Section~\ref{sec:designConsiderationsHardware}) call for new protocols and methods of network control and management.
Naturally, there is a continuous flow of systems papers proposing new networking architectures, e.g. for SDN~\cite{bremlerbarr_sigcomm_2016}, data center networks~\cite{handley_2017_sigcomm}, content delivery networks~\cite{chen_sigcomm_2015} or cloud computing~\cite{zheng_sigcomm_2015}, to name a few. Yet, we are unaware of any system-level papers proposing a quantum network stack including protocols for concrete hardware implementations.

\section{Design Considerations for Quantum Network Architectures}\label{sec:designArch}
\label{sec:design_arch}

We first discuss design considerations of quantum networks themselves, followed by considerations specific to the physical and link layer (Section~\ref{sec:designLinkLayer}). These can be roughly subdivided into three categories:
(i) fundamental considerations due to quantum entanglement, 
(ii) technological limitations of near-term quantum hardware, 
and (iii) requirements of quantum protocols themselves.

\subsection{Qubits and Entanglement}\label{sec:entanglement}\label{sec:qubits}
We focus on properties of entanglement as relevant for usage and control (see Appendix and~\cite{Nielsen2010}).
Teleportation~\cite{bennett1993teleporting} allows entanglement to be used to send qubits (see Figure~\ref{fig:useEntanglement}).
We will hence also call two entangled qubits an \emph{entangled link} or \emph{entangled pair}. 
Teleportation consumes the entangled link, and requires two additional classical bits to be transmitted per qubit teleported. 
Already at the level of qubit transmission we hence observe the need for a close integration between a quantum and classical communications. Specifically, we will need to match quantum data stored in quantum devices, with classical control information that is sent over a separate physical medium, akin to optical control plane architectures for classical optical networks~\cite{strand_commag_2001}.
To create long-distance entanglement, we can first attempt to produce short-distance entangled links, and then connect them to form longer distance ones~\cite{briegel1998quantum, munro2015inside} via an operation known as entanglement swapping (see Figure~\ref{fig:useEntanglement}).
This procedure can be used iteratively to create entanglement along long chains, where we remark that the swapping operations can in principle be performed in parallel.
%notifying only the final two end points. 
From a resource perspective, we note that to store entanglement, both nodes need to store one qubit per entangled link. 
Proposals for enabling quantum communication by forward communication using quantum error correction also exist, which avoid entanglement swapping~\cite{munro2012quantum}.
%However, they have arguably more resource demands than preparing a few entangled pairs at a time: the proposal of~\cite{muralidharan2014ultrafast} uses $10^2$ simultaneously 
%entangled photons to convey one qubit (only $10$ realized today~\cite{gao2010experimental}), bringing them liley into a technologically more distant future.
However, these have arguably much more stringent requirements in terms of hardware, putting them in a technologically more distant future: they require the ability to create entangled states consisting of a large number of photons (only $10$ realized today~\cite{gao2010experimental}) and densely placed repeater stations performing near perfect operations~\cite{muralidharan2014ultrafast}.

\begin{figure}
	    \centering
	        \includegraphics[width=0.45\textwidth]{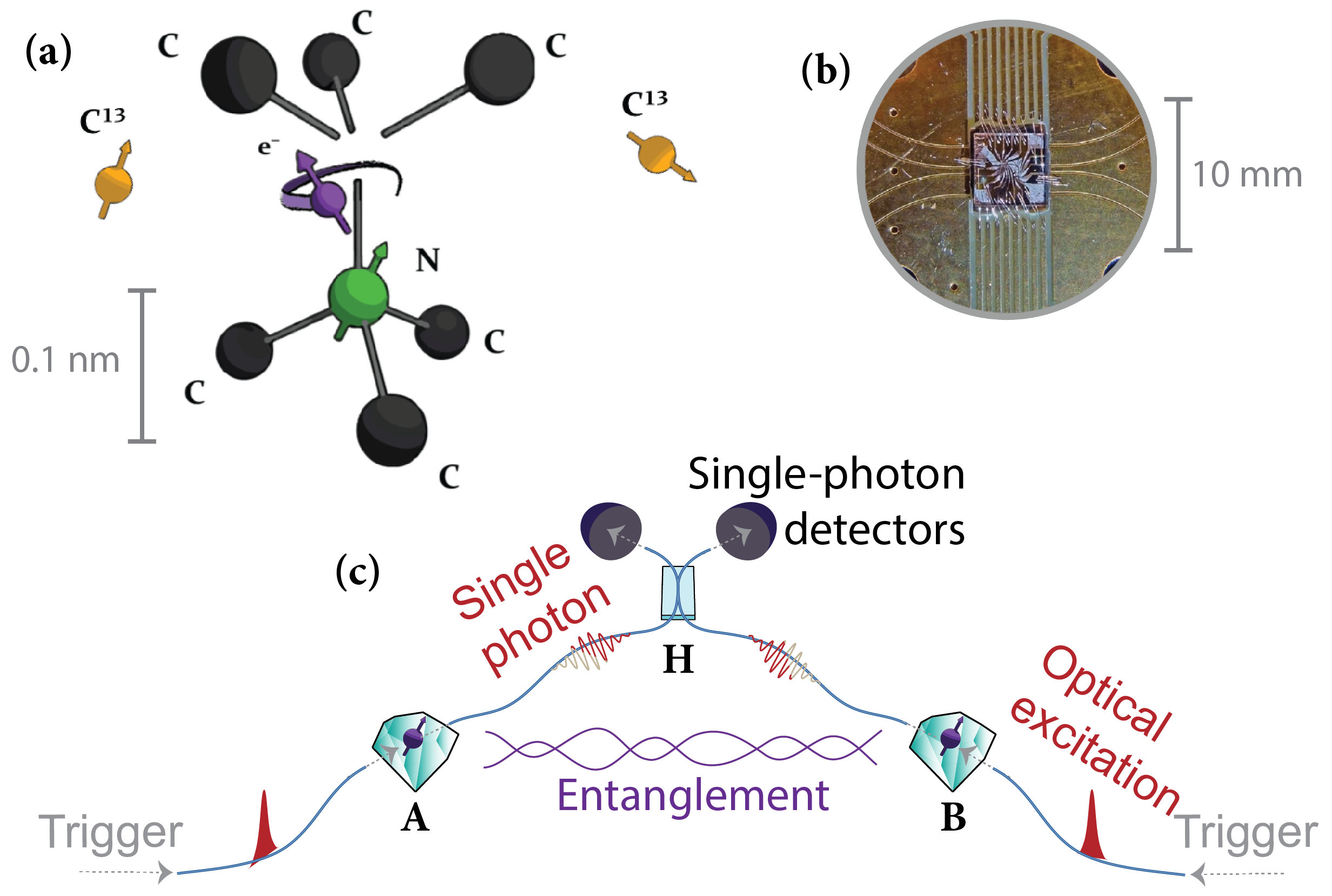}
			\caption{Heralded entanglement generation on the NV platform.  (a) NV centres are point defects in diamond with an electronic spin as a communication qubit (purple) and carbon-13 nuclear spins as memory qubits (yellow), realized in custom chips (b). (c) A trigger produces entanglement between the communication qubits of $A$ and $B$ (diamonds) and two qubits (photons) traveling over
				fiber to the heralding station $H$. $H$ measures the photons by observing clicks in the left or right detector giving the \emph{heralding signal $s$}: [failure] (none or both click), [success,$\ket{\Psi^+}$] (left clicks), [success,$\ket{\Psi^-}$] (right clicks). Success confirms one of two types of entangled pairs $\ket{\Psi^+}$ or $\ket{\Psi^-}$ (wiggly purple line). $H$ sends $s$ to $A$ and $B$ (not pictured).}
				    \label{fig:NV}
				    \vspace*{-10pt}
\end{figure}

Producing heralded entanglement does however allow long-distance quantum communication without the need to create entanglement consisting of many qubits.
Here, the heralding signal (see Figure~\ref{fig:NV}) provides a confirmation that an entanglement generation attempt has succeeded. Such heralding allows long-distance quantum communication without exponential overheads~\cite{briegel1998quantum}, and without the need for more complex resources~\cite{barrett2005efficient, cabrillo1999creation}.
Creating long-distance links between two controllable nodes by means of entanglement swapping (Section~\ref{sec:quantum_network_devices}), and executing
complex applications requires both nodes to know the state of their entangled links (which qubits belong to which entangled link, and who holds the other qubit of the entangled pair).
As illustrated in Figure~\ref{fig:useEntanglement}, remote nodes ("$B$" in the figure) can change the state of such entangled links ("$A$" and "$C$" in the figure).
Entanglement is an inherently connected element at the lowest physical level, whereas classical communications deal with unidirectional forward communication that abstracts the notion of a connection between a sender and receiver.  In order to make use of entanglement for a quantum network special devices capable of producing entanglement and manipulating local qubits are required.

\subsection{Quantum Network Devices} \label{sec:quantum_network_devices}
We focus on a high level summary of devices in a quantum network without delving into detailed physics (for more details, see~\cite{Wehner2018,awschalom2018quantum, sangouard2011quantum} and Section~\ref{sec:NV}).
Qubits can be sent optically through standard optical fiber using a variety of possible encodings, such as polarization~\cite{bb84,mattle1996dense}, time-bin~\cite{brendel1999pulsed}, or absence and presence of a photon~\cite{cabrillo1999creation}. 
Such qubits can be emitted from the devices in quantum nodes~\cite{bernien2013heralded,blinov2004observation, ritter2012elementary}, but in principle also transferred~\cite{nemoto2016photonic, ritter2012elementary, 
kalb2015heralded} from optical fiber into the node's local quantum memory.
Present day quantum memories have very limited lifetimes, making it highly desirable to avoid the exchange of additional control information before the entanglement can be used.

% Platforms such as NV in diamond~\cite{ronald} (Section~\ref{sec:NV}) or Ion Traps~\cite{chrisMonroe} cannot only store qubits in a quantum memory, but also perform arbitrary quantum gates and measurements. 
%Gate operations and measurements can then be used to perform deterministic entanglement swapping on any pair of entangled links available in memory. Such platforms hence allow deterministic control to perform 
%decisions which entangled links to swap, next to enabling the execution of complex quantum applications. Note that there at present no strict separation between a memory, and a CPU as in classical computing systems. 
%It could be that technological progress will change this in the next decades, but our discussion will not depend on this.
%Simpler systems, such as for example quantum memories based on rare earth ion doped crystal~\cite{xxx} do not allow gates and measurements to be performed on the memory itself. Limited control is possible
%using auxiliary optics to perform probabilistic entanglement swapping operations. These can be heralded, i.e. confirmed, but do not succeed with unit probability. 
%In such systems, it is also not always possible to retrieve the qubits at any given time from memory, but qubits are auto released after a specific pre-set time. 

We distinguish two classes of quantum nodes. One, which we will call a \emph{controllable quantum node}, offers the possibility to perform controllable quantum operations as well as storing qubits.
Specifically, these nodes enable decision making, e.g. which nodes to connect by entanglement swapping. Such nodes can act as quantum repeaters and decision making routers in the network (e.g. NV platform or other quantum memories combined with auxiliary optics), and --- if they support the execution of gates and measurements --- function as \emph{end nodes}~\cite{Wehner2018} on which we run applications (e.g. NV centre in diamond or Ion Traps).
Others, which we call \emph{automated quantum nodes}, are typically only timing controlled, i.e. they perform the same preprogrammed action in each time step. Such nodes also support quantum operations and measurements, but only those necessary to perform their preprogrammed tasks.
The latter is still very useful, for example, to establish entanglement along a chain of quantum repeaters performing the entanglement swapping operations~\cite{briegel1998quantum, munro2015inside} (see again Figure~\ref{fig:useEntanglement}). In Section~\ref{sec:NV} we give a concrete example of such a timing controlled node.

%From the perspective of a quantum network stack, 
%the distinction between controllable and automated nodes indicates at which layer we envision these elements fit, where automated nodes are dealt with at the physical layer (Section~\ref{sec:stack}). 

%At present, the rate at which entanglement has been produced between two quantum memories is low (Hz)~\cite{citations}, each node can store only very few (XXX~\cite{XXX}) qubits, and lifetimes are short (Numbers\cite{XX}).
%The maximum number of qubits in a node that has been entangled remotely is XXX~\cite{isThisRonald?}, with a lifetime ofXXX. Progress is to be expected since locally (not yet remotely) chips based on NV in diamonds have already shown 20? qubits, with a lifetime of XXX seconds~\cite{timLatestPaper}. Yet it is to be expected that the lifetimes of quantum memories and the number of qubits remains a precious resource for many years to come.

\subsection{Use Cases}
%When designing a quantum network stack, and indeed a link layer protocol for producing entanglement between controllable quantum nodes, 
%It is useful to reflect on how entanglement will be used.
We distinguish four use cases of a quantum network: one related to producing long-distance entanglement, and three that come from application demands. Since no quantum network has been realized to date, we cannot
gain insights from actual usage behavior. Instead we must resort to properties of application protocols known today.
% As quantum networks will initially be very small also application use cases are relevant for us here, since we intend to use our link layer protocol also at directly connected end nodes.
We desire flexibility to serve all use cases, including supporting multiple applications at the same time.

\emph{Measure Directly (MD) Use Case:}
The first application use case comes from application protocols that produce many ($\geq 10^4$) pairs of entangled qubits sequentially, where both qubits are immediately measured to produce classical correlations. 
As such, no quantum memory is needed to store the entanglement and it is not necessary to produce all entangled pairs at the same time.  
%Here, we also do not need to wait for entanglement production to be heralded before measurement can commence, since we may later discard the classical measurement outcomes if the heralding signal is negative (see Section~\ref{sec:heraldedProduction}). 
It follows that applications making use of this use case may tolerate fluctuating delays in entanglement generation.
Additionally, it is not essential to deliver error free correlations obtained from entanglement to the application.  Such applications will thus already anticipate error fluctuation across the many pairs.
This contrasts with classical networking where errors are often corrected before the application layer. 
%The reason for this difference is that the analgous form of error correction would be quantum error correction at the qubit/entanglement level, while the application can employ classical error correction which is technologically much easier to realize.
Examples of such applications are QKD~\cite{e91}, secure identification~\cite{damgaard2007secure} and other two-party cryptographic protocols~\cite{chailloux2011optimal, aharonov2000quantum, damgaard2008cryptography, wehner2008cryptography, ribeiro2015tight} at the prepare-and-measure network stage~\cite{Wehner2018}, 
and device-independent protocols at the entanglement network stage~\cite{Wehner2018}.

\emph{Create and Keep (CK) Use Case:}
The second application use case stems from protocols that require genuine entanglement, possibly even multiple entangled pairs to exist simultaneously. Here, we may wish to  perform joint operations on multiple qubits, and perform quantum gates that depend on back and forth communication between two nodes while keeping the qubits in local quantum storage. While more applications can be realized with more qubits, this use case differs substantially in that we want to create relatively few (even just one) pairs, but want to store this entanglement. 
%When producing multiple pairs we typically want these pairs to be available at the same time. Since memory lifetimes are short this means that we want to avoid delay producing multiple pairs, because we need to store all previous entangled pairs of qubits until the last one if produced. 
Since we typically want these pairs to be available at the same time, and memory lifetimes are short, we want to avoid delay between producing consecutive pairs, which is superficially similar to constraints in real time classical traffic.
Also for CK, many applications can perform well with noisy entangled links and the amount of noise forms a performance metric (Section~\ref{sec:perf_metrics}).
Examples of such protocols lie in the domain of sensing~\cite{gottesman2012longer}, metrology~\cite{komar2014quantum}, and distributed systems~\cite{ben2005fast, denchev2008distributed} which lie in the quantum memory network stage and beyond~\cite{Wehner2018}.

\emph{Send Qubit (SQ) Use Case:}
While many application protocols known to date consume entanglement itself, some --- such as distributed quantum computing applications --- ask for the transmission of (unknown) qubits. 
This can be realized using teleportation over any distance as long as entanglement is confirmed between the sender and the receiver. For the link layer, this does not differ from CK, where we want to produce one entangled pair per qubit to be sent. 
%Teleportation forms an example where the entanglement must be stored, and we perform gates and a measurement at the sender on the data qubit and entangled link, followed by a quantum gate at the receiver whose choice depends on a classical signal (the measurement outcome) sent by the sender.

\emph{Network Layer (NL) Use Case: } \label{sec:use_cases}
In analogy to the classical notion of a link layer, we 
take the link layer to refer to producing entanglement between neighboring nodes 
(see Section~\ref{sec:stack}).
The network layer will be responsible for producing entanglement between more distant ones. While usage behavior of quantum networks is unknown, it is expected (due to technological limitations) that routing decisions, i.e. how to form long-distance links from pairwise links, will not be entirely dynamic. One potential approach would be to 
%if multiple pairs should be produced end-to-end, then 
first determine a path, and reserve it for some amount of time such that pairwise entanglement can be produced.  Producing pairwise entanglement concurrently enables simultaneous entanglement swapping along the entire path with minimal delay to combat limited memory lifetimes. 
For this, the network layer needs to be capable of prioritizing entanglement production between neighboring nodes.
% in order to facilitate long-distance entanglement generation.   

\subsection{Network Stack}\label{sec:stack}
Based on these considerations, we propose an initial functional allocation of a quantum network stack (see Figure~\ref{fig:stack}). 
In analogy to classical networking, we refer to the lowest element of the stack as the physical layer. This layer is realized by the actual quantum hardware devices and physical connections such as fibers. 
We take the physical layer to contain no decision making elements (comparable to e.g. ISP path tunneling architectures~\cite{peter_sigcomm_2014}) and keep no state about the production of entanglement (or the transmissions of qubits). The hardware at the physical layer is responsible for timing synchronization and other synchronization, such as laser phase stabilization~\cite{Humphreys2018}, required to make attempts to produce heralded entanglement (Section~\ref{sec:NV}). 
A typical realization of the physical layer involves two controllable quantum nodes linked by a (chain of) automated quantum nodes that attempt entanglement production in well defined time slots.
% forms a simple example. 
%Physical layer protocols would typically ask the higher layer whether entanglement production is desired in a specific time slot, and if so with what parameters, in order to avoid keeping
%state themselves.  

The task of the link layer is then to turn the physical layer making entanglement attempts into a robust entanglement generation service, that can produce entanglement between controllable quantum nodes connected by a (chain of) automated quantum nodes. Requests can be made by higher layers to the link layer to produce entanglement,  where robust means that the link layer
endows the physical system with additional guarantees: a request for entanglement generation will (eventually) be fulfilled or result in a 
time-out.
This can be achieved by instructing the physical layer to perform many attempts to produce entanglement until success.
% is achieved. 
%We discuss the focus of this paper - the link layer  - in more detail in Section~\ref{sec:LL}.

Built on top of the link layer rests the network layer, which is responsible
for producing long-distance entanglement between nodes that are not connected directly or by automated quantum nodes.  This may be achieved by means of entanglement swapping, using the link layer to generate entanglement between neighboring controllable nodes. 
%Its interface to the higher layer is currently envisioned to be similar as to the link layer itself 
%(Section~\ref{sec:LL}).
 %but additional functionality, including an entanglement manager. 
In addition, it contains an entanglement manager that keeps track of entanglement in the network,
%, i.e. who the nodes is entangled with, and 
and which may choose to pre-generate entanglement to service later requests from higher layers. 
It is possible that the network layer and entanglement manager may eventually be separated.
% as development progresses.

A transport layer takes responsibility for transmitting qubits deterministically (e.g. using teleportation). One may question why this warrants a separate layer, rather than a 
library.  Use of a dedicated layer allows two nodes to pre-share entanglement that is used as applications of the system demand it.
Here, entanglement is not assigned to one specific application (purpose ID, Section~\ref{sec:reqEnt}). This potentially increases the throughput of qubit transmission via teleportation as teleportation requires no additional connection negotiation, but only forward communication from a sender to the receiver.  Implementing such functionality in a library would incur delays in application behavior as entanglement would need to be generated on-demand rather than supplying it from pre-allocated resources.
  %Given pre-shared entanglement teleportation can be achieved using only forward communication from the sender to the receiver.%, who anyways needs to wait for this communication to process its qubit. 
%\steph{I feel this section could be a bit more clear}

\section{Link Layer Design Considerations}\label{sec:designLinkLayer}

\subsection{Desired Service}
The link layer offers a robust entanglement creation service between a pair of controllable quantum nodes $A$ and $B$ that are connected by a quantum link, which may include
automated nodes along the way. This service 
allows higher layers to operate independently of the underlying hardware platform, depending only on high-level parameters capturing the hardware capabilities.

\subsubsection{Requesting entanglement}\label{sec:reqEnt}
Our use cases bring specific requirements for such a service.
Entanglement creation can be initiated at either $A$ or $B$ by a CREATE request from the higher layer with parameters:
\begin{enumerate}
    \item \emph{Remote node} with whom entanglement generation is desired if the node is connected directly to multiple others.
    \item \emph{Type of request}  - create and keep (K), and create and measure (M). The first type of request (K) stores entanglement, addressing the use cases
	    CK and NL (see Section \ref{sec:use_cases}. The second (M) leads to immediate measurement, supporting the use case MD. The reason for distinguishing these two scenarios is twofold: first, we will show later (Section \ref{sec:phys_ent_gen} that a higher throughput can for some implementations be achieved for M than for K on the same system. Second, simple photonic quantum hardware  without a quantum memory and sophisticated processing  capabilities~\cite{scarani2009security} only supports the M mode of operation.
    \item \emph{Number} of entangled pairs to be created. Allowing the higher layer to request several pairs at once can increase
throughput by avoiding additional processing delays due to increased inter-layer communication (comparing to classical networks~\cite[Table 2]{liu_sigcomm_2014}).
It also helps the CK use case where an application actually needs several pairs concurrently. 
    \item \emph{Atomic} is a flag that assists the requirement from the CK use case, by ensuring that the request must be fulfilled as one, and all pairs be made available at the same time.
% - as opposed to allowing any scheduling decisions at the link layer level to optimize amongst several requests at once. We remark that the size of the nodes quantum memory, also immediately allows a judgement of whether the request can be fullfilled at all.
    \item \emph{Consecutive} is a flag indicating an OK is returned for each pair made for a request (typical for NL use case). Otherwise, an OK is sent only when the entire request is completed (more common in application use cases).
    \item \emph{Waiting time, $t_{\rm max}$} can be used to indicate the maximum time that the higher layer is willing to wait for completion of the request. This allows a general timeout to be set, and 
enables the NL and CK use case to specify strict requirements since the requested pairs may no longer be desirable if they are delivered too late.
% to - for example - perform an entanglement swapping operation.
    \item A \emph{purpose ID} can be specified which allows the higher layer to tag the entanglement for a specific purpose. For an application use case, this purpose ID may be considered analagous to a port number found in the TCP/IP stack and including it in the CREATE request
allows both nodes to immediately pass the entanglement to the right application to proceed with processing at both ends without incurring further communication delays. A purpose ID is also useful to identify entanglement
created by the NL use case for a specific long-distance path. 
We envision that an entanglement manager who may decide to pre-generate entanglement would use a special tag to indicate ``ownership`` of the requested pairs. 
For the NL use case for example, if the entanglement request does not correspond to a pre-agreed path, then the remote node may refuse to engage
in entanglement generation.
Finally, a purpose ID enables rejection of 
generation by the remote node based on scheduling or security policies, e.g. for NL when no path is agreed upon. 
    \item  A \emph{priority} that may be used by a scheduler.
% to prioritze some types of requests. 
Here we use only three priorities (use cases NL, MD and CK).
% but more complex priorities may be used.
    \item Finally, we allow a specification of a purely quantum parameter (refer to Appendix~\ref{app:quantum101}), the \emph{desired minimum fidelity}, $F_{\min}$, of the entanglement~\cite{Nielsen2010}. 
Here, it is sufficient to note that the fidelity $0 \leq F \leq 1$ measures the quality of entanglement, where a higher value of $F$ means higher entanglement quality. The ideal target state has $F=1$,
while $F \geq 1/2$ is often desirable~\cite{horodecki1999reduction}. 
\end{enumerate}
%Using more quantum background, we node that the 
%ideal target state is a maximally entangled state of two qubits, which are all equivalent up to local quantum gates to
%\begin{align}
%\ket{\Psi^+} = \frac{1}{\sqrt{2}}\left(\ket{0}_A \ket{1}_B + \ket{1}_A \ket{0}_B\right)\ . 
%\end{align}
%The fidelity of the state $\rho$ actually produced has fidelity $F = \bra{\Psi^+}\rho\ket{\Psi^+}$ to the target
%state where $F=1$ means we are exactly in the ideal target state. We demand that with high confidence the link layer produces
%states with $F \geq F_{\min}$. \steph{Less detail may be fine}
The reason for allowing different $F_{\rm min}$ instead of fixing one for each hardware platform is that the same platform 
can be used to produce higher or lower fidelity pairs, where a higher fidelity pair costs more time to prepare. Examples of this are the use of entanglement distillation~\cite{dur2007entanglement,Kalb2017} where
two lower quality pairs are combined into one higher quality one, and the bright state population $\alpha$ (see Appendix \ref{app:NV}) used to generate entanglement.
% , and the parameter $\alpha$ (Section~\ref{sec:phys_ent_gen}).
% allowing a direct tradeoff between fidelity and generation rate.

\subsubsection{Response to entanglement requests}\label{sec:ok}
If entanglement has been produced successfully, an OK message should be returned. In addition, the use cases specified in Section \ref{sec:use_cases} desire several other pieces of information, which may also be tracked at higher layer:
%we desire several other pieces of information that higher lyaers can store for each pair. Both the applications and the NL use case, benefit from 
\begin{enumerate}
    \item An entanglement identifier $Ent_{ID}$ unique in the network during the lifetime of the entanglement.
%One way to achieve this is to include the IDs of both nodes involved in generation, and a sequence number. 
This allows both nodes to immediately process the entanglement without requiring an additional round of communication degrading the entanglement due to limited memory lifetimes.
    \item A qubit ID for $K$-type (create and keep) requests which identifies where the local qubit is in the quantum memory device.
    \item The ``Goodness`` $G$, which for $K$ requests is an estimate (Section~\ref{app:testing}) of the fidelity --- where $G \geq F_{\rm min}$ should hold --- and for $M$ an estimate of the quantum bit error rate (QBER, see again Appendix~\ref{app:quantum_101}).
    \item The measurement outcome for $M$ type requests.
    \item The time of entanglement creation.  
    \item The time the goodness parameter was established. The goodness may later be updated given fixed information about the underlying hardware platform. 
\end{enumerate}

Evidently, there are many possibilities of failure resulting in the return of error messages. 
This includes:

\begin{itemize}
    \item Timeout when a request could not be fulfilled in a specific time frame (TIMEOUT).
    \item An immediate rejection of the request because the requested fidelity is not achievable in the given time frame (UNSUPP).
    \item The quantum storage is permanently (MEMEXCEEDED) or temporarily (OUTOFMEM) too small to simultaneously store all pairs of an atomic request.
    \item Refusal by the remote node to participate (DENIED). 
\end{itemize}

Finally, we allow an EXPIRE message to be sent, indicating that the entanglement is no longer available. This in principle can be indicated by a quantum memory manager (see Appendix, Section~\ref{sec:QMM}) instead of the protocol, but we will that show that it allows for recovery from unlikely failures.

\subsubsection{Fixed hardware parameters}
Not included in these request or response messages are parameters that are fixed for the specific hardware platform, or change only very infrequently. As such, these may be obtained by high-level software by querying the low level system periodically, similarly to some classical network architectures (e.g.~\cite{marinos_sigcomm_2014}).
Such parameters include:
\begin{itemize}
    \item The number of available qubits. 
    \item The qubit  memory lifetimes.
    \item Possible quantum operations.
    \item Attainable fidelities and generation time.
    \item The class of states that are produced.
\end{itemize}
The latter refers to the fact that more information than just the fidelity allows optimization at layers above the link layer.
%We will see an example of such a class of states in Section~\ref{sec:NV}.

\subsection{Performance Metrics}\label{sec:perf_metrics}
Before designing any protocols that adhere to these requirements for entanglement generation, we consider the performance metrics that such protocols may wish to optimize. 
Standard metrics from networking
% ~\cite{WHAT?}
also apply here, such as \emph{throughput} (entangled pairs/s), and the \emph{latency}.  We distinguish between:
\begin{enumerate}
    \item Latency per request (time between submission of a CREATE request and its successful completion at a requesting node).
    \item Latency per pair (time between CREATE and OK at requesting node).
    \item Latency per request per number of requested pairs (which we denote as the \emph{scaled latency}).
\end{enumerate}
	    % The latency here refers to the time between submission of a CREATE request and its successful completion at the requesting node\steph{Or the worst case?} (not exceeding $t_{\max}$).
Given requests may originate at both $A$ and $B$, we also demand \emph{fairness}, i.e., the metrics should be roughly independent of the origin of the request.
Here, we also care about genuinely quantum quality metrics, specifically the fidelity $F$ (at least $F_{\rm min}$).
%, but it is possible that a higher fidelity is achieved. 
%In an actual implementation the fidelity is not constant, but may fluctuate in time (see Section~\ref{ref:FEU}).
%Recall that the desired fidelity is also not necessarily fixed for a specific hardware implementation, but some implementations (see e.g. Section~\ref{sec:NV}) offer the choice of a higher throughput at the expense of a lower fidelity.

The non-quantum reader may wonder about the significance of  $F$, and why we do not simply maximize throughput (e.g.~\cite{singh_sigcomm_2018,bremlerbarr_sigcomm_2016}) or minimize latency (e.g.~\cite{dogar_sigcomm_2014,chen_sigcomm_2015}).
% E.g. in QKD, a MD use case we measure the qubits, and are interested in the quantum bit error rate (QBER) between measurement outcomes at $A$ and $B$ (related to $F$, Appendix~\ref{app:quantum_101}). 
For instance, QKD (a MD use case as listed in Section \ref{sec:use_cases}), requires a minimum quantum bit error rate (QBER) between measurement outcomes at $A$ and $B$ (related to $F$, see Appendix~\ref{app:quantum_101}).
A lower $F$ results in a larger QBER, allowing less key to be produced per pair.
We may thus achieve a higher throughput, but a lower number of key bits per second, or key generation may become impossible.
%\steph{plot later in terms of expected error probabilities not correlators}

%If we measure both qubits in three different bases labelled $Z$, 
%(standard basis) 
%$X$,
% (Hadamard basis) 
%and $Y$ (see Appendix), then the error rate is defined as probability of observing unequal measurement outcomes when measuring both qubits in the same
%basis. For a large number $n_Z$ of pairs measured in the $Z$ basis, we thus have
%\begin{align}
%e_Z \approx \frac{\#\{i_A \neq i_B\}}{n_Z}\,  
%\end{align}
%where $i_A$ and $i_B$ correspond to the measurement outcome obtained by measuring pair $i$ in basis $Z$ at node $A$ and $B$ respectively (and analogously for $X$ and $Z$).
%Error rates for $X$ and $Y$ may be defined analogously. 
%For perfect measurements, the fidelity can be directly expressed in terms of the 
%expected number of errors in each of the three bases $\mathbb{E}[Err_Z]$, $\mathbb{E}[Err_X]$, $\mathbb{E}[Err_Y]$ and vice versa (see Appendix). 
%For realistic measurement errors, we will however observe 
%higher error rates. 

\subsection{Error Detection}
% Link layer protocols for classical communication typically aim to correct~\cite{wifiCitations?} or detect errors~\cite{ethernet}, e.g. using a CRC~\cite{ethernetPapers}.
Link layer protocols for classical communication typically aim to correct or detect errors, e.g. using a CRC.
In principle, there exists an exact analogy at the quantum level: We could use a checksum provided by a quantum error correcting code (QECC)~\cite{Nielsen2010,terhal2015quantum} to detect errors. 
This is technologically challenging 
%proposals for using codes during transmission requires X qubits (photons) per Y qubit sent~\cite{XXX}, 
and experimental implementations of QECC are in very early stages~\cite{cramer2016repeated, riste2015detecting, corcoles2015demonstration}.
%This, however, would mean introducing redundancy at the qubit level, which currently is technologically highly challenging, because it requires the simutaneously generation of entanglement between very many qubits (numbers from all photonic repeater paper). Only very small scale QECCs that outperform doing nothing have locally been realized~\cite{XXX}. No code has yet been realized during transmission (CHECK). The closest thing of realizing QECC during transmission to date is probabilistic entanglement distillation~\cite{ronaldDistill} where two pairs are converted into one higher fidelity with a large (specify) probability of failure. This, however, does not enable error detection and we may instead use it to obtain higher quality pairs to begin with.
Yet, apart from technological limitations, future link layer protocols may not use quantum checksums due to different use case requirements:
We typically only demand some minimum fidelity $F_{min}$ with high confidence that may also fluctuate slightly for pairs produced within a time window. 
  
%Of course, one may ask whether we may still wish to use QECC to perform error detection to gain confidence in a minimum fidelity $F$. 
%Yet, apart from technological challenges, there are several reasons why also quantum link layer protocols of the future 
%nevertheless may not use QECC: Instead, they may use a mechanism to 
%ensure high confidence in $F$ for pairs produced within certain time window. 
As we focus primarily on fidelity, we instead use a different mechanism: we intersperse test rounds during entanglement generation (for details, refer to Appendix~\ref{app:testing}) to verify the quality of the link. Such test rounds are easy to produce without the need for complex
gates or extra qubits. 
%since for each test only one pair is generated that is immediately measured, avoiding complex gates and or extra qubits. 
%. This thus avoids complicated gates to create a QECC, and also requires no additional storage qubits. 
Evidently, there exists an exact analogy in the classical networking world, where we would transmit test bits to measure the current quality of transmission, e.g. a direct analogy to network profiling~\cite[Section 4.3]{marinos_sigcomm_2014} to gain confidence that
the non-test bits are also likely to be transmitted with roughly the same amount of error. Yet, there we typically care about correctness of a specific data item, rather than an enabling
resources like entanglement.
% (see Appendix).
 %This of course does not allow the same level of confidence as error detection
%in the quality of transmission.
%Nevertheless, for many quantum protocols an estimate of $F$ is indeed sufficient.
%Since many protocols for our use cases are probabilistic, or make many pairs (especially NL and MD use cases), we often do not require more confidence on the exact quality of 
%a single pair. Indeed, we can pass errors all the way up to the application level (such as for example in QKD~\cite{XX}), where errors are then corrected using classical instead of quantum error correction. Fluctuations in quality are thus often expected at the application level. 
%Finally, we remark that using a QECC for error detection may require an additional exchange
%of communication between the two nodes to convey that an error has occured, which imposes additional demands on memory lifetimes. 

\subsection{Physical Entanglement Generation}\label{sec:phys_ent_gen}\label{sec:NV}
Let us now explain how heralded entanglement generation is actually performed between two controllable nodes $A$ and $B$ (see Appendix~\ref{app:NV} for details). 
As an example, we focus on the hardware platform available to us 
(NV in diamond, Figure~\ref{fig:NV}), but analogous implementations have been performed using remote Ion Traps~\cite{moehring2007entanglement} and Neutral Atoms~\cite{hofmann2012heralded}. 

Nodes $A$ and $B$ are few-qubit quantum processors, capable of storing and manipulating qubits. 
They are connected to an intermediate station called the \emph{heralding station} $H$ over optical fibers. This station is a much simpler automated node, built only from linear optical elements. 
Each node can have two types of qubits: \emph{memory qubits} as a local memory, and 
%Such qubits can be \emph{memory qubits}, yielding in addition a quantum memory. 
\emph{communication qubits} with an optical interface.
% that allows a trigger signal to initiate the probabilistic generation of entanglement between the communication qubit and 
%a traveling qubit encoded as a photon that can be send over an optical fiber. 
%We refer to such qubits as \emph{communication qubits}. 
To produce entanglement, a time synchronized trigger is used at both $A$ and $B$ to create entanglement between each communication qubit, and a corresponding traveling qubit (photon). These photons are sent from $A$ and $B$ to $H$ over fiber. 
When both arrive at $H$, $H$ performs an automatic entanglement swapping operation which succeeds with some probability.
Since $H$ has no quantum memory, both photons must arrive at $H$ at the same time to succeed. Success or failure is then transmitted back from $H$ to the nodes $A$ and $B$ over a standard classical channel (e.g. 100Base-T).
In the case of success, one of several entangled states may be produced, which can however be converted to one other using local quantum gates at $A$ and $B$. 
After a generation attempt, the communication qubit may be moved to a memory qubit, in order to free the communication qubit to produce the next entangled pair.
Many parameters influence the success and quality of this process, such as the quality of the qubits themselves, the probability of emission of a photon given a trigger signal, losses in fiber, and quality of the optical elements such as detectors used at $H$ (Figure~\ref{fig:NV}).

%\edit{
%Many parameters influence the success and quality of this process, such as the quality of the qubits themselves, the probability of emission of a photon given a trigger signal, losses in fiber, quality of the optical elements such as detectors used at the heralding station, and many others.

For further information on this process see~\cite{Humphreys2018}.  For an overview on NV centres in diamond see~\cite{childress_hanson_2013}.
Two different schemes for producing entanglement have been implemented, that differ in how the qubits are encoded into photons (time-bin~\cite{barrett2005efficient}, or presence/absence of a photon~\cite{cabrillo1999creation}).
While physically different, both of these schemes fit into the framework of our physical and link layer protocols.

To evaluate the performance of the protocol (Section~\ref{sec:eval}) and provide intuition of timings, % and its design choices depending on the underlying physical implementation, 
we compare to data from the setup~\cite{Humphreys2018} which uses presence/absence of a photon as encoding. 
A microwave pulse prepares the communication qubit depending on a parameter $\alpha$, followed by a laser pulse to trigger photon emission (total duration $5.5 \mu s$).
A pair ($\ket{\Psi^+}$ or $\ket{\Psi^-}$) is successfully produced with fidelity $F \approx 1-\alpha$ (ignoring memory lifetimes and other errors, see Appendix~\ref{app:NV}) 
with probability $p_{\rm succ} \approx 2 \alpha p_{\rm det}$, where $p_{\rm det} \ll 1$ is the probability of emitting a photon followed by heralding success. 
The parameter $\alpha$ thus allows a trade-off between the rate of generation ($p_{\rm succ}$), and the quality metric $F$. 
For K type requests, we may store the pair in the communication qubit, or move to a memory qubit (duration of $1040 \mu s$ for the qubit considered).
The quality of this qubit degrades as we wait for $H$ to reply.
%(here $F(t) \approx XXX original F$ for comm. qubit, and $F \approx XXX $ for memory qubit).
For M type requests, we may choose to measure immediately before receiving a reply (here readout $3.7\mu s$).
Important is the time of an attempt $t_\mathrm{attempt}$ (time preparing the communication qubit until receiving a reply from $H$, and completion of  any post-processing such as moving to memory),  and the maximum attempt rate $r_\mathrm{attempt}$ (maximum number of attempts that can be performed per second not including waiting for a reply from H or post-processing).
% For M type requests, $r_\mathrm{attempt}$ can in fact be larger than $1/t_\mathrm{attempt}$ if the communication qubit is measured before receiving the reply from $H$ and thus allowing for multiple to overlap.
The rate $r_\mathrm{attempt}$ can be larger than $1/t_\mathrm{attempt}$: (1) for M the communication qubit is measured before receiving the reply from $H$ and thus allows for multiple attempts to overlap and (2) for K, if the reply from $H$ is failure, then no move to memory is done.

%Errors in this process stem from losses in the optical elements ($4\cdot 10^{-4}$~\cite{Humphrey2018}), optional frequency conversion to telecom (0.3~\cite{}), losses in the fiber itself ($<12$ dB without frequency conversion and $<0.5$ with),
% timing jitter causing imperfect interference (mention errors? or just refer appendix error model?), Axel: Matteo said this is not relevant
% losses in the optical elements (value?),
%and detector dark counts causing unwanted detector clicks (20 Hz~\cite{Humphrey2018}).

%     \begin{center}
%         \includegraphics[width=0.3\textwidth]{figures/fid_vs_comm.png}
%     \end{center}
%     \caption{caption}
%     \label{fig:fid_vs_comm}
% \end{wrapfigure}

% \begin{figure}[H]
%     \vspace{-10pt}
%     \begin{center}
%         \includegraphics[width=0.4\textwidth]{figures/fid_vs_comm.png}
%     \end{center}
%     \vspace{-5pt}
%     \caption{\edit{caption}}
%     \label{fig:fid_vs_comm}
%     \vspace{-15pt}
% \end{figure}

For performance evaluation we consider two physical setups as an example (see Appendix~\ref{app:NV}) with additional parameters hereafter referred to as the \Lab\ scenario and the \Qlink\ scenario.
The \Lab\ scenario already realized~\cite{Humphreys2018} with distance to the station $\SI{1}{\meter}$ 
from both $A$ and $B$
(communication delay to $H$ negligible), $p_{\rm succ} \approx \alpha \cdot 10^{-3}$ ($F$ vs. $\alpha$, Figure~\ref{fig:physical-modelling}).
For $M$ requests, we act the same for \Lab\ and \Qlink\ and always measure immediately before parsing the response from $H$ 
to ease comparison (thus $t_\mathrm{attempt} = 1/r_\mathrm{attempt} = 10.12$ $\mu$s which includes electron readout 3.7 $\mu$s, photon emission 5.5 $\mu$s and a 10 \% extra delay to avoid race conditions). For $K$ requests in \Lab, $t_\mathrm{attempt}=1045$ $\mu$s but $1/r_\mathrm{attempt}\approx 11$ $\mu$s as memory qubits need to be periodically initialized (330 $\mu$s every $3500$ $\mu$s).
The \Qlink\ scenario has not been realized and is based on a targeted implementation connecting two European cities by the end of 2020 ($\approx 10km$ from $A$ to $H$ with a communication delay of $48.4 \mu s$ in fiber, and $\approx 15 km$ from $B$ to $H$ with a $72.6 \mu s$ delay).
Frequency conversion of 637nm to 1588nm is performed on the photons emitted in our modeled NV centre while fiber losses at 1588nm are taken to be \SI{0.5}{\decibel/\kilo\meter} (values for deployed \Qlink\ are fibers 0.43-0.47 db/km).
We model the use of optical cavities to enhance photon emission~\cite{riedel2017deterministic, bogdanovic2017design} giving a probability of success $p_{\rm succ} \approx \alpha \cdot 10^{-3}$.
$F$ is worse due to increased communication times from $H$ (Figure~\ref{fig:fid_vs_comm}).
For \Qlink\ $t_\mathrm{attempt} = 145$ $\mu$s for M (trigger, wait for reply from $H$) and $t_\mathrm{attempt} = 1185$ $\mu$s for K (trigger, wait for reply from $H$, swap to carbon). Maximum attempt rates are $1/r_\mathrm{attempt} = 10.120$ $\mu$s (M) and $1/r_\mathrm{attempt} \approx 165$ $\mu$s (K).

% Total time to make an attempt $10 \mu s$ for $M$ (measure: trigger, readout), and $165 \mu s$ for $K$ (store: trigger, wait for reply $H$ and re-initialize memory qubit every $3500 \mu s$. If successful, a move to memory qubit takes additional $XXX \mu s$), see Appendix~\ref{}.

\subsection{Hardware Considerations}\label{sec:designConsiderationsHardware}
Quantum hardware imposes design considerations for any link layer protocol based on top of such experiments.

\emph{Trigger generation:} Entanglement can only be produced if both photons arrive at the heralding station at the same time. This means that the low level system requires tight timing control; such control (ns scale) is also 
required to keep the local qubits stable.
This imposes hard real time constraints at the lowest level, with dedicated timing control (AWG) and software running on a dedicated microcontroller (Adwin ProII). 
When considering a functional allocation between the physical and link layer, this motivates taking all timing synchronization to happen at the physical layer. 
At this layer, we may then also timestamp classical messages traveling to and from $H$, to form an association between classical control information and entangled pairs.
%\steph{In the appendix we should explain the question of two lasers, via one laser and local shutters, and why we like local control}

\emph{Scheduling and flow control:}
Consequently, we make the link layer responsible for all higher level logic, including scheduling, while keeping the physical layer as simple as possible.
An example of scheduling other than priorities,
% of different types of requests, 
is flow control which controls the speed of generation, depending on the availability of memory on the remote node to store such entanglement.

Note that depending on the number of communication qubits, and parallelism of quantum operations that the platforms allows, a node also has a global scheduler for the entire system and not only the actions of the link layer.
%. It may however also take care of other scheduling decisions such as demanded by the network layer, 
%or applications running on the local node. 

%\emph{Limited memory lifetimes:} Since memory lifetimes are presently limited, we would like to avoid any delay due to additional discussion between $A$ and $B$ before the 
%entanglement can be used. This means that prior agreement before communication (entanglement generation), 
%while typically affecting throughput, is beneficial here in order to obtain a high fidelity. Lifetimes also explain our desire for unique entanglement IDs (Section~\ref{XX}) which allow the high layers to process the entangled pairs immediately.

\emph{Noise due to generation: }
One may wonder why one does not continuously trigger entanglement generation locally whenever the node wants a pair, or why one does not continuously produce pairs and then this entanglement is either discarded or otherwise made directly available.
%  to the higher layer. 
%Two reasons motivated by the physical implementation considered (NV in diamond) make this undesirable (except if we were to use the NV as an automated quantum node):
%First, triggering an attempt to produce entanglement causes additional noise on the memory qubits~\cite{noirtbert}. 
In the NV system, triggering entanglement generation causes the memory qubits
to degrade faster~\cite{Reiserer2016, Kalb2018}.
As such we would like to achieve agreement between nodes to avoid triggering unless entanglement it is indeed desired.
% Second, there are only a small number of memory qubits. If we produce entanglement quickly, by for example triggering an attempt and then immediately transfering the state of the communication qubit to the storage qubit if the time of swap operations are favorable (examples), and then proceeding with the next attempt before having heard back from the heralding station, then several storage qubits are needed to support this, making the memory unavailable for other purposes. 

This consideration also yields a security risk: if an attacker could trick a node into triggering entanglement generation, without a matching request on the other side, 
this would allow a rapid destruction of contents of the nodes' local quantum memory. For this reason, we want classical communication to be authenticated which can be achieved using standard methods.

\emph{Memory allocation:}
Decisions on which qubits to use for what purpose lies in the domain of higher level logic, where more information is available.
We let such decisions be taken by a global quantum memory manager (QMM), which can assist the link layer to make a decision on which qubits to employ. It can also translate logical qubit IDs into physical qubit IDs in case multiple qubits are used to redundantly form one logical storage qubit. 

\section{Protocols}
We now present our protocols satisfying the requirements and considerations set forth in Sections~\ref{sec:designArch} 
and~\ref{sec:designLinkLayer}. The entanglement generation protocol (EGP) at the link layer, uses the midpoint heralding protocol (MHP) at the physical layer.
Classical communication is authenticated, and made reliable using 
standard methods (e.g. 802.1AE~\cite{ae}, authentication only).

\subsection{Physical Layer MHP}
Our MHP is a lightweight protocol built directly on top of physical implementations of the form of Section~\ref{sec:phys_ent_gen}, supplementing them
with some additional control information. 
%While we focus on physical implementations of the form given in Section~\ref{sec:heralding} (specifically, the NV platform for concreteness), 
%we note that 
With minor modifications this MHP can be adapted to other forms of heralded entanglement generation between controllable nodes, even using multiple automated middle nodes~\cite{guha2015rate}.

\begin{figure}%[h!]
\centering
%\begin{sequencediagram}
%	\newinst[0]{mema}{EGP $A$}
%	\newinst[1]{a}{MHP $A$}
%	\newinst[1]{mid}{Midpoint $M$}
%
%	\begin{call}{a}{trigger?}{mema}{y/n, info}
%	\end{call}
%
%	\prelevel
%	\mess[1]{a}{{$GEN$, $p$}}{mid}
%	\prelevel
%	\mess[1]{mid}{{$REPLY$}}{a}
%
%	\prelevel
%	\mess{a}{{$f(r_A)$}}{mema}
%\end{sequencediagram}
	\includegraphics[width=0.45\textwidth]{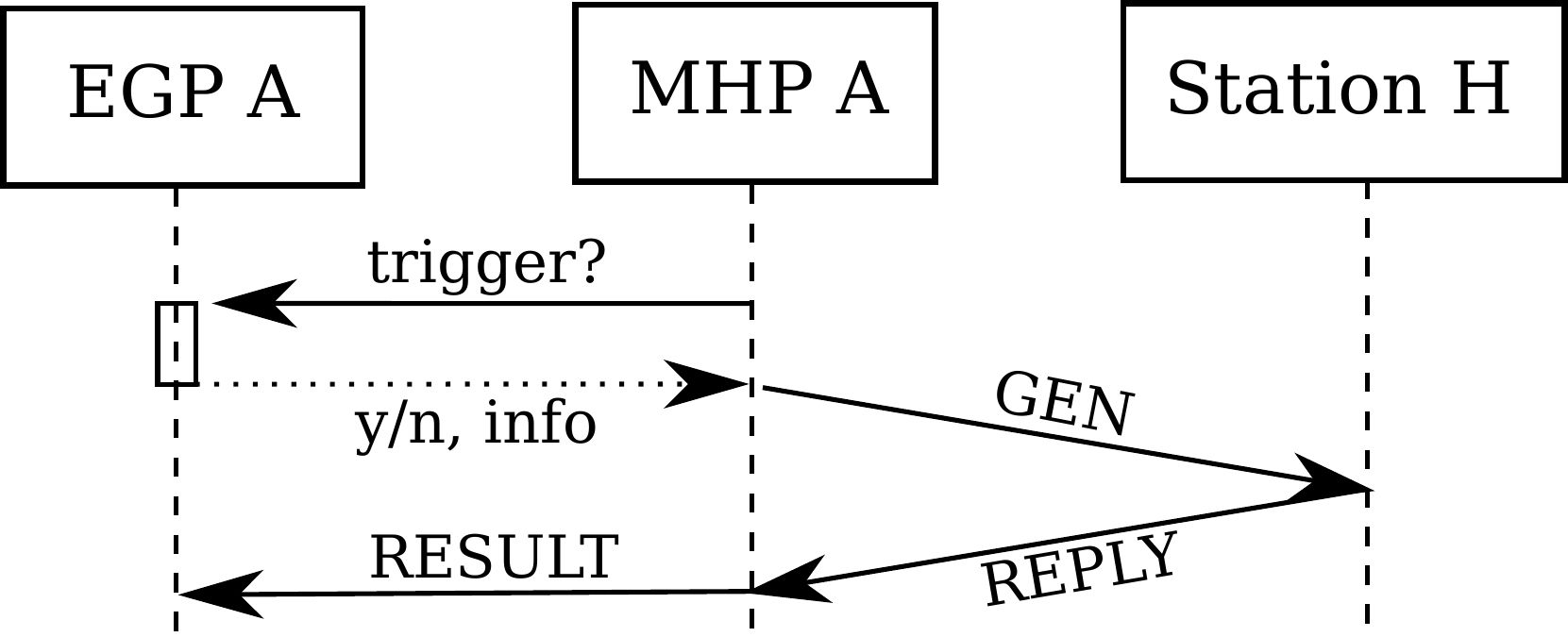}
	\caption{Timeline of the MHP polling higher layers to see if entanglement should be produced.}
	\label{fig:mhp_polling_egp}
    % \vspace{-12pt}
\end{figure}

The MHP is meant to be implemented directly at the lowest level subject to tight timing constraints, which is why we let the MHP poll higher layer (Figure~\ref{fig:mhp_polling_egp}, the link layer EGP) at each timestep to determine whether entanglement generation is required, and keep no state. A batched operation is possible, should the delay incurred by the polling exceed the minimum time to make one entanglement generation attempt - \emph{the MHP cycle} - and hence dominate the throughput.
%This is in contrast to the EGP sending a signal to the MHP to produce a pair. 
%Since the EGP does not deal with timing synchronization, it cannot know when the actual physical trigger should be produced and hence the MHP could then only save the request until pair production is timely. We remark that this would mean that the MHP would need to keep state of how many outstanding triggers there are, which is not desirable as higher layers may decide they do not want to generate entanglement anymore. 
Upon polling, the higher layer may respond ``no`` in which case no attempt will be made or with ``yes``, additionally providing parameters to use in the attempt. These parameters include the type of request (M, measure) or (K, store) passed on from the higher layer, for which the MHP takes the following actions.
%Here we describe two ``yes`` responses that each fulfill a desired use case by the higher layer, namely, the create and keep/measure use cases.

\subsubsection{Protocol for Create and Keep (K)}
The parameters given to the MHP with a ``yes`` response contain the following:
\begin{itemize}
    \item An ID for the attempt that is forwarded to $H$
    \item Generation parameters ($\alpha$, Section~\ref{sec:phys_ent_gen})
    \item The device qubits for storing the entanglement
    \item A sequence of operations to perform on the device memory~\footnote{Less abstractly, by specifying microwave and laser pulse sequences controlling the chip (see Appendix).}.
\end{itemize}
The higher layer may instruct the MHP to perform a gate on the communication qubit
depending on the heralding signal from $H$ allowing the conversion from the $|\Psi^- \rangle$ state to the $|\Psi^+ \rangle$ state. 
%The controllable nodes trigger entanglement generation using the locally specified bright state population and transmit GEN messages along with the photon to $H$.
Entanglement generation is then triggered at the start of the next time interval, including generation parameter $\alpha$, and a GEN message is sent to $H$ which includes a timestamp, and the given ID. The motivation for including the ID is to protect against errors in the classical control, for example losses.  

%The GEN message is a request to the midpoint to attempt establishing entanglement between the controllable node and contains a timestamp along with the ID provided in the ``yes`` response from the higher layer.  
The station $H$ uses the timestamp to link the message to a detection window in which the accompanying photons arrived.
Should messages from both nodes arrive, the midpoint verifies that the IDs transmitted with the GEN messages match, and checks the detection counts (Figure~\ref{fig:NV}) from the corresponding detection window. The midpoint will then send a REPLY message indicating success or failure, and in the case of success which of the two states $\ket{\Psi^+}$ and $\ket{\Psi^-}$ was produced.
%This measurement is used by the nodes to determine whether their qubits form a $|\Psi^+ \rangle$ or $|\Psi^- \rangle$ state.  
The REPLY additionally contains a sequence number uniquely identifying the generated pair of entangled qubits chosen by $H$, 
which later enables the EGP to assign unique entanglement identifiers. This REPLY and the ID is forwarded to the link layer for post-processing.
Note that the REPLY may be received many MHP cycles later, allowing the potential for emission multiplexing (Section~\ref{sec:egp}).
% and for future nodes with many memory qubits not arrive in order.

\subsubsection{Protocol for Create and Measure (M)}
Handling M type requests is very similar,  differing only in two ways: Instead of performing a gate on the communication qubit, the ``yes`` message requests the MHP to perform a measurement on the communication qubit in a specified basis once the photon has been emitted, even before receiving the response from $H$ . The outcome of the measurement and the REPLY are passed back to the EGP.
In practice, the communication time from transmitting a GEN message to receiving a REPLY may currently exceed the duration of such a local measurement (3.7 $\mu s$ vs. 
communication delay \Lab\ 9.7 ns, and \Qlink\ 145 $\mu$s), and the MHP may choose to perform the measurement only after a successful response is received.

\subsection{Link Layer EGP}\label{sec:egp}
\label{sec:protoErrors}
Here we present an implementation of a link layer protocol, dubbed EGP, satisfying the service requirements put forth in Section~\ref{sec:designLinkLayer} (see Appendix~\ref{app:protocol} for details and message formats). 
We build up this protocol from different components:

\subsubsection{Distributed Queue}
Both nodes that wish to establish entangled link(s) must trigger their MHP devices in a coordinated fashion (Section~\ref{sec:phys_ent_gen}).  
%Thus, it is important that both nodes agree to create an established link.  
To achieve such agreement, the EGP employs a distributed queue comprised of synchronized local queues at the controllable nodes. These local queues can be used to 
separate requests based on priority, where here we employ 3 for the different use cases (CK, NL, MD). Due to low errors in classical communication (estimated $<4\times10^{-8}$ on \Qlink, see Appendix~\ref{app:classical})), we let one node hold the master copy of the queue, and use a simple two-way handshake for enqueing items, and a windowing mechanism to ensure fairness. Queue items include a $min\_time$ that specifies the earliest possible time a request is deemed ready for processing by both nodes (depending on their distance). Specifying $min\_time$ prevents either node from beginning entanglement generation in different timesteps. 
%The remote node may request additions to the queue, based on the policy given the purpose ID.

\subsubsection{Quantum Memory Management (QMM)}\label{sec:QMM}
The EGP uses the node's QMM (Section~\ref{sec:designConsiderationsHardware}) to determine which physical qubits to use for generating or storing entanglement.
%must be able to provide the MHP information, which qubits in the communication device should be used for generating entanglement.  While not part of the actual protocol, the QMM of the node (Section~\ref{sec:hardwareReq}) is be asked to (smartly) allocate a pairs of qubits to used for the MHP.
% as well as convert logical qubit IDs to physical Qubit IDs and vice versa.

\subsubsection{Fidelity estimation unit (FEU)}\label{sec:feu}
In order to provide information about the quality of entanglement, the EGP employs a fidelity estimation unit. This unit is given a desired quality parameter $F_{\rm min}$, 
and returns generation parameters (such as $\alpha$) along with an estimated minimal completion time. 
%The EGP rejects the CREATE request immediately, if the minimal required time exceeds 
%$t_{\rm max}$.
Such a fidelity estimate can be calculated based on known hardware capabilities such as the quality of the memory and operations.
To further improve this base estimate the EGP intersperses test rounds. 
%In such a round, a test pair is produced and measured to obtain a fidelity estimate (see Appendix~\ref{app:feu}).

%\subsubsection{Resource Management Unit (RMU)}
%\steph{Let's discuss how we can formulate this}
%A K type request demanding storage, can only be fullfilled if both nodes have available quantum memory. The RMU engages in periodic discussions between $A$ and $B$ to update each other on the demands on their local quantum storage, to cut down on unncessary triggering of entanglement generation (Section~\ref{sec:hardwareLL}) if no memory is available at the remote node. This information is held locally, avoiding additional communication rounds for every request.
%Evidently, one approach would be to immediately 
%let the local node reject requests to produce entanglement when submitted, or the remote node when adding them to the queue. Since we allow many items in the queue, memory may however be available by the time this request can be served. The RMU ensures both nodes periodically update each other on the state of their local quantum storage, and stores such information to make an estimate of whether the remote node is capable of producing a pair. This cuts down on unnecessary triggering of generation, which can adversely affect the nodes local quantum memory (Section~\ref{sec:hardwareLL}).
%\steph{It is possible here to produce out of sync triggers}

\subsubsection{Scheduler}
The EGP scheduler decides which request in the queue should be served next. In principle, any scheduling strategy is suitable 
as long as it is deterministic, ensuring that both nodes select the same request locally. This limits two-way communication, which 
adversely affects entanglement quality due to limited memory lifetimes.
%If the nodes would assign entanglement created to different requests, then identifiers passed to higher layers locally may be forwarded to different applications and knowledge of the entangled link would not be consistent. The EGP scheduler tracks the number of outstanding pairs left to be generated for a given request. We remark that despite initial investigations~\ref{sec:scheduling}, determining
%a scheduling strategy that is globally (i.e, also at the network layer) efficient is outside the scope of this work. We fully expect that in the future a division will be made between application and network scheduling, and that a global scheduler at each node will assist the link layer and network layer, and hence the EGP scheduler will interact with a larger scheduling system. 
%\steph{I dont think we are very clear here. If there is also a local scheduler then the two schedulers may not act consistently}

\subsubsection{Protocol}
\begin{figure}
% \vspace{-12pt}
    \centering
    \includegraphics[width=\linewidth]{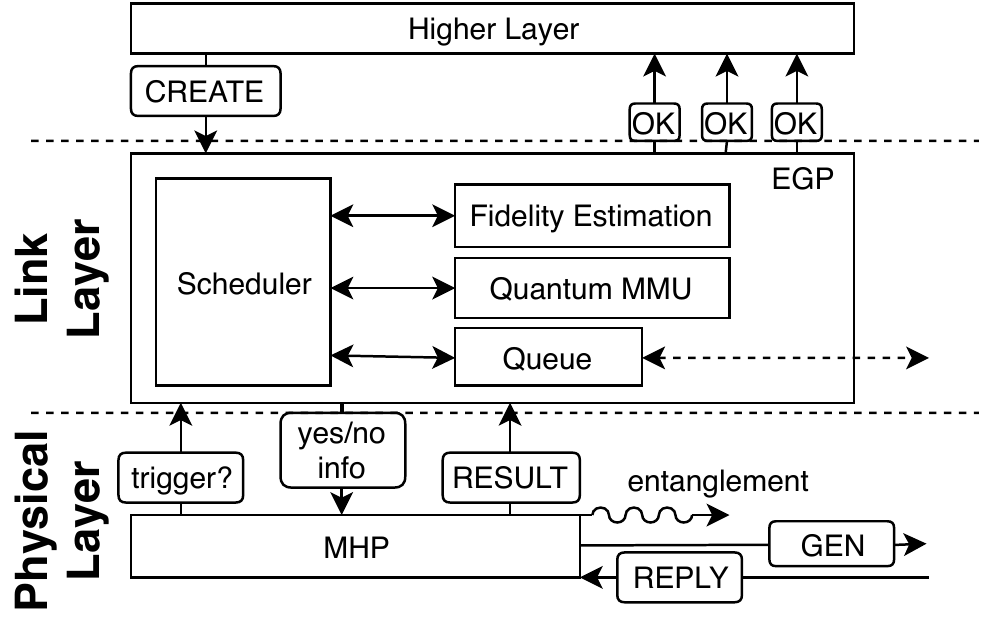}
    % \vspace{-10pt}
    \caption{Flow diagram of the MHP and EGP operation.  The EGP handles CREATE requests and schedules entanglement generation attempts are issued to the MHP.  Replies from the midpoint are parsed and forwarded to the EGP from request management.}
    \label{fig:egp_architecture}
% \vspace{-12pt}
\end{figure}

Figure \ref{fig:egp_architecture} presents an architecture diagram visualizing 
the operation.  The protocol begins when a higher layer at a controllable node issues a CREATE operation to the EGP specifying a desired number of entangled pairs along with $F_{min}$ and $t_{max}$.  Upon receipt of a request the EGP will 
query the FEU to obtain hardware parameters ($\alpha$), and a minimum completion time. If this time is larger than $t_{\rm max}$, the EGP immediately rejects the request (UNSUPP).
Should the request pass this evaluation, the local node will compute a fitting $min\_time$ specifying the earliest MHP polling cycle the request may begin processing. 
The node then adds the request into the distributed queue shared by the nodes. This request may be rejected by the peer should the remote node have queue rules that do not accept the specified purpose ID.  Then, the EGP locally rejects the request (DENIED).
%\steph{I dont understand why you want to do this before adding it to the queue now?}

The local scheduler selects the next request to be processed, given that there exists a ready one (as indicated by $min\_time$).
% If no response to the MHP is current awaiting polling, the local scheduler selects the next ready (as indicated by $min\_time$) request to process.
%When a request is selected (and ready by the specified $min\_time$) the scheduler will consult the local RMU 
%to determine if both nodes are capable of proceeding with an attempt at entanglement. \steph{lets be careful and discuss}
The QMM is then used to allocate qubits needed to fulfill the specified request type (create and keep K or create and measure M).  
%A state population is also selected to use for photon emission that satisfies the desired $F_{min}$ and has high probability of fulfilling the full request within the specified $t_{max}$.  
% If there is a ready request, 
The EGP will then again ask the FEU to obtain a current parameter $\alpha$ due to possible fluctuations in hardware parameters during the time spent in the queue. 
The scheduler then constructs a ``yes'' response to the MHP containing $\alpha$ from the FEU, along with an ID containing the unique queue ID of the request in the 
distributed queue, and number of pairs already produced for the request.
%\edit{and the number of pairs generated for this request thus far? $(qid, qseq, i)$}.  
%\steph{I dont understand your comment}
%This queue ID fulfills the roll of the attempt ID in the MHP and allows $H$ to check if the nodes are fulfilling entanglement for the same request. 
This response is then forwarded to the local MHP upon its next poll to the EGP. If no request is ready for processing, a ``no'' response is returned to the MHP .
%, else a defaui. Should no request be available for processing the scheduler will construct a default response 
%that instructs the MHP to not trigger. 
%Upon a poll from the MHP the EGP conveys the constructed response to the MHP.  
At this point the MHP behaves as described in the previous section and an attempt at generating entanglement is made. 

Whenever a REPLY and ID is received from the MHP, the EGP uses the ID to match the REPLY to an outstanding request, and evaluates the REPLY for correctness.
Should the attempt be successful, the number of pairs remaining to be generated for the request is decremented and an OK message is propagated to higher layers containing the information specified in Section~\ref{sec:ok}, where the Goodness is obtained from the FEU.
%\steph{we have now asked the FEU three times. shall we streamline? }

In the Appendix, we consider a number of examples to illustrate decisions and possible pitfalls in the EGP.
An example is the possibility of
\emph{emission multiplexing~\cite{van2017multiplexed}:} The EGP can be polled by the MHP before receiving a response from the MHP for the previous cycle. 
This allows the choice to attempt entanglement generation multiple times in succession before receiving a reply from the midpoint, e.g., in order to increase the throughput for the MD use case.  
%If no memory qubits are in use, there is no harm in causing additional noise to them due to attempting entanglement generation.
Errors such as losses on the classical control link can lead to an inconsistency of state (of the distributed queue) at $A$ and $B$ from which we need to recover.
Inconsistencies can also affect the higher layer. Since the probability of e.g. losses is extremely low, we choose not
to perform additional two way discussion to further curb all inconsistencies at the link layer. Instead, the EGP can issue an EXPIRE message 
for an OK already issued if inconsistency is detected later, e.g.  when the remote node never received an OK for this pair.  

\section{Evaluation}\label{sec:eval}
We investigate the performance of our link layer protocol using a purpose built discrete event simulator for quantum networks (NetSquid~\cite{netsquid}, Python/C++) based on DynAA~\cite{dynaa1} (see Appendix~\ref{sec:app-simulation} for details and more simulation results). Both the MHP and EGP are implemented in full in Python running on simulated nodes
that have simulated versions of the quantum hardware components, fiber connections, etc.
All simulations were performed on the supercomputer \textit{Cartesius} at SURFsara~\cite{surfsara}, 
in a total of 2578 separate runs using a total of 94244 core hours, and 
707 hours time in the simulation ($\sim$250 billion MHP cycles).

We conduct the following simulation runs:
\begin{itemize}
    \item Long runs: To study robustness of our protocol, we simulate the 169 scenarios described below for an extended period of time. Each scenario was simulated twice for 120 wall time hours, simulating  $502 - 13437$ seconds. We present and analyze the data from these runs in sections~\ref{sec:robustness},~\ref{sec:performance_metrics} and~\ref{app:simulation_data}.
    \item Short runs: We perform the following simulations for a shorter period of time (24 wall time hours, reaching $67 - 2356$ simulated seconds):
        \begin{itemize}
            \item Performance trade-offs: To study the trade-off between latency, throughput and fidelity we sweep the incoming request frequency and the requested minimum fidelity, see Figure~\ref{fig:combined_fid_req_freq}.
            \item Metric fluctuations: To be able to study the impact of different scheduling strategies on the performance metrics, we run 4 scenarios, 102 times each. The outcomes of theses simulation runs are discussed in section~\ref{sec:simSchedule}.
        \end{itemize}
\end{itemize}

To explore the performance at both short and long distances, the underlying hardware model is based on the \Lab\ and \Qlink\ scenarios, 
where we validate the physical modeling 
of the simulation against data collected from the quantum hardware system of the \Lab\ scenario already realized (Figure~\ref{fig:validate}).
For the quantum reader we note that while our simulations can also be used to predict behavior of physical implementations (such as \Qlink), the focus here is on the performance and behavior of the link layer protocol. 
%This includes a first investigation of how different scheduling strategies can affect traditional performance metrics (such as throughput) in relation of genuinely quantum ones (the fidelity) for different use cases.

We structure the evaluation along the three different use cases (NL, CK, MD), 
leading to a total of 169 different simulation scenarios.
First, we use different 
kinds of requests: (1) \textit{NL} (K type request, consecutive flag, priority 1 (highest), store qubit in memory qubit), (2) 
\textit{CK}, an application asking for one or more long-lived pairs (K type request, immediate return flag, priority 2 (high), store qubit in memory qubit)
and (3) (\textit{MD}) measuring directly (M type request, consecutive flag, priority 3 (lowest)). For an application such as QKD, one would not set the immediate return flag in practice for efficiency, but we do so here to ease comparison to the other two scenarios. Measurements in $\textit{MD}$ are performed in randomly chosen bases $X$, $Z$ and $Y$ (see Appendix~\ref{app:quantum_101}).

% We conduct $2\times 169$ long runs to study long term behaviour ($2 \times 169 \times 120$ core hours, $502 - 13437$ s simulated time for each scenario), and 1280 short runs with $93 - 2355$ s simulated time studying the tradeoff between latency, throughput and fidelity. 

In each MHP cycle, we randomly issue a new CREATE request for a random number of pairs $k$ (max $k_\mathrm{max}$), and random kind $P\in\{\textit{NL},\textit{CK},\textit{MD}\}$ with probability $f_P \cdot p_{\rm succ}/(E \cdot k)$, where $p_{\rm succ}$ is the probability of one attempt succeeding (Section~\ref{sec:NV}), $f_P$ is a fraction determining the load in our system of kind $P$, and $E$ is the expected number of MHP cycles to make one attempt ($E=1$ for MD and $E\approx 1.1$ for NL/CK in \Lab\ due to memory re-initialization and post-processing). $E \approx 16$ for NL/CK in \Qlink\ due to classical communication delays with $H$ ($145 \mu s$)).
In the long runs, we first study single kinds of requests (only one of MD/CK/NL), with $f_P = 0.7$ (\textproc{Low}), $0.99$ (\textproc{High}) or $1.5$ (\textproc{Ultra}).
For the long runs, we fix one target fidelity $F_{\rm min} = 0.64$ to ease comparison.
For each of the 3 kinds (MD/CK/NL), we examine (1) $k_\mathrm{max}=1$, (2) $k_\mathrm{max}=3$, and (3) only for \textit{MD}, $k_\mathrm{max}=255$.
For \textproc{Ultra} the number of outstanding requests intentionally grows until the queue is full (max 256), to study an overload of our system.
To study fairness, we take 3 cases of CREATE origin for each single kind (MD/CK/NL) scenario: (1) all from $A$ (master of the distributed queue), (2) all from $B$, (3) $A$ or $B$ are randomly chosen with equal probability. 
To examine scheduling, we additionally consider long runs with mixed kinds of requests (Appendix, e.g. Figure~\ref{fig:latency_vs_time}).

% We will denote scenarios with only a single type of request by \textit{phys}\_\textit{type}\_\textit{freq}\_\textit{node}, where \textit{phys} is the physical setup, \textit{type} is the request type, \textit{freq} is the request frequency and \textit{node} is the origin node of the request, for example \Lab\_\textit{MD}\_\textproc{High}\_AB.
% For mixed scenarios, \textit{type} and \textit{freq} is replaced by the name in table~\ref{tab:req_mixes} and the type of scheduling will be specified last, for example \Qlink\_\textproc{Uniform}\_\textproc{FCFS}.

\subsection{Robustness}\label{sec:robustness}
To study robustness, we artificially increase the probability of losing classical control messages (100 Base T on $\Qlink$ fiber $< 4\times10^{-8}$, Appendix~\ref{app:classical}), which can lead to an inconsistency of state of the EGP but also at higher layers (Section~\ref{sec:protoErrors}). We ramp up loss probabilities up to $10^{-4}$ (Appendix~\ref{app:losses})
and observe our recovery mechanisms work to ensure stable execution in all cases (35 runs, 281 - 3973 s simulated time), 
with only small impact to the performance parameters (maximum relative differences~\footnote{Relative difference between $m_1$ and $m_2$ is $\abs{m_1-m_2}/\max(\abs{m_1},\abs{m_2})$} to the case of no losses, fidelity (0.005), throughput (0.027), latency (0.629), number of OKs (0.026) with no EXPIRE messages).
We see a relatively large difference for latency, which may however
be due to latency not reaching steady state in the course of this simulation 
($70 \times 70$ core hours).

\subsection{Performance Metrics}\label{sec:performance_metrics}
\begin{figure}
    % \vspace{-8pt}
    \begin{center}
        \includegraphics[width=0.5\textwidth]{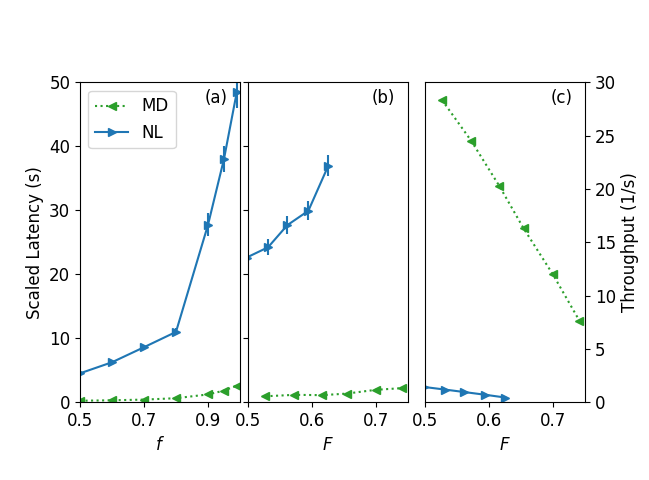}
    \end{center}
    % \vspace{-10pt}
    \caption{Performance trade-offs. (a) Scaled latency vs. $f_P$ determining fraction of throughput 
	(b) Scaled latency vs. fidelity $F_{min}$. Demanding a higher $F_{min}$ lowers the probability of an attempt being successful (Section~\ref{sec:NV}), meaning (c) throughput directly scales with $F_{min}$ (each point averaged over $40$ short runs each $24$ h, $93-2355$ s simulated time, \Qlink, $k_\mathrm{max}=3$, for (b,c) $f_P=0.99$). Higher $F_{min}$ not satisfiable for NL in (b).}
    \label{fig:combined_fid_req_freq}
    % \vspace{-12pt}
\end{figure}

\begin{figure}[h!]
	    % \vspace{-12pt}
	    \begin{center}
		            \includegraphics[width=0.5\textwidth]{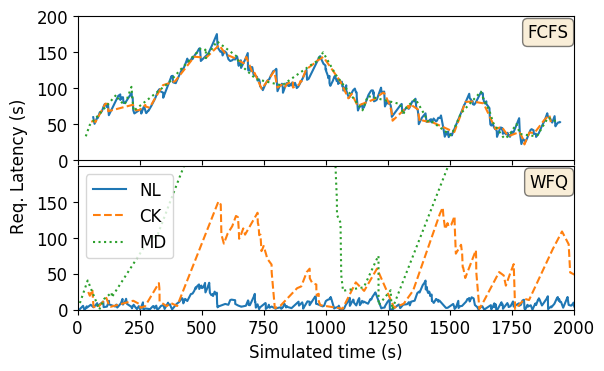}
			        \end{center}
				    % \vspace{-10pt}
				        \caption{Request latency vs. time for two scheduling scenarios (long runs simulated $120$ h wall time). As expected the max. latency for \textit{NL} is decreased due to strict priority. In this scenario, there are more incoming \textit{NL} requests ($f_\textit{NL}= 0.99\cdot4/5$ ,  $f_\textit{CK}=0.99\cdot1/5$ and  $f_\textit{MD}=0.99\cdot1/5$).}
					    \label{fig:latency_vs_time}
					        % \vspace{-5pt}
\end{figure}

\begin{figure}[h!]
		\includegraphics[width=0.45\textwidth]{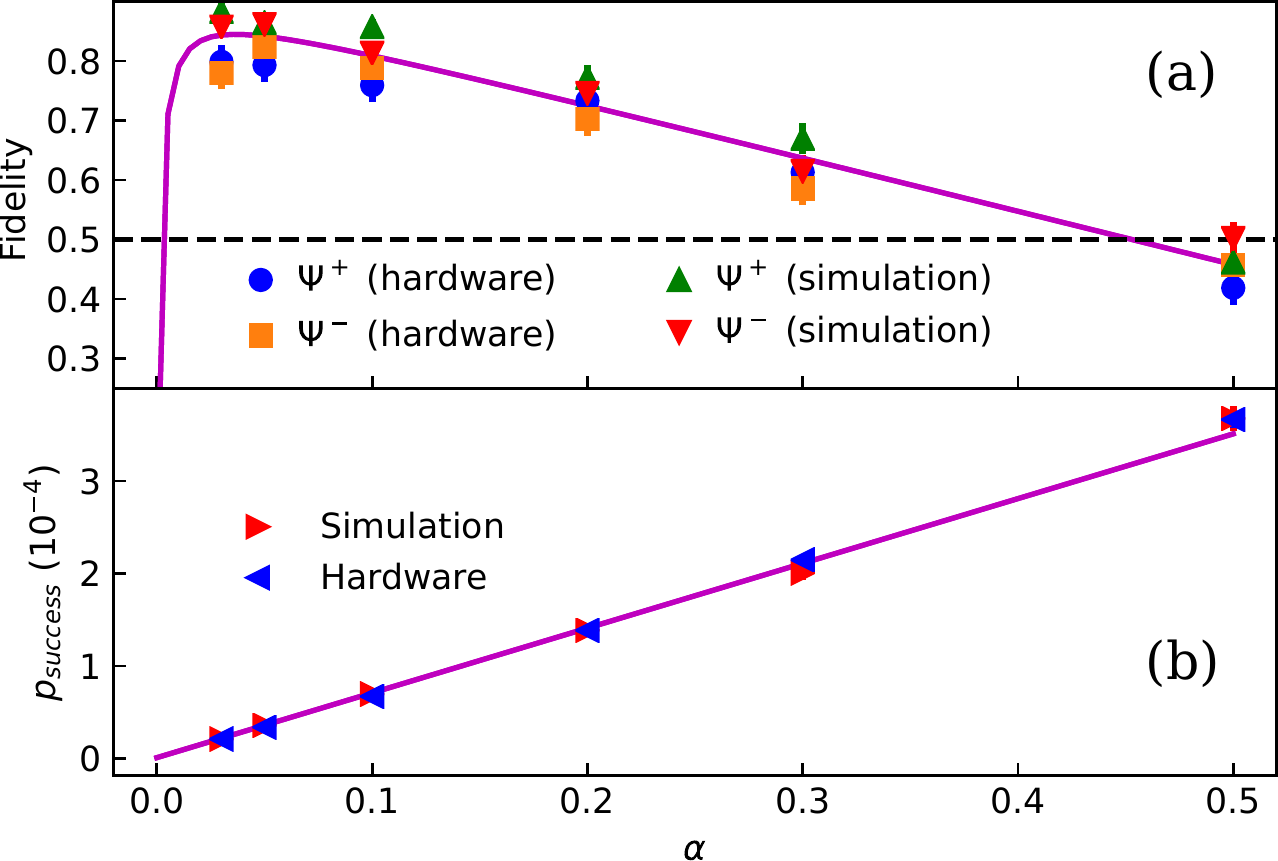}
		    % \vspace{-10pt}
		    	\caption{Validation against data from NV hardware (\Lab\ scenario). Fidelity (a) and probability an attempt succeeds (b) in terms of $\alpha$ (Section~\ref{sec:NV}) shows good agreement between hardware and simulation points (each at least $300$ pairs averaged, $5$s$-117$s simulated time, $500$k$-10.000$k attempts, $122$ hours wall time). Theoretical model~\cite{Humphreys2018} as visual guide (solid line).}\label{fig:validate}

			%of sufficiently close modelling of the the NV-center physics by comparison to experimental results from \cite{Humphreys2018}, in which heralded entanglement is generated and directly measured with the \Lab{ }scenario setup (i.e. nodes $2$m apart). 
			%The figure shows (a) fidelities and (b) probability that a single entanglement generation attempt succeeds, both as a function of $\alpha$ which can be freely set in the NV-center setup. Solid line refers to the theoretical model from \cite{Humphreys2018}. Each simulation data point was computed by averaging over at least 600 links, versus \textcolor{red}{XXX} links for the experimental data. Error bars indicate: (a) $\sigma(F) / \sqrt{N}$, where $\sigma(F)$ is the standard deviation of the fidelity and $N$ the number of data points that was averaged over, and (b)$\mbox{ }\sigma(A) / \left(\bar{A} ^ 2 \cdot \sqrt{N}\right)$, which equals the relative standard deviation with respect to $p_{success} = 1/\bar{A}$ where $\bar{A}$ denotes the average number of attempts until a successful heralding signal. The index 0/1 (legend) refers to the heralding detector. The fact that simulation fidelities are slightly higher than the experimental data indicates the presence to undetermined sources of noise in the experiments, as already noted in \cite{Humphreys2018}; these were thus not included in the simulated models. See Appendix~\ref{sec:app-simulation} for more details.
				 \label{fig:physical-modelling}
				     % \vspace{-10pt}
\end{figure}

We first consider runs including only a single kind of request (MD/CK/NL). In addition to the long runs, we conduct specific short runs examining the trade-off between latency and throughput for fixed target fidelity $F_{\min}$ (Figure~\ref{fig:combined_fid_req_freq}(a)), and the trade-off between latency (throughput) and the target fidelity in Figure~\ref{fig:combined_fid_req_freq}(b) (Figure~\ref{fig:combined_fid_req_freq}(c)).

Below we present the metrics extracted from the long runs with single kinds of requests:

\textit{Fidelity:} As a benchmark, we began by recording the average fidelity $F_\mathrm{avg}$ in all 169 scenarios with fixed minimum fidelity. We observe that $F_\mathrm{avg}$ is independent of the other metrics but does depend on the distance, and whether we store 
or measure: $0.745<F_\mathrm{avg}<0.757$ \textit{NL}/\textit{CK} \Lab, $0.626<F_\mathrm{avg}<0.653$ \textit{NL}/\textit{CK} \Qlink, $0.709<F_\mathrm{avg}<0.779$ \textit{MD} \Lab, $0.723<F_\mathrm{avg}<0.767$ \textit{MD} 
\Qlink\ (Fidelity \textit{MD} extracted from QBER measurements, Appendix~\ref{app:quantum_101}). This is to be expected since (1) we fix one $F_{\rm min}$ and (2) we consider an NV platform with only 1 available memory qubit (\Lab).
% We expect that for nodes with a larger memory, memory lifetimes and the effect of triggering on memory quality (Section~\ref{sec:phys_ent_gen}), we will in laster see an impact on fidelity and such metrics.

\textit{Throughput:} All scenarios \textproc{High} and \textproc{Ultra} in \Lab\ reach an average throughput $th_\mathrm{avg}$ (1/s) of $6.05<th_\mathrm{avg}<6.47$ \textit{NL}/\textit{CK} and $6.51<th_\mathrm{avg}<7.09$ for \textit{MD}. It is expected that \textit{MD} has higher throughput, since no memory qubit needs to be initialized. The time to move to memory ($1040 \mu s$) is less significant since many MHP cycles are needed to produce one pair, but we only move once. 
As expected for \textproc{Low} the throughput is slightly lower in all cases, $4.44<th_\mathrm{avg}<4.72$ \textit{NL}/\textit{CK}, and $4.86<th_\mathrm{avg}<5.22$ \textit{MD}.
For \Qlink, the throughput for \textit{NL}/\textit{CK} is about 14 times lower, since we need to wait ($145 \mu s$) for a reply from $H$ before MHP can make a new attempt.
%starting a new generation attempt (photon emission and measurement takes a bit over \SI{10}{\micro\second} and communication round trip time to the midpoint from the furthest node is about \SI{145}{\micro\second}).\axel{Check that we state these numbers also earlier}

\textit{Latency:}
% We examine (1) latency per request (time from CREATE until total request completed), (2) latency per pair (time from CREATE until OK) and (3) latency per request divided by number of requested pairs. 
% Here, we focus on (3), \emph{scaled latency}.
The scaled latency highly depends on the incoming request frequency as the queue length causes higher latency.
However, from running the same scenarios many (102) times for a shorter period (24 wall time hours), see section~\ref{sec:simSchedule}, we see that the average scaled latency fluctuates a lot, with a standard deviation of up to 6.6 s in some cases.
For \Qlink\ with \textit{NL} requests specifying 1-3 pairs from both nodes, we observe an average scaled latency of 10.97 s \textproc{Low}, 142.9 s \textproc{High} and 521.5 s \textproc{Ultra}.
For \textit{MD} requests, 0.544 s \textproc{Low}, 3.318 s \textproc{High} and 32.34 s \textproc{Ultra}.
The longer scaled latency for \textit{NL} is largely due to longer time to fulfill a request, and not that the queues are longer (average queue length for \textit{NL}: 3.83 \textproc{Low}, 56.3 \textproc{High}, 
214 \textproc{Ultra}), and for \textit{MD}: 3.23 \textproc{Low}, 22.4 \textproc{High} and 219 \textproc{Ultra}).

\textit{Fairness:} For 103 scenarios of the long runs (single kinds of requests (MD/CK/NL) randomly from $A$ and $B$), we see only slight differences 
in fidelity, throughput or latency between requests from $A$ and $B$.
Maximum relative differences do not exceed: fidelity 0.033, throughput 0.100, latency 0.073, number of OKs 0.100 (which occurs for \textproc{Ultra}).
% We find this to be 0.033, 0.100, 0.073 and 0.100 for fidelity, throughput, latency and number of OKs respectively.
% \axel{Say something about "even in \textproc{Ultra}"?}

\subsection{Scheduling}\label{sec:simSchedule}

% \axel{The motivation for WFQ can maybe be made shorter if it's moved to an earlier section but I think we need it still}
%We take a first step in investigating the impact on the metrics above when using different scheduling strategies for the incoming requests of entanglement.
We take a first step studying the effect of scheduling strategies on the performance when using mixed kinds of requests. 
Part of simulating the performance of a scheduling strategies can certainly be done without implementing all details of 
the physical entanglement generation. However, since we do simulate these details we can first 
confirm that different scheduling strategies below do not affect the average fidelity in these scenarios.
Here, we examine two simple scheduling strategies: (1) first-come-first-serve (\textproc{FCFS}) and (2)
a strategy where \textit{NL} of priority 1 has a strict highest priority, and use a weighted fair queue (WFQ) 
for \textit{CK} (priority 2) and \textit{MD} (priority 3), where \textit{CK} has 10 times the weight of \textit{MD}.
With these scheduling strategies, we simulate two different request patterns ((i) uniform and (ii) no \textit{NL} more \textit{MD}), 102 times over 24 wall time hours each and extract the performance metrics of throughput and scaled request latency, see table~\ref{tab:many_runs}.
%Imagine for instance that the network layer decides to generate entanglement over a chain of links across the network in order to generate long distance entanglement between non-adjacent end-nodes.
%Each of these links are running link layer protocols which are unaware of each other.
%On the other hand adjacent links need to generate entanglement close in time, since stored entanglement have a finite lifetime due to noisy quantum memory storage of qubits.
%Thus, it is of importance that the network layer protocol can submit requests to the link layer protocol which are serviced as fast as possible in order to generate long distance entanglement.
% For \textproc{WFQ} we use two different sets of weights for the \textit{CK} and \textit{MD} queues: (1) \textit{CK} has twice the weight of \textit{MD} (\textproc{LowerWFQ}), (2) \textit{CK} has 10 times the weight of \textit{MD} (\textproc{HigherWFQ}).

As expected we see a drastic decrease of the average scaled latency for \textit{NL} when giving it strict priority: 10.3 s with \textproc{FCFS} and 3.5 s with \textproc{WFQ}.
For \textit{CK} there is similarly a decrease in average scaled latency, however smaller than for \textit{NL}, 10.1 s and 6.5 s for \textproc{FCFS} and \textproc{WFQ} respectively.
For \textit{MD} the average scaled latency goes up in both cases when using \textproc{WFQ} instead of \textproc{FCFS}, by factors of $2.49$ (uniform) and $1.28$ (no \textit{NL} more \textit{MD}).

We observe that the throughput gets less affected by the scheduling strategy than the latency for these scenarios.
The maximal difference between the throughput for \textproc{FCFS} and \textproc{WFQ} is by a factor of $1.16$ (for \textit{MD} in the scenario of no \textit{NL} and more \textit{MD}).
Furthermore, we see that the total throughput for all requests goes down from $2.75$ ($5.99$) 1/s for \textproc{FCFS} to $2.44$ ($5.92$) 1/s for \textproc{WFQ} in the case of uniform (no \textit{NL} more \textit{MD}).
% For \textit{CK} we see a slightly higher scaled latency (63.9 s) with \textproc{LowerWFQ}, but drastically lower (8.22 s) with \textproc{HigherWFQ} compared to (50.9 s) \textproc{FCFS}.

\begin{table}
    \caption{Throughput (T) and scaled latency (SL) using scheduling strategies \textproc{FCFS} and \textproc{WFQ} for two request patterns: (i) with $f_\textit{NL}=f_\textit{CK}=f_\textit{MD}=0.99\cdot \sfrac{1}{3}$, i.e. a uniform load of the different priorities and (ii) with $f_\textit{NL}=0$, $f_\textit{CK}=0.99\cdot \sfrac{1}{5}$ and $f_\textit{MD}=0.99\cdot \sfrac{4}{5}$, i.e. no \textit{NL} and more \textit{MD}.
    The physical setup: \Qlink\ and number of pairs per request: $2$ (\textit{NL}), $2$ (\textit{CK}), and $10$ (\textit{MD}).
    Each value average over $102$ short runs each $24$ h, with standard error in parentheses.
}
    \label{tab:many_runs}
    \begin{center}
        \begin{tabular}{|l|ccc|}
            \hline
            T (1/s)              & \textit{NL}   & \textit{CK}   & \textit{MD}   \\ \hline
            (i) \textproc{FCFS}  & 0.146 (0.003) & 0.144 (0.003) & 2.464 (0.056) \\ \hline
            (i) \textproc{WFQ}   & 0.154 (0.003) & 0.156 (0.003) & 2.130 (0.063) \\ \hline
            (ii) \textproc{FCFS} & -             & 0.086 (0.003) & 5.912 (0.033) \\ \hline
            (ii) \textproc{WFQ}  & -             & 0.096 (0.003) & 5.829 (0.049) \\ \hline
        \end{tabular}
    \end{center}
% \end{table}

% \begin{table}[H]
%     \caption{Scaled request latency}
%     \label{tab:many_run_latency}
\hspace{1ex}

    \begin{center}
        \begin{tabular}{|l|ccc|}
            \hline
            SL (s)               & \textit{NL}    & \textit{CK}    & \textit{MD}   \\ \hline
            (i) \textproc{FCFS}  & 10.272 (0.654) & 10.063 (0.631) & 1.740 (0.120) \\ \hline
            (i) \textproc{WFQ}   & 3.520 (0.085)  & 6.548 (0.361)  & 4.331 (0.336) \\ \hline
            (ii) \textproc{FCFS} & -              & 5.659 (0.313)  & 0.935 (0.062) \\ \hline
            (ii) \textproc{WFQ}  & -              & 2.503 (0.100)  & 1.194 (0.093) \\ \hline
        \end{tabular}
    \end{center}
\end{table}

\section{Conclusion}
Our top down inventory of design requirements, combined with a bottom up approach based on actual quantum hardware allowed us to take quantum networks a step further on the long path towards their large-scale realization.
Our work readies \Qlink, and paves the way towards the next step, a robust network layer control protocol. The link layer may now be used as a robust service without detailed knowledge of the physics of the devices. 
We expect that at the network layer, and when considering larger quantum memories, smart scheduling strategies will be important not only to combat memory lifetimes but also to coordinate actions of different nodes in time, calling for significant effort in computer science and engineering.  

\section*{Acknowledgements}
We thank Kenneth Goodenough for comments in earlier drafts.  This work was supported by ERC Starting Grant (Stephanie Wehner), ERC Consolidator Grant (Ronald Hanson), EU Flagship on Quantum Technologies, Quantum Internet Alliance, NWO VIDI (Stephanie Wehner), Marie Sk\l odowska-Curie actions --- Nanoscale solidstate spin systems and emerging quantum technologies --- Spin-NANO, grant agreement number 676108.

\newpage
\bibliographystyle{ACM-Reference-Format}
\bibliography{reference}

\newpage
\onecolumn
\appendix

\section*{Appendix}
% \axel{Check this after possible re-ordering}
In this Appendix, we provide further background and details, as well as more in-depth simulation results.
\begin{itemize}
    \item In Section~\ref{app:quantum101}, we provide a very brief introduction to quantum information, including concepts like entanglement, fidelity, QBER and provide an intuition on how the fidelity is related to memory lifetimes.
    \item In Section~\ref{app:testing}, we explain how to estimate the fidelity by interspersing test rounds.
    \item In Section~\ref{app:simulationResults}, we provide further simulation results to illustrate protocol performance and further validate our simulation against hardware.
    \item In Section~\ref{app:NV}, we provide further details of the NV platform as relevant for the design considerations of the proposed protocol.
        Furthermore, we provide details the physical modeling as well as numerical methods used to conduct the simulation.
    \item In Section~\ref{app:protocol}, we provide a complete description of the proposed protocol.
\end{itemize}

\section{Quantum prelude}
\label{app:quantum_101}
\label{app:quantum101}
This section provides a very short introduction to quantum information and in particular quantum states, entanglement, fidelity and decoherence.
For a deeper introduction to quantum information, see for example~\cite{Nielsen2010}.

\subsection{Qubits and states}
A quantum bit (\emph{qubit}) is a two-level system, where the two levels are usually denoted $\ket{0}$ and $\ket{1}$ respectively (``ket''-notation) and called the \emph{basis states} of the qubit.
These levels can for example be two energy levels of an electron spin or - when considering transmitting qubits - vertical and horizontal polarization of a photon, presence or absence of a photon, or a time-bin of early and late.
Compared to a ``classical'' bit $\ket{0}$ \emph{or} $\ket{1}$, a qubit can be in \emph{superpositions} thereof.
Mathematically, a state $\ket{\phi}$ of a qubit is written as
\begin{equation}\label{eq:qubit_state}
    \ket{\phi} = \alpha \ket{0} + \beta \ket{1}
\end{equation}
where $\alpha$ and $\beta$ are arbitrary complex numbers with the constraint that $\abs{\alpha}^2 + \abs{\beta}^2 = 1$,
and 
\begin{align}
\ket{0} = \left(\begin{array}{c} 1 \\ 0\end{array}\right)\ , &
\ket{1} = \left(\begin{array}{c} 0 \\ 1\end{array}\right)\ .
\end{align}
Note that $\ket{0}$ and $\ket{1}$ form a basis for $\mathbb{C}^2$.
%More formally, a qubit is a normalized vector in the two-dimensional complex vectorspace $\mathcal{C}_2$.
Some common states are
\begin{align}
    \ket{X,0} &= \frac{1}{\sqrt{2}}(\ket{0} + \ket{1})           & \ket{X,1} &= \frac{1}{\sqrt{2}}(\ket{0} - \ket{1}) \\
    \ket{Y,0} &= \frac{1}{\sqrt{2}}(\ket{0} + \mathrm{i}\ket{1}) & \ket{Y,1} &= \frac{1}{\sqrt{2}}(\ket{0} - \mathrm{i}\ket{1}) \\
    \ket{Z,0} &= \ket{0}                                         & \ket{Z,1} &= \ket{0}
\end{align}
corresponding to a '0' or '1' in the three different bases labeled $X$, $Y$, and $Z$. The label $Z$ also refers to the \emph{standard basis}.
We also use $\bra{\phi} = (\ket{\phi}^*)^T$ to denote the conjugate transpose of $\ket{\phi}$.
%which are the eigenstates of the Pauli matrices
%\begin{equation}
    %X=\begin{pmatrix} 0 & 1 \\ 1 & 0 \end{pmatrix}, \quad Y=\begin{pmatrix} 0 & -\mathrm{i} \\ \mathrm{i} & 0 \end{pmatrix}, \quad Z=\begin{pmatrix} 1 & 0 \\ 0 & -1 \end{pmatrix}.
%\end{equation}

Measuring a qubit in the standard ($Z$) basis ($\ket{0}$, $\ket{1}$), gives measurement outcomes '0' (i.e. $\ket{0}$) or '1' (i.e. $\ket{1}$). 
Measuring a qubit which is in the state $\ket{\phi}$ as in equation~\ref{eq:qubit_state} in the \emph{standard basis}, yields the outcomes $0$ or $1$ with the following probabilities
\begin{equation}
    P[\text{"measuring 0"}|Z\text{-basis}] = \abs{\alpha}^2, \quad P[\text{"measuring 1"}|Z\text{-basis}] = \abs{\beta}^2.
\end{equation}
which is why $\ket{\phi}$ needs to be normalized. Measuring a qubit in the standard basis collapses it to $\ket{0}$ or $\ket{1}$.
Measuring a qubit in the $X$- or $Y$-basis yields outcomes with probabilities
\begin{equation}
    P[\text{"measuring 0"}|X/Y\text{-basis}] = |\braket{X/Y,0}{\psi}|^2\ , \quad 
    P[\text{"measuring 1"}|X/Y\text{-basis}] = |\braket{X/Y,1}{\psi}|^2 \, 
\end{equation}
where $\braket{\cdot}{\cdot}$ is the inner product.

%In general a (projective) measurement on a qubit is described by two projectors $P_0$ and $P_1$, such that $P_i*P_j=\delta_{ij}$.
%The probabilities for outcome $i$ (0 or 1) is then given as
%\begin{equation}\label{eq:born}
    %P[\text{"measuring i"}] = \bra{\phi}P_i\ket{\phi}
%\end{equation}
%where $\bra{\phi}=(\ket{\phi})^\dagger$ is the complex conjugate of $\ket{\phi}$.
%Measurements with projectors $\ket{X,i}\bra{X,i}$, $\ket{Y,i}\bra{Y,i}$ or $\ket{Z,i}\bra{Z,i}$ are called measurements in the $X$-, $Y$- or $Z$-basis.

\subsection{Entangled states}\label{app:entanglement}
If qubit $A$ is in a state $\ket{\phi_1}$ and qubit $B$ is in the state $\ket{\phi_2}$, then their joint state (at possibly remote nodes) 
is given by the \emph{tensor product} of the individual states $\ket{\phi_1}_A$ and $\ket{\phi_2}_B$, i.e. as
\begin{equation}\label{eq:sep_state}
    \ket{\text{"joint state"}} = \ket{\phi_1}_A \otimes \ket{\phi_2}_B.
\end{equation}
Importantly, for the discussion here is that not all joint states can be factorized into single qubit states $\ket{\phi_1}_A$ and $\ket{\phi_2}_B$ in this way. These 
are called \emph{entangled} states.
For example, consider the state
\begin{equation}
    \ket{\Phi^+}=\frac{1}{\sqrt{2}}\left(\ket{0}_A\otimes\ket{0}_B + \ket{1}_A\otimes\ket{1}_B\right),
\end{equation}
which is a superposition of (1) both qubits being in the state $\ket{0}$ and (2) both qubits being in the state $\ket{1}$. This is an entangled state, i.e., it cannot be factorized
into two individual states, giving rise to genuinely quantum correlations between $A$ and $B$ that have no classical analogue.
%Thus measuring the two qubits of the state $\ket{\Phi^+}$ in the standard basis gives either the outcome $00$ or $11$ with equal probability ($1/2$).
The state $\ket{\Phi^+}$ is one of the so called \emph{Bell states}. These are entangled states, where the other three are given as
\begin{align}
    \ket{\Phi^-} &= \frac{1}{\sqrt{2}}(\ket{0}_A\otimes\ket{0}_B - \ket{1}_A\otimes\ket{1}_B), \\
    \ket{\Psi^+} &= \frac{1}{\sqrt{2}}(\ket{0}_A\otimes\ket{1}_B + \ket{1}_A\otimes\ket{0}_B), \\
    \ket{\Psi^-} &= \frac{1}{\sqrt{2}}(\ket{0}_A\otimes\ket{1}_B - \ket{1}_A\otimes\ket{0}_B).
\end{align}
Measurement outcomes of measuring the two qubits in any of the Bell-states in the bases $X$, $Z$ and $Y$ are either perfectly correlated or perfectly anti-correlated.
For example, for $\ket{\Phi^+}$ the measurement outcomes are perfectly correlated in the $X$ and $Z$ bases but perfectly anti-correlated in the $Y$ basis.
On the other hand, for $\ket{\Psi^-}$ the measurement outcomes are perfectly anti-correlated in all three bases.

Relevant to understand the generation of entanglement is that all the Bell-states can be transformed to one another by only applying \emph{local} quantum gates to one of the qubits.
Two useful gates are the bit flip $X \ket{x} = \ket{x + 1 \mod 2}$ and phase flip $Z \ket{x} = (-1)^x \ket{x}$. 
Applying them on qubit $A$ (at node $A$ only) allows one to transform:
\begin{equation}\label{eq:bell_relations}
    \ket{\Phi^-} = Z^A\ket{\Phi^+}, \quad \ket{\Psi^+}=X^A\ket{\Phi^+}, \quad \ket{\Psi^-}=Z^AX^A\ket{\Phi^+}\ ,
\end{equation}
where we added the index $A$ to emphasize the gates are applied to qubit $A$. We could also apply such gates to qubit $B$ to have the same effect.

In the heralded entanglement generation (Section~\ref{sec:NV}), we can obtain either failure, or else success. In the case of success, an additional bit indicates whether we produced the state $\ket{\Psi^+}$ or $\ket{\Psi^-}$.
From equation~\eqref{eq:bell_relations}, these two states can be transformed between each other by simply applying a $Z$-gate to one of the qubits.

\subsection{Fidelity and QBER}
In any real implementation of a quantum network, the generated entangled states will always differ from the perfect Bell states above due to noise in the system.
When writing noisy states, it is convenient to express the state as a \emph{density matrix}. For a perfectly prepared state $\ket{\Psi^-}$, the density matrix is
$\rho = \ketbra{\Psi^-}$. 
This allows one to express noise. For example, the analogue of applying a classical bit flip error $X$ with some probability $p_{\rm err}$ can be written as
\begin{align}
\rho_{\rm noisy} = (1-p_{\rm err}) \rho + p_{\rm err} X \rho X\ .
\end{align}

The \emph{fidelity} $F$ measures how close a realized state $\rho$ is to an ideal target state $\ket{\Psi^-}$. The fidelity of a state $\rho$ with the target state $\ket{\Psi^-}$ can be written as 
\begin{align}
F[\Psi^-] = \bra{\Psi^-}\rho\ket{\Psi^-}\, 
\end{align}
where $F=1$ if $\rho$ is identical to the target state. We have $0\leq F \leq 1$, where a larger value of $F$ means we are closer to the target state.
%Useful entanglement~\cite{XXX} desires $F=1/2$ to one of the Bell states above.

It is important to note that one cannot measure the fidelity of a single instance of a quantum state. However, if we produce the same state many times in succession, we can estimate its fidelity.
One way to do this is to measure the qubit-error-rate (\emph{QBER}).
Consider $\ket{\Psi^-}$ above and recall that measurement outcomes in the $X$, $Z$ and $Y$ bases are always perfectly anti-correlated in this case.
I.e. we always get different measurement outcome for qubit $A$ and for qubit $B$.
In case the state is noisy, this is no longer the case.
For a fixed basis (say $Z$) the QBER (here QBER$_Z$) is the probability of receiving equal\footnote{QBER for the other Bell states is defined in a similar manner, taking into account that measurement outcomes are always equal in some bases for the other ideal Bell states.} measurement outcomes, when measuring qubit $A$ and qubit $B$ in the $Z$ basis.
Similarly, we can define QBER$_X$ and QBER$_Y$ for measurements in the $X$ and $Y$ bases.
One can show that the fidelity and QBER of the Bell state state $\ket{\Psi^-}$ are related as
\begin{equation}\label{eq:F_vs_QBER}
    F[\Psi^-] = 1 - \frac{\mathrm{QBER}_X + \mathrm{QBER}_Y + \mathrm{QBER}_Z}{2}.
\end{equation}

\subsection{Decoherence}\label{sec:decoherence}
\begin{figure}[H]
    \centering
    \begin{subfigure}{0.4\textwidth}
        \includegraphics[width=\textwidth]{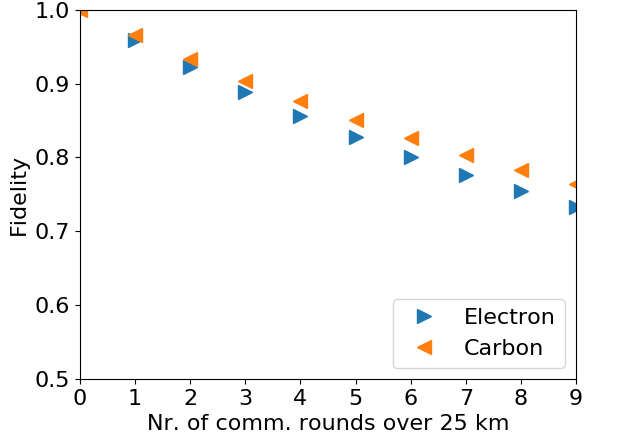}
        \caption{Reduction in fidelity $F$ when storing a perfect entangled state $\ket{\Psi^+}$ in the communication (blue, left triangles) and memory (orange, right triangles) qubit in terms of the number of communication rounds in the \Qlink\ scenario (over 25 km). Noise parameters listed in Table~\ref{tab:gate_noise} ($T_1=2.68$ ms and $T_2=1.00$ ms for communication and $T_1=\infty$ and $T_2=3.5$ ms for memory qubit).
	    \label{fig:fid_vs_comm_short_dec}
\label{fig:T1T2plot}
	    }
        % \label{label}
    \end{subfigure}
    ~
    \begin{subfigure}{0.38\textwidth}
        \includegraphics[width=\textwidth]{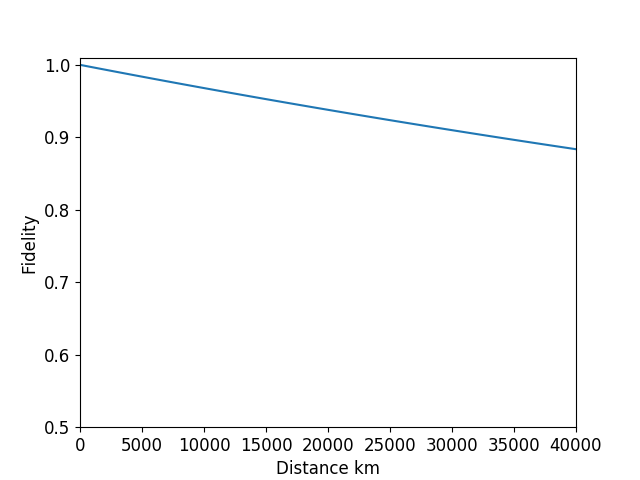}
        \caption{Illustration of an improved communication qubit by means of dynamical decoupling with $T_2=1.46$ s ($T_1=\infty$). If such a qubit was used in an NV platform connected to a network, the qubit could be kept 
alive while waiting for classical control communication over long distances. \\ \\  
	    % \label{fig:fid_vs_comm}
	    }
        % \label{label}
    \end{subfigure}
    \caption{}
    \label{fid_vs_comm}
		\label{fig:fid_vs_comm}
\end{figure}

Quantum memories are inherently noisy and the amount of noise a qubit experiences depends on how long it stays in the memory.
How long a qubit state is preserved is usually captured by the two numbers T1 (energy/thermal relaxation time) and T2 (dephasing time) of the qubit~\cite{Nielsen2010}, as well as free-induction decay $T_2^*$ (see e.g.~\cite{Kalb2017}).
%$T1$ is how long it takes for a qubit to reach thermal equilibrium from being initialized in the standard basis\axel{Check with Matteo if this is more correct}.
%Similarly for $T2$, however here the qubit is initialized in a equal superposition of $\ket{0}$ and $\ket{1}$.
Here, we focus on our performance metric (the fidelity) and illustrate in Figure~\ref{fig:fid_vs_comm_short_dec} how it behaves as a function of time, where to highlight
the actual effect of limited memory lifetimes we show the timescales in terms kms in fiber, where $c=206753$ km/s is the speed of light in fiber.
Figure~\ref{fig:fid_vs_comm} shows the fidelity of an entangled state stored in two electron states with a coherence time (T2) of 1.46 s as a function of the time it takes to communicate over a certain distance.

% \begin{figure}[H]
%     \centering
%     \includegraphics[width=0.5\textwidth]{figures/fid_vs_comm.png}
%     \caption{caption}
%     \label{fig:fid_vs_comm}
% \end{figure}

\section{Testing}
\label{app:testing}
\label{app:FEU}
We now explain the test used to gain confidence in transmission quality, 
specifically to estimate the quality parameter fidelity $F$ used in the FEU (Section~\ref{sec:feu}).
The following is a standard procedure in quantum information to estimate the fidelity $F = F[\ket{\Psi^-}]$ 
of a state $\rho$ to the entangled target state $\ket{\Psi^-}$. We emphasize that it is not possible to measure $F$ from a single copy of the state $\rho$. 
The matrices $\rho$ are a mathematical description of an underlying quantum system, and not a matrix that one can read or access like classical information.

We first describe the standard procedure in the way that it is normally used. We then outline how this protocol can be extended to the case of interest here, and why we can draw conclusions even in real world scenarios in which we can experience arbitrary correlated errors. 

Let us first assume a simpler scenario, in which $n$ identical noisy entangled states $\rho$ are produced in succession and we want to estimate $F$. 
We remark that when using imperfect quantum devices it is evidently a highly idealized situation that all states $\rho$ are exactly identical.
We can see from Eq.~\eqref{eq:F_vs_QBER} that we can express $F$ in terms of the quantum bit error rates QBER$_X$, QBER$_Z$, and QBER$_Y$, which immediately suggests a protocol: specifically, we will estimate the QBERs in bases $X$, $Z$ and $Y$ to obtain $F$. We sketch such a protocol in a specific way to build intuition for the more general procedure below:
\begin{itemize}
\item Node $A$ randomly chooses an $n$ element string $r = r_1,\ldots,r_n \in \{X,Z,Y\}$ and sends it to Node $B$.
\item Nodes $A$ and $B$ now perform the following procedure for $1 \leq j \leq n$ rounds:
\begin{itemize}
\item Node $A$ produces one entangled pair $\rho$ with Node $B$.
\item Nodes $A$ and $B$ both measures their respective qubits in the basis $r_j$ and record outcomes $x_j^A$ (Node $A$) and $x_j^B$ (Node $B$) respectively.
\end{itemize}
\item Node $B$ ($A$) transmits the outcome string $x^B = x^B_1,\ldots,x^B_n$ ($x^A = x^A_1,\ldots,x^A_n$) to Node $A$ ($B$). 
\item Both nodes estimate the error rates
\begin{align}
QBER_Z &\approx \frac{\#\{j\mid x^A_j = x^B_j, r_j = Z\}}{\#\{j\mid r_j = Z\}}\ ,\\
QBER_X &\approx \frac{\#\{j\mid x^A_j = x^B_j, r_j = X\}}{\#\{j\mid r_j = X\}}\ , \\
QBER_Y &\approx \frac{\#\{j\mid x^A_j = x^B_j, r_j = Y\}}{\#\{j\mid r_j = Y\}}\ , \\
\end{align}
where $\#\{j\mid condition\}$ is the number of indices $1 \leq j \leq n$ satisfying the stated condition. 
\end{itemize}
Using Eq.~\eqref{eq:F_vs_QBER} then yields an estimate of $F$.

Before we continue it may be instructive to compare the procedure above to the classical world. Evidently, classically, one way to gain confidence in a channels ability to transmit classical bits would be rather similar: Instead of preparing states $\rho$, we choose $n$ random bits and send them. In the end, we estimate the error rate. 
Translated to the quantum setting, we would be preparing random bits $\ket{0}$ and $\ket{1}$ at node $A$, and sending them to node $B$ which measures them in the $Z$ bases to obtain an estimate of the bit error rate, similarly to $QBER_Z$.
Such an estimate can give us confidence that 
also future bits are likely to be transmitted with roughly the same amount of errors as the test bits. This of course does \emph{not} allow the same level of confidence as error detection in the quality of transmission. Specifically, a CRC is a check for a \emph{specific} piece of data (e.g. one frame in 100 Base T), whereas such a test only yields a confidence in transmission quality.

Creating an analogous quantum CRC is theoretically possible by using a quantum error correcting code~\cite{Nielsen2010}, but technologically highly challenging and highly infeasible for many years to come. Yet, we remark that also in a future in which such methods would become feasible we may not want to employ them because the requirements of our use cases are different. 
Since many protocols for our use cases are probabilistic, or make many pairs (especially NL and MD use cases), we often do not require more confidence on the exact quality of 
a single pair. Indeed, we can pass errors all the way up to the application level (such as for example in QKD~\cite{bb84}), where errors are then corrected using classical instead of quantum error correction. In such cases, fluctuations in quality are indeed expected at the application level. Here, using fast and easy to produce test rounds may remain preferable over more time consuming quantum CRCs.

The protocol above is limited in two ways: (1) all states were assumed to be both identical and independent from each other. I.e., there are no memory effects in the noise. Such memory effects are non-trivial in the quantum regime since they may inadvertently create (some amount of) entanglement not only between $A$ and $B$, but between subsequent pairs produced. (2) We measured all $n$ rounds, consuming all entangled pairs. Instead, we would like a protocol in which only test rounds are interspersed, and we can draw an inference about the pairs which we did \emph{not} measure. Again, the possibility of quantum correlated noise between subsequent rounds makes this non-trivial.

To achieve this, we use a slight variant of the above as in~\cite{capacityEstimation}. Precise statistical statements are relatively straightforward - but very lengthy - to obtain using the techniques in~\cite{tomamichel:QKDpaper,capacityEstimation} and are out of scope of this paper. Here, we focus on the practical protocol and intuition without the need for mathematical tools from quantum information, which is a direct extension of the one above:
\begin{itemize}
\item Nodes $A$ and $B$ agree on a sampling window $N$.
\item Nodes $A$ and $B$ randomly pick an $N$ bit string $t=t_1,\ldots,t_N$ where $\Pr[t_j = 1] = q$ for some parameter $q$ determining the frequency of using test rounds. 
$A$ and $B$ periodically refresh $t$ as needed.
\item Nodes $A$ and $B$ randomly pick an $N$ element basis string $r = r_1,\ldots,r_N \in \{X,Z,Y\}$. $A$ and $B$ periodically refresh $r$ as needed.
\item The EGP uses $t$ to determine when to intersperse a test round. When producing the $j$-th response to the MHP, the EGP checks whether $t_j = 1$. If so, it
uses a standard test response instead to attempt to produce a test pair $\rho$, and takes as the measurement basis the next available in the random basis string $r$.
\item $A$ and $B$ record their measurement outcomes. 
\item $A$ and $B$ estimate QBER$_X$, QBER$_Z$, QBER$_Y$ over the past $N$ rounds of producing entanglement (tested and untested 'data' rounds)
\end{itemize}
The key insight in the analysis of this procedure is that we can (with some amount of confidence depending on $N$ and $q$) use the QBER measured on the test rounds
to determine the QBER on the untested - i.e. data - rounds~\cite[Inequality 1.3]{tomamichel:QKDpaper}. Using Eq.~\eqref{eq:F_vs_QBER}, then again allows one to draw conclusions about the average fidelity of the untested rounds to inform the FEU.

\section{Simulation and modeling \label{sec:app-simulation}}
\label{app:simulationResults}
We here provide additional simulation results, and further verification against the quantum hardware.

For our simulations we make use of a purpose built discrete event simulator for quantum networks: NetSquid\footnote{NetSquid is an acronym for Network Simulator for Quantum Information using Discrete events.}. By utilizing the discrete event paradigm NetSquid is capable of efficiently simulating the transmission and decay of quantum information in combination with the complex and stochastic nature of communication protocols. NetSquid can simulate both arbitrary quantum operations and Clifford-only gates, the former allowing for a precise simulation of small networks, while the latter allowing for networks containing thousands of nodes and qubits to be studied. Complete libraries of base classes enable users to simulate protocols and model physical devices at different levels of abstraction; for instance, (quantum) channels with modular noise, loss and delay models, or quantum processing devices with configurable gate topologies. NetSquid thus provides an ideal tool to validate network design choices and verify the performance of quantum network protocols in a physically-realistic setting.

The core simulation engine used by NetSquid is based on DynAA\cite{dynaa1,dynaa2,dynaa3}, a computer-aided analysis and design tool for the development of large, distributed, adaptive, and networked systems.  It combines the best of network and system simulation technologies in a discrete-event modeling framework. DynAA provides a simple, yet powerful language able to describe large and complex system architectures, and innovative constructs to express adaptation mechanisms of the system, such as dynamic parameterization, and functional and architectural reconfiguration.  A DynAA model can be simulated and/or analyzed to reveal system wide performance indicators, such as -- but not limited to -- throughput, power consumption, connectivity, reliability, and availability.  

\subsection{Validation of simulation}\label{sec:app-validation}
We compare our simulation model against further data gathered from the NV platform \Lab scenario.
Node A rotates its qubit over the Z-axis of the Bloch sphere by a fixed angle, followed by measuring its communication qubit in a basis ($X,Y$ or $Z$) that the nodes agreed upon beforehand. Node B only performs the measurement on its communication qubit, in the same pre-agreed basis. Regardless of the signal from the heralding station, both nodes initialize their qubit in $\ket{0}$ before the start of the next round.

\begin{figure}[H]
     \centering
	 \includegraphics[width=\textwidth]{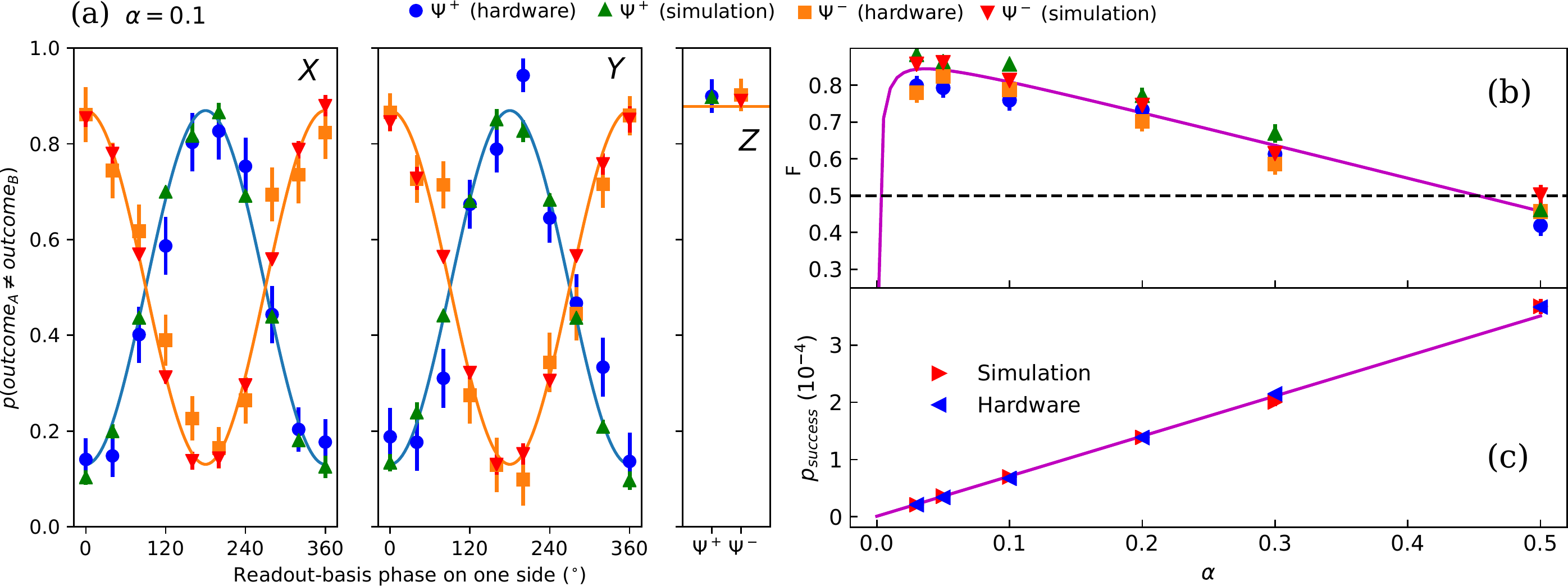}
	 \caption{Comparison of simulation results with data from NV hardware from \cite{Humphreys2018} (\Lab{ }scenario), showing good agreement. 
	 (a) Probability of success that the two nodes' measurements in basis $X/Y/Z$ on the state after a one-sided $Z$-rotation are unequal, at $\alpha = 0.1$ and (b) fidelities, both computed from correlations in the measurement outcomes (see text). (c) Probability that a single generation attempt succeeds, which is computed as $1/\bar{N}$ where $\bar{N}$ is the average number of runs up to and including successful heralding of entanglement. Solid line is the theoretical model from \cite{Humphreys2018}. Error bars indicate 1 standard deviation.
	The simulation data was extracted by running our model implemented in NetSquid on the supercomputer \textit{Cartesius} at SURFsara\cite{surfsara} for 122 hours of wall clock time using 63 
cores. A single data point is the average over at least (a) 100 pairs, (b) 300 pairs and (c) 600 pairs, which took between 500k (for $\alpha=0.5$) and 10.000k (for $\alpha=0.03$) entanglement generation attempts, with elapsed simulated time between 5 and 117 seconds.
     \label{fig:app-physical-modelling}
	 }
 \end{figure}

We compute the correlations of the measurement outcomes ($m_A, m_B = \pm 1$)as shown in Figure~\ref{fig:app-physical-modelling} using
\[
	\Pr(m_A \neq m_B) = \frac{1 - \langle \mathcal{B}\otimes \mathcal{B}\rangle}{2}
	\]
where $\langle \mathcal{B}\otimes \mathcal{B} \rangle$ is the expectation value of the product of joint measurement outcomes $m_A \cdot m_B$ with $\mathcal{B} \in \{X, Y, Z\}$ the measurement basis after rotation. 

The fidelity with the target state $\ket{\Psi^{\pm}}$ (where $\pm$ denotes the heralding detector) can be expressed as a function of the correlations as
\[
	\frac{1}{4} \left[ 1 \pm \langle X \otimes X \rangle \pm \langle Y \otimes Y \rangle - \langle Z\otimes Z\rangle \right].
	\]
	Assuming independence between the different rounds, propagation of standard deviations can be computed using standard techniques.

\subsection{Simulation data}\label{app:simulation_data}
In this section we present further results from the simulations of our proposed link layer protocol.
Simulation data will be made available upon request.
In total 1618 simulation runs were performed: $2\times169$ long runs (120 h wall time each) with 169 scenarios and 1280 shorter runs (24 h wall time each) with varying request load and minimal requested fidelity.
Out of the 169 scenarios used in the long runs, $2\times 5 \times 3$ concerned scenarios where the entanglement generation requests where a mix of the priorities \textit{NL}, \textit{CK} and \textit{MD}.
For these mixed scenarios we considered (1) two physical setups, \Lab\ and \Qlink, (2) five usage patterns (described below) and (3) three different schedulers, \textproc{FCFS}, \textproc{LowerWFQ} and \textproc{HigherWFQ}.

We implement different usage patterns of the link layer by, in every MHP cycle, issuing a new CREATE request for a random number of pairs $k$ (max $k_\mathrm{max}$) with probability $f \cdot p_{\rm succ}/(E \cdot k)$, where $p_\mathrm{succ}$ is the probability of an attempt being successful, $f$ is a fraction determining load of our system and $E$ is the expected number of MHP cycles to make one attempt.
For \Lab (\Qlink) $E=1$ ($E=1$) for M requests and $E\approx 1.1$ ($E\approx16$) for K requests.
We consider five different use patterns with $f$ and $k_\mathrm{max}$ defined in table~\ref{tab:scenarios}.

\begin{table}[H]
    \caption{}
    \label{tab:scenarios}
    \begin{center}
        \begin{tabular}{|l|ccc|}
            \hline
            Usage pattern          & \textit{NL}    & \textit{CK}    & \textit{MD}    \\ \hline
            \textproc{Uniform}     & $f=0.99\cdot\sfrac{1}{3}$, $k_\mathrm{max}=1$ & $f=0.99\cdot\sfrac{1}{3}$, $k_\mathrm{max}=1$ & $f=0.99\cdot\sfrac{1}{3}$, $k_\mathrm{max}=1$ \\
            \textproc{MoreNL}      & $f=0.99\cdot\sfrac{4}{6}$, $k_\mathrm{max}=3$ & $f=0.99\cdot\sfrac{1}{6}$, $k_\mathrm{max}=3$ & $f=0.99\cdot\sfrac{1}{6}$, $k_\mathrm{max}=256$ \\
            \textproc{MoreCK}      & $f=0.99\cdot\sfrac{1}{6}$, $k_\mathrm{max}=3$ & $f=0.99\cdot\sfrac{4}{6}$, $k_\mathrm{max}=3$ & $f=0.99\cdot\sfrac{1}{6}$, $k_\mathrm{max}=256$ \\
            \textproc{MoreMD}      & $f=0.99\cdot\sfrac{1}{6}$, $k_\mathrm{max}=3$ & $f=0.99\cdot\sfrac{1}{6}$, $k_\mathrm{max}=3$ & $f=0.99\cdot\sfrac{4}{6}$, $k_\mathrm{max}=256$ \\
            \textproc{NoNLMoreCK}  & $f=0$,                     $k_\mathrm{max}=3$ & $f=0.99\cdot\sfrac{4}{5}$, $k_\mathrm{max}=3$ & $f=0.99\cdot\sfrac{1}{5}$, $k_\mathrm{max}=256$ \\
            \textproc{NoNLMoreMD}  & $f=0$,                     $k_\mathrm{max}=3$ & $f=0.99\cdot\sfrac{1}{5}$, $k_\mathrm{max}=3$ & $f=0.99\cdot\sfrac{4}{5}$, $k_\mathrm{max}=256$ \\
            \hline
        \end{tabular}
    \end{center}
\end{table}

We make use of the following three scheduling strategies:
\begin{itemize}
    \item \textproc{FCFS}: First-come-first-serve with a single queue.
    \item \textproc{LowerWFQ}: \textproc{NL} are always service first (strict priority) and a weighted fair queue (\textproc{WFQ}) is used between \textit{CK} (weight 2) and \textit{MD} (weight 1).
    \item \textproc{HigherWFQ}: \textproc{NL} are always service first (strict priority) and a weighted fair queue (\textproc{WFQ}) is used between \textit{CK} (weight 10) and \textit{MD} (weight 1).
\end{itemize}

Figures~\ref{fig:lat_Uniform}-\ref{fig:lat_NoNLMoreMD} show scaled latencies and request latencies as functions of simulated time for 20 of the first long simulations runs using the different scenarios of mixed request types.
Furthermore, Figures~\ref{fig:th_Uniform}-\ref{fig:th_NoNLMoreMD} show the throughput as a function of simulated time for the same runs.
We also ran a second batch with the same scenarios which produced similar plots.
When using \textproc{FCFS} the request latency for the different requests are highly correlated, which is to be expected since all requests are in the same queue.
On the other hand the scaled latency for the different request priorities, in particular \textit{MD}, deviates from the others, which is due to the varying number of number of pairs in the requests.
From Figures~\ref{fig:th_Uniform}-\ref{fig:th_NoNLMoreMD} one can observe that the type of scheduler, at least for these simulated scenarios, have a relative small affect on the throughput.
In table~\ref{tab:throughput} and~\ref{tab:latency} the average throughputs, scaled latency and request latencies are collected for the same 20 simulation runs.

% Latencies
\foreach \mix in {Uniform, MoreNL, MoreCK, MoreMD, NoNLMoreCK, NoNLMoreMD}
{
    \begin{figure}[H]
        \centering
        \foreach \phys in {QL2020, Lab}
        {
            \foreach \latType in {scaled, req}
            {
                \begin{subfigure}{0.45\textwidth}
                    \includegraphics[width=\textwidth]{\latType_latency_vs_time_\phys_\mix.png}
                    \caption{\textproc{\phys}}
                    \label{fig:\phys_\mix_\latType}
                \end{subfigure}
                ~
            }

        }
        \caption{Latencies for \textproc{\mix}}
        \label{fig:lat_\mix}
    \end{figure}
}

% Throughputs
\foreach \mix in {Uniform, MoreNL, MoreCK, MoreMD, NoNLMoreCK, NoNLMoreMD}
{
    \begin{figure}[H]
        \centering
        \foreach \phys in {QL2020, Lab}
        {
            \begin{subfigure}{0.45\textwidth}
                \includegraphics[width=\textwidth]{throughput_vs_time_\phys_\mix.png}
                \caption{\textproc{\phys}}
                \label{fig:\phys_\mix_th}
            \end{subfigure}
            ~
        }
        \caption{Throughputs for \textproc{\mix}}
        \label{fig:th_\mix}
    \end{figure}
}

\newpage

\begin{table}[H]
    \caption{Average throughputs for 20 simulation scenarios, as defined above, of requests with mixed priorities.
    Each value is computed as $(\#\text{pairs}/t_\mathrm{sim\_max})$ where $t_\mathrm{sim\_max}$ is the reached simulated time (93 - 2355 s) of a single simulation run (24 h wall time).}
    \label{tab:throughput}
    \begin{center}        \begin{tabular}{|l|ccc|}
            \hline
            Scenario & Throughput\_\textit{NL} (1/s) & Throughput\_\textit{CK} (1/s) & Throughput\_\textit{MD} (1/s) \\ \hline
			\Lab\_\textproc{MoreCK}\_\textproc{FCFS} & 0.976 & 4.126 & 1.187 \\ \hline
			\Lab\_\textproc{MoreCK}\_\textproc{HigherWFQ} & 1.046 & 3.719 & 1.408 \\ \hline
			\Lab\_\textproc{MoreMD}\_\textproc{FCFS} & 1.025 & 0.905 & 4.771 \\ \hline
			\Lab\_\textproc{MoreMD}\_\textproc{HigherWFQ} & 0.981 & 1.058 & 4.659 \\ \hline
			\Lab\_\textproc{MoreNL}\_\textproc{FCFS} & 3.975 & 0.950 & 1.375 \\ \hline
			\Lab\_\textproc{MoreNL}\_\textproc{HigherWFQ} & 4.447 & 0.986 & 1.117 \\ \hline
			\Lab\_\textproc{NoNLMoreCK}\_\textproc{FCFS} & - & 4.696 & 1.366 \\ \hline
			\Lab\_\textproc{NoNLMoreCK}\_\textproc{HigherWFQ} & - & 5.101 & 0.916 \\ \hline
			\Lab\_\textproc{NoNLMoreMD}\_\textproc{FCFS} & - & 1.044 & 4.600 \\ \hline
			\Lab\_\textproc{NoNLMoreMD}\_\textproc{HigherWFQ} & - & 1.300 & 5.408 \\ \hline
			\Lab\_\textproc{Uniform}\_\textproc{FCFS} & 2.066 & 2.035 & 2.170 \\ \hline
			\Lab\_\textproc{Uniform}\_\textproc{HigherWFQ} & 2.210 & 2.186 & 1.908 \\ \hline
			\Qlink\_\textproc{MoreCK}\_\textproc{FCFS} & 0.064 & 0.302 & 1.398 \\ \hline
			\Qlink\_\textproc{MoreCK}\_\textproc{HigherWFQ} & 0.078 & 0.329 & 1.146 \\ \hline
			\Qlink\_\textproc{MoreMD}\_\textproc{FCFS} & 0.075 & 0.078 & 4.139 \\ \hline
			\Qlink\_\textproc{MoreMD}\_\textproc{HigherWFQ} & 0.066 & 0.073 & 4.793 \\ \hline
			\Qlink\_\textproc{MoreNL}\_\textproc{FCFS} & 0.312 & 0.066 & 1.667 \\ \hline
			\Qlink\_\textproc{MoreNL}\_\textproc{HigherWFQ} & 0.292 & 0.084 & 1.374 \\ \hline
			\Qlink\_\textproc{NoNLMoreCK}\_\textproc{FCFS} & - & 0.355 & 1.480 \\ \hline
			\Qlink\_\textproc{NoNLMoreCK}\_\textproc{HigherWFQ} & - & 0.374 & 1.180 \\ \hline
			\Qlink\_\textproc{NoNLMoreMD}\_\textproc{FCFS} & - & 0.084 & 6.756 \\ \hline
			\Qlink\_\textproc{NoNLMoreMD}\_\textproc{HigherWFQ} & - & 0.091 & 5.036 \\ \hline
			\Qlink\_\textproc{Uniform}\_\textproc{FCFS} & 0.175 & 0.143 & 2.538 \\ \hline
			\Qlink\_\textproc{Uniform}\_\textproc{HigherWFQ} & 0.154 & 0.166 & 2.483 \\ \hline
       \end{tabular}
    \end{center}
\end{table}

\begin{table}[H]
    \caption{Average scaled latencies (SL) and request latencies (RL) for 20 simulation scenarios, as defined above, of requests with mixed priorities of a single simulation run (93 - 2355 s simulated time and 24 h wall time).
    Values in parenthesis are estimates of standard errors, computed as $s_n/\sqrt{n}$ where $s_n$ is the sample standard deviation and $n$ is the number of data points used for averaging.}
    \label{tab:latency}
    \begin{center}
        \begin{tabular}{|l|cccccc|}
            \hline
            Scenario & SL\_\textit{NL} (s) & SL\_\textit{CK} (s) & SL\_\textit{MD} (s) & RL\_\textit{NL} (s) & RL\_\textit{CK} (s) & RL\_\textit{MD} (s) \\ \hline
			\Lab\_\textproc{MoreCK}\_\textproc{FCFS} & 40.18 (0.90) & 41.09 (0.42) & 19.64 (5.52) & 55.50 (0.30) & 55.58 (0.14) & 55.11 (2.83) \\ \hline
			\Lab\_\textproc{MoreCK}\_\textproc{HigherWFQ} & 0.30 (0.01) & 25.63 (0.61) & 24.95 (15.20) & 0.46 (0.02) & 33.88 (0.66) & 264.28 (35.75) \\ \hline
			\Lab\_\textproc{MoreMD}\_\textproc{FCFS} & 40.25 (1.15) & 41.70 (1.28) & 13.87 (2.91) & 55.63 (1.00) & 57.24 (1.00) & 62.55 (2.43) \\ \hline
			\Lab\_\textproc{MoreMD}\_\textproc{HigherWFQ} & 0.23 (0.01) & 2.09 (0.29) & 27.53 (6.00) & 0.36 (0.02) & 2.65 (0.35) & 129.30 (3.51) \\ \hline
			\Lab\_\textproc{MoreNL}\_\textproc{FCFS} & 45.59 (0.47) & 46.44 (0.95) & 14.70 (4.65) & 60.36 (0.21) & 60.59 (0.42) & 61.80 (2.76) \\ \hline
			\Lab\_\textproc{MoreNL}\_\textproc{HigherWFQ} & 0.69 (0.02) & 83.79 (2.64) & 98.34 (46.15) & 0.97 (0.02) & 114.89 (2.57) & 299.28 (37.25) \\ \hline
			\Lab\_\textproc{NoNLMoreCK}\_\textproc{FCFS} & - & 13.05 (0.27) & 3.55 (1.36) & - & 17.58 (0.30) & 23.04 (3.21) \\ \hline
			\Lab\_\textproc{NoNLMoreCK}\_\textproc{HigherWFQ} & - & 6.70 (0.16) & 26.14 (9.05) & - & 9.04 (0.19) & 76.02 (12.51) \\ \hline
			\Lab\_\textproc{NoNLMoreMD}\_\textproc{FCFS} & - & 23.45 (1.65) & 10.97 (2.51) & - & 30.86 (1.94) & 39.33 (4.09) \\ \hline
			\Lab\_\textproc{NoNLMoreMD}\_\textproc{HigherWFQ} & - & 2.26 (0.31) & 44.33 (11.92) & - & 3.34 (0.43) & 204.78 (9.54) \\ \hline
			\Lab\_\textproc{Uniform}\_\textproc{FCFS} & 11.41 (0.27) & 11.46 (0.27) & 12.38 (0.26) & 11.41 (0.27) & 11.46 (0.27) & 12.38 (0.26) \\ \hline
			\Lab\_\textproc{Uniform}\_\textproc{HigherWFQ} & 0.35 (0.01) & 0.73 (0.02) & 61.19 (0.97) & 0.35 (0.01) & 0.73 (0.02) & 61.19 (0.97) \\ \hline
			\Qlink\_\textproc{MoreCK}\_\textproc{FCFS} & 40.65 (4.43) & 37.46 (1.93) & 12.22 (3.82) & 52.20 (5.21) & 51.34 (2.33) & 43.41 (5.79) \\ \hline
			\Qlink\_\textproc{MoreCK}\_\textproc{HigherWFQ} & 4.11 (0.24) & 26.66 (0.95) & 76.29 (16.50) & 5.72 (0.34) & 35.91 (1.05) & 238.09 (21.00) \\ \hline
			\Qlink\_\textproc{MoreMD}\_\textproc{FCFS} & 25.45 (3.03) & 28.34 (3.01) & 9.30 (1.63) & 32.94 (3.43) & 38.90 (3.65) & 37.97 (2.67) \\ \hline
			\Qlink\_\textproc{MoreMD}\_\textproc{HigherWFQ} & 2.62 (0.42) & 3.04 (0.42) & 14.44 (1.92) & 3.79 (0.60) & 4.51 (0.62) & 47.64 (2.75) \\ \hline
			\Qlink\_\textproc{MoreNL}\_\textproc{FCFS} & 65.92 (1.93) & 64.05 (4.18) & 26.16 (5.84) & 89.83 (1.79) & 85.99 (3.70) & 85.98 (5.24) \\ \hline
			\Qlink\_\textproc{MoreNL}\_\textproc{HigherWFQ} & 7.04 (0.35) & 45.87 (3.73) & 50.55 (12.63) & 9.78 (0.41) & 59.92 (4.15) & 236.03 (21.62) \\ \hline
			\Qlink\_\textproc{NoNLMoreCK}\_\textproc{FCFS} & - & 15.98 (0.64) & 4.63 (0.87) & - & 21.90 (0.72) & 21.27 (1.63) \\ \hline
			\Qlink\_\textproc{NoNLMoreCK}\_\textproc{HigherWFQ} & - & 39.31 (1.64) & 70.64 (16.03) & - & 55.94 (1.93) & 277.08 (19.97) \\ \hline
			\Qlink\_\textproc{NoNLMoreMD}\_\textproc{FCFS} & - & 60.12 (6.95) & 22.28 (4.04) & - & 102.23 (4.17) & 104.88 (2.69) \\ \hline
			\Qlink\_\textproc{NoNLMoreMD}\_\textproc{HigherWFQ} & - & 3.01 (0.35) & 6.21 (1.58) & - & 5.18 (0.70) & 38.91 (4.89) \\ \hline
			\Qlink\_\textproc{Uniform}\_\textproc{FCFS} & 49.13 (1.29) & 50.85 (1.38) & 47.39 (0.33) & 49.13 (1.29) & 50.85 (1.38) & 47.39 (0.33) \\ \hline
			\Qlink\_\textproc{Uniform}\_\textproc{HigherWFQ} & 4.16 (0.26) & 8.22 (0.58) & 34.39 (0.61) & 4.16 (0.26) & 8.22 (0.58) & 34.39 (0.61) \\ \hline
       \end{tabular}
    \end{center}
\end{table}

\begin{table}[H]
    \caption{Relative difference (Rel. Diff.) for the metrics: fidelity (Fid.), throughput (Throughp.) scaled latency (Laten.) and number of generated entangled pairs (Nr. pairs), between two identical scenarios except that the probability of losing a classical message ($p_\mathrm{loss}$) is zero for one and between $10^{-10}$ and $10^{-4}$ for the other.
    The relative difference is maximized over three simulation runs (281 - 3973 s simulated time and 70 h wall time each) with requests of priority either \textit{NL}, \textit{CK} or \textit{MD} ($f=0.99$, $k_\mathrm{max}=3$), with equal $p_\mathrm{loss}$.}
    \label{tab:high_c_loss}
    \begin{center}
        \begin{tabular}{|l|cccc|}
            \hline
            $p_\mathrm{loss}$ & Max Rel. Diff. Fid. & Max Rel. Diff. Throughp. & Max Rel. Diff. Laten. & Max Rel. Diff. Nr pairs \\ \hline
			$10^{-4}$ & 0.005 & 0.027 & 0.629 & 0.026 \\ \hline
			$10^{-5}$ & 0.004 & 0.012 & 0.469 & 0.008 \\ \hline
			$10^{-6}$ & 0.016 & 0.037 & 0.332 & 0.047 \\ \hline
			$10^{-7}$ & 0.040 & 0.026 & 0.576 & 0.020 \\ \hline
			$10^{-8}$ & 0.007 & 0.023 & 0.623 & 0.020 \\ \hline
			$10^{-9}$ & 0.004 & 0.026 & 0.338 & 0.021 \\ \hline
			$10^{-10}$ & 0.018 & 0.075 & 0.742 & 0.077 \\ \hline
       \end{tabular}
    \end{center}
\end{table}

\def\nvzero{$\textnormal{NV}^0$}
\def\nvminus{$\textnormal{NV}^-$}

\section{Under the hood}
\label{app:NV}
We now provide some more details on the simulation, numerical methods and the underlying NV platform. We remark that physical models for different parts of the NV platform are
well established and validated~\cite{Bernien2014}. The validation performed here is thus only about how the combined system performing entanglement generation
matched with our simulation.

\subsection{The simulated network}
To understand our simulation we perform a full implementation of the MHP and EGP, running on simulated quantum hardware. To achieve this, we start by defining basic components in NetSquid, 
which, inspired by NS-3, is entirely modular and can be used to construct complex simulation scenarios by combining component elements. 
The components in the simulation are as follows, and our simulation could easily be configured to examine the performance of our protocol on other underlying hardware platforms such as Ion Traps.
\begin{itemize}
\item A QuantumProcessingDevice, which is a general component we defined in NetSquid. Abstractly, such a QuantumProcessingDevice is described by the following:
\begin{enumerate}
\item A number of communication and memory qubits. Each such qubit is associated with a noise-model that describes how quantum information decoheres over time when kept in the memory itself.
Concrete parameters for the NV platform are given in Section~\ref{app:memoryNoise}.
\item Possible one or two-qubit quantum gates to be performed on each (pair of) qubit(s), the time required to execute the gate, as well as a noise-model associated with each such gate that may differ from qubit to qubit. For the NV platform we will only need to consider the gates given in Section~\ref{app:NVops}. 
\item With each communication qubit we associate a trigger that allows the generation of entanglement between this communication qubit and a traveling qubit (a photon). 
Such an operation only succeeds with some probability of success, requires a certain amount of time, and can also be noisy. For the NV platform, we explain this in Section~\ref{app:emit}.
\item Readout - i.e. measurement of a qubit. This takes a certain amount of time, and does again carry a noise-model. For NV, we explain this in Section~\ref{app:readout}.
\end{enumerate}
\item A FiberConnection, which is a general NetSquid component that allows us to simulate optical fibers, including photon loss per km.
\item A HeraldingStation, which automatically measures incoming photons in a certain time interval. This process is again subject to several possible errors explained in Section~\ref{app:heralding}.
\item ClassicalConnection, which we use to transmit classical messages allowing us to simulate, for example, losses on the channel. Section~\ref{app:classical} explains the model considered here.
\item Node, which includes a QuantumProcessingDevice, and enjoys fiber connections with the HeraldingStation or other nodes. Each Node can run programs in the same way that they could be run on e.g. a computer or microcontroller, allowing these Programs to make use of - for example - the QuantumProcessing Device. This allows us to perform a full implementation (here in Python) of the MHP and EGP including all subsystems in the same way as the program will later run on an actual microcontroller (however, in C). 
\end{itemize}
 We briefly review properties of the nitrogen-vacancy (NV) centre in diamond~\cite{Bernien2014} and how they affect the design and performance of our protocol. We will also highlight how this can be modeled in simulation.
Important to the design and performance of our protocol is how long operations on qubits stored in the NV-center take.
Additionally, the coherence time, i.e. how long a qubit can be stored, has an impact on our protocol.
% Coherence times are usually captured as the two numbers: $T1$ (energy relaxation time) and $T2$ (dephasing time)~\cite{Nielsen2010}.
% $T1$ is how long it takes for the probability of measuring a qubit in the excited state to be $\mathrm{e}^{-1}$, when initialized in the excited.
% Similarly for $T2$, however here the qubit is initialized in a uniform superposition of the excited and ground state.
% \axel{$T2^*$?}
We summarize typical values for the noise-level and execution time of the allowed operations of a NV-center together with coherence times.
These are the values we used in our simulation.
Note however that these values can vary significantly between samples.
% This section provides a brief description of how a nitrogen-vacancy center in diamond can be used to store and manipulate qubits.

\begin{table}
    \begin{center}
        \begin{tabular}{| l | c | c | l|}
        \hline
            & (Unsquared) fidelity & Duration/time & Experimentally realized\\\hline
            Electron $T_1$& - &$2.86$ ms   &  $>1$h\cite{Abobeih2018}\\\hline
            Electron $T_2^*$& - &$1.00$ ms  &  $1.46$ s\cite{Abobeih2018} \\\hline
            % Electron $T_{\textnormal{coh}}$& -& $2.86$ ms & Using dynamical decoupling  \\\hline
		Carbon $T_1$& - & $\infty$ & $>6$m \cite{bradley2019solidstate} \\\hline
		Carbon $T_2^*$& - & $3.5$ ms & $\approx 10$ms \cite{bradley2019solidstate} \\\hline
            Electron single-qubit gate& $1.0$ & $5$ ns &  $>0.995$ (100 ns)~\cite{Kalb2017}\\\hline
            % &&& time is the time of a $\pi$-pulse \\\hline
            E-C controlled-$\sqrt{\X}$-gate (E=control) & $0.992$ & $500$ $\mu$s & $0.992$ (500-1000 $\mu$s) fig 2 in~\cite{Kalb2017}\\\hline
            % Carbon $\sqrt{\X}$-gate & $0.95$ & $500-1000$ $\mu$s & Private comm. with Arien Stolk\\\hline
            Carbon Rot-$Z$-gate & $0.999$ & $20$ $\mu$s & $1.0$ (20 $\mu$s)~\cite{Taminiau2014}\\\hline
            Electron initialization in $\ket{0}$ & $0.95$ & $2$ $\mu$s & $0.99$ (2 $\mu$s)~\cite{Reiserer2016} \\\hline
            Carbon initialization in $\ket{0}$ & $0.95$ & $310$ $\mu$s & $0.95$ (300 $\mu$s)~\cite{Cramer2016}\\\hline
            Electron readout & $0.95$ ($\ket{0}$), $0.995$ ($\ket{1}$) & $3.7$ $\mu$s & $0.95$ ($\ket{0}$), $0.995$ ($\ket{1}$) (3-10 $\mu$s)~\cite{Humphreys2018} \\\hline
    \end{tabular}
    \end{center}
    \caption{Gates and coherence times used in simulation. Values used in the simulation corresponding to \Lab. We remark that since these are custom chips, no two are exactly identical. Individual values have since 
seen significant improvements (Experimentally realized), but not been realized simultaneously for producing entanglement that would allow a direct comparison to simulation. We have thus
focused in simulation only what enables a comparison to data gathered from entanglement generation on hardware.}
    \label{tab:gate_noise}
\end{table}

\subsection{Qubits on the NV Platform}\label{app:memoryNoise}
A NV centre is formed by replacing a carbon atom in a diamond lattice with a nitrogen atom and removing a neighboring carbon (vacancy).
This structure traps electrons which together form a spin-1 system.
Two of the levels of the collective spin-1 state can be used as a communication qubit in a quantum network.
Around the NV centre there is also a natural abundance of carbon-13 atoms which interact with the communication qubit (electron spin).
The surrounding carbon spins can be addressed using the communication qubit and can thus be used as memory qubits.
We here consider a situation with only one communication qubit, and one memory qubit.

\subsubsection{Noise model - Free evolution of the electronic and nuclear spins} 
% \subsubsection{Characteristics of the electronic spin}
Noise in experimental implementations is described in terms of $T_1$, $T_2$, $T_2^*$ times, where Section~\ref{sec:decoherence} serves to provide intuition on how our quantity of interest - the fidelity to the maximally entangled target state $\ket{\Psi^+}$ - depends on their values. Table~\ref{tab:gate_noise} lists values used in simulation (reflecting \Lab), and 
state of the art for the communication qubit (Electron), and memory qubit (Carbon).
%For the electron spin: $T_2^* = 3.3(1) \mu$s~\cite{Taminiau2014}.
% But with initialization of the electron spin, the electronic dephasing becomes $T_2^* = 4.0(2) \mu$s~\cite{Taminiau2014}.
%However, using \emph{dynamical decoupling}, see below, the coherence time of the electron can increase by many orders of magnitude.
%In~\cite{Abobeih2018}, a coherence time of 1.46 s is experimentally realised.
%Similarly a \edit{XXX} s coherence time for a nuclear spin has been shown in~\cite{}.

% \subsubsection{Characteristics of the nuclear spin}
% Dephasing $T_2* = 2.7(2)$ ms for spin 1 and $4.4(5)$ ms for spin 2 \cite[Suppl. p.19]{Taminiau2014}. Other research gives the $T_2*$ times for each of the two states $\ket{0/1}$: $3.4(1)$ ($3.4(1)$) for first nuclear spin examined and $19.4(3)$ ($16.2(3)$) for the second, for the $\ket{m_S = 0}$ state ($\ket{m_S = \pm 1}$ state) \cite[Suppl. Table S1]{Kalb2017}.

\subsubsection{Quantum gates}\label{app:NVops}

\paragraph{Procedure and parallelism constraints}
Quantum gates can be realized by applying microwave pulses. Of specific interest that affects the throughput is the duration of such operations given in Table~\ref{tab:gate_noise}.
While not absolutely necessary for the understanding of the simulation, we briefly sketch how operations are performed also on the carbons to illustrate one feature of this system that is 
relevant for the performance of our protocols - namely the parallelism of the allowed gate operations.
We remark that operations on the carbon spins are performed using the following pulse sequence
\begin{equation}
	\label{eq:pulse-sequence}
(\tau - \pi - 2\tau - \pi - \tau)^{N/2} ,
\end{equation}
where $\pi$ is a microwave-$\pi$-pulse on the electron spin, $2\tau$ is the time between the pulses and $N$ is the total number of pulses.
% Here, the time $\tau$ determines the target carbon spin.
The target carbon spin can be chosen by picking $\tau$ such that it is precisely in resonance with the target carbon spin's hyperfine interaction with the electron spin.
If the electron spin is in the state $\ket{0}$ ($\ket{1}$) the target carbon spin will rotate around the $X$-axis of the Bloch sphere in the positive (negative) direction, with an angle $\theta$ which depends on the total number of pulses $N$. This means that one can perform quantum gates on the carbon that are controlled by the state of the electron spin.
The effective unitary operation (E=control) on the electron and the target carbon spin is then given as
\begin{equation}
    \begin{pmatrix}
        R_X(\theta) & 0 \\
        0           & R_X(-\theta)
    \end{pmatrix},
\end{equation}
where $R_X(\theta) = \exp(\mathrm{i}\theta/2 X)$ denotes a rotation around the $X$-axis of an angle $\theta$.
% One can compute the interpulse delay $2\tau$ using $\tau = 2\pi/\omega_L$, where $\omega_L = 2\pi\times 431$ kHz is the Larmor frequency of the carbon nuclear spins~\cite{Taminiau2014}. The total number of pulses $N$ determines the total rotation angle (see Methods section in \cite{Taminiau2014}
% Arian Stolk expects that a twice as large $N$ will probably double the rotation angle, but we should check this).
Not only does the pulse sequence \eqref{eq:pulse-sequence} manipulate the carbon spin, but it also decouples the electron from its environment, thereby prolonging its coherence time and is thus also called \emph{dynamical decoupling}, allowing longer memory lifetimes (Figure~\ref{fig:T1T2plot}).
We thus see a limit to the amount of parallelism when operating on the carbon and the electron spin.

% Gates on the nuclear spins correspond to pulse sequences from eq.~\eqref{eq:pulse-sequence}.\\

% Not only do the pulse sequences apply a gate on the nuclear spins, but simultaneously, they decouple the electron from its environment.
% Therefore, such gates are called ``decoherence-protected'' and these allow the full electron-dephasing time to be exploited; the gates are thus not limited by the electron spin dephsing time $T_2^*$ or the Hahn echo time $T_2$ and do not require strong coupling.
% (see also subsection below. Quotes from \cite{taminiau2014universal}).\\

Other quantum gates are however simpler: 
Since the carbon continuously precess around the $Z$ axis of the Bloch sphere, rotations around $Z$ (Carbon Rot-$Z$) are simply done by waiting a correct amount of time.
Thus, also controlled rotations around other axes than $X$ can be performed by correctly interspersed waiting times during the pulse sequence above.

% Furthermore, $Z$-gates on the carbon nuclear spins requires one to just wait: in the rotating frame of the electron spin, the carbons precess, so one just has to wait until the carbon has picked up the correct phase.
% In Arian's words: ``a controlled-$\Z$ is difficult'' (I interpret this as: we don't do these).

% Furthermore, rotations of a carbon spin over the $\X$-axis are always performed as conditional operations with the electron spin as control qubit. A single-qubit rotation of a carbon spin is thus performed by first initializing the electron in $\ket{1}$, after which a conditional rotation over the $\X$-axis is performed.\\

% \section{Experimental parameter values}
\subsection{Gates and their noise}
% \axel{There is also a gate section just above}
In this section we collect parameters for noise and delays of gates used in our simulation.
Table~\ref{tab:gate_noise} summarize the possible gates that can be performed on the electron and carbon spins in the NV system, together with decoherence times.
Section~\ref{sec:EG_noise} describes how the noise occurring from entanglement generation attempts is modeled.

% Most of the realistic parameters used are taken from \cite{taminiau2014universal}, which describes a quantum-error-correction experiment which involves the electron spin and two nuclear spins. This experiment was performed at room temperature, whereas most of the remaining literature describes experiments performed at a temperature of $4$ Kelvin.
% \textbf{Note} beforehand: the notions of `dephasing' and `depolarization' might have a different meaning to an experimentalist than a theorist (I haven't figured out the former yet).

% \axel{Where does 340 and 280 come from, for carbon INIT? Also private comm.?}

% Final choice for noise model parameters.
% Here, $S = R_{\Z}(\pi/2)$ is the phase gate and $\sqrt{\X} = R_{\X}(\pi/2)$.
Here the E-C controlled-$X$ rotates the carbon spin around the $X$-axis in the positive (negative) direction if the electron is in the $\ket{0}$ ($\ket{1}$) state.
Furthermore, note that there is an asymmetry in reading out the $\ket{0}$-state and the $\ket{1}$-state of the electron.

\subsubsection{Modeling noisy operations}
Noise in gate operations is modeled by applying noise after a perfect gate (a standard method):
\[
	U_{\textnormal{noisy}} \left(\rho\right) = \mathcal{N}^f_{\textnormal{dephas}} \circ  U_{\textnormal{perfect}}\left(\rho\right),
\]
where
\[
    \mathcal{N}^p_{\textnormal{dephas}} : \rho \mapsto f\rho + (1-f)\Z\rho\Z
\]
is the dephasing channel in $\Z$ and $f$ is the gate fidelity as given in table~\ref{tab:gate_noise}.
States are initialized as $\mathcal{N}^p_{\textnormal{depol}} (\dyad{0})$, where
\[
	\mathcal{N}^p_{\textnormal{depol}} : \rho \mapsto f\rho + \frac{1-f}{3}\left[\X\rho\X + \Y\rho\Y + \Z\rho\Z\right]
\]
denotes the depolarization channel by.

\subsubsection{How the electron spin is initialized}
Initialization of the electron spin means setting the state to $\ket{0}$ \cite{Reiserer2016}. Initialization of the electron spin is done by performing \textit{optical pumping}, in which light shines onto the electron, thereby bringing its quantum state in a higher energy level, given it was in $\ket{1}$, after which it falls back to either $\ket{0}$ or $\ket{1}$ with a given probability.
If the electron falls down to the state $\ket{0}$ is will stay there, thus after many repetitions of this process, the electron is with high confidence in the state $\ket{0}$.
For our discussion here, it will be relevant to remark that this operation takes time (Table~\ref{tab:gate_noise}), and we will need to perform it repeatedly as the first step
in producing entanglement.

\subsubsection{Moving a qubit to memory}
When moving a qubit from the communication qubit to the memory qubit, the memory qubit needs to already be initialized.
Initialization of the carbon is done by effectively swapping the $\ket{0}$ state from the electron to the carbon and cannot therefore be performed while having an entangled state in the electron.
For this reason, initialization of the carbon ($310$ $\mu$s) needs to be performed before a photon is emitted from the electron during an entanglement generation attempt.
However, it is not necessary to re-initialize the carbon before every entanglement generation attempt but simply periodically depending on the coherence time.
In our simulation we assumed $T1$ to be $3500$ $\mu$s and thus re-initialize the carbon every $3500$ $\mu$s (every 350th MHP cycle).

Swapping a state in the electron to the carbon can be done by 2 E-C controlled-$\sqrt{\X}$-gates and single qubit gates (total time $1040$ $\mu$s)~\cite{Kalb2017}.

% Again, this operation takes time (Table~\ref{tab:gate_noise}).
% To illustrate the timings of this, consider initializing the carbon nuclear spin via the electron spin.
% \begin{enumerate}
% \item prepare the electron spin in $m_S=0$ by optical pumping
% \item swap the electron state onto the nuclear spin % (note that we can use a more efficient scheme for this than a general swap! Namely with only 2 conditional gates instead of 3; see fig.~\ref{fig:nuclear-initialization}.)
% \item re-initialize the electron spin 
% \end{enumerate}

%\begin{figure}[h!]
	%\centering
	%\includegraphics[width=0.5\textwidth]{figures/nuclear_spin_initialization_TCS+14.png}
	%\caption{
		%\label{fig:nuclear-initialization}
        %Procedure for nuclear spin initialization (reproduced from \edit{taminiau's paper}, fig 1c). 
		%}
%\end{figure}

\subsubsection{How a measurement (readout) is performed}\label{app:readout}

\paragraph{Readout of the communication qubit}
First, we are again interested in the time to perform this operation given in Table~\ref{tab:gate_noise} which will be relevant in the MD use case.
%The electron spin is measured through state-dependent fluorescence (`single-shot readout').
% Readout fidelity is, on average, $0.95$ \cite[p.1]{abobeih2018one}. Readout takes $3.7 \mu$s \cite[Fig.2]{hensen2015loophole}.
Evidently, also a readout can be noisy, where we here remark that the noise is asymmetric in that the probability of incorrectly obtaining measurement outcome $0$ is much lower than incorrectly getting outcome $1$. %; bright state fidelity ($\ket{0}$) is less than dark state fidelity ($\ket{1}$). (\edit{values}~\cite{})
% Fidelities for electron readout are given in \cite[Extended Data Table 1]{humphreys2018deterministic} as $0.959(3)$ and $0.905(3)$ for $\ket{0}$ and $0.995(1)$ and $0.996(1)$ for $\ket{1}$ for the two nuclear spins examined. Older results \cite[Suppl.p.1]{kalb2017entanglement} give $0.9379(6)$ and $0.89857$ for $\ket{0}$ and $0.9932(2)$ and $0.9962(1)$ for $\ket{1}$ for the two nuclear spins involved in the experiment.\\

% Also, there is a nonzero probability that the electron spin state remains in $\ket{m_S = 0}$ even though outcome $1$ is measured: this error probability is $1-0.985 = 0.015$ \cite[Suppl. Table S3]{kalb2017entanglement}.

% \begin{figure}[h]
% 	\centering
% 	\includegraphics[width=0.4\textwidth]{figures/readout_fidelities.png}
% 	\caption{Readout fidelities as a function of readout duration. Reproduced from \cite[Fig.2c]{hensen2015loophole}, in which the duration is set at $3.7 \mu$s.
% 		\label{fig:readout-fidelities}
% 		}
% \end{figure}

% \begin{figure}[h]
% 	\centering
% 	\includegraphics[width=0.9\textwidth]{figures/residual_pulse_photons.png}
% 	\caption{Choice of time window start and duration. The decay of the electron and thus the emission of a photon may already happen at the start of the pulse (green line), while the time window has not been set yet. This effect reduces the entanglement generation rate. The NV-center decay constant is approximately $12$ ns. Reproduced from \cite[Extended data Fig. 3]{humphreys2018deterministic}.
% 		\label{fig:residual-pulse-photons}
% 		}
% \end{figure}

\paragraph{Reading out a memory qubit}
We again remark that next to timing constraints (Table~\ref{tab:gate_noise}), we have limited parallelism on the current NV platform, since we need the electron spin to readout the memory qubit. Reading out the nuclear spin is done by performing the following steps: 
\begin{enumerate}
\item initialize the electron spin,
\item apply an effective controlled NOT operation with the nuclear spin as control (consisting of one E-C controlled-$\sqrt{\X}$-gate and single-qubit gates),% \footnote{Possibly use the fact here that a CNOT with control and target qubit can be implemented by first applying Hadamards on both qubits, followed by a regular CNOT, followed by again Hadamards on both qubits.}
\item measure (readout) the electron spin.
\end{enumerate}
The reason why a controlled NOT is sufficient, rather than a full swap, is the following: If the nuclear spin is in state $\alpha\ket{0} + \beta\ket{1}$, then after the CNOT, the combined state is $\alpha\ket{00}_{EC} + \beta\ket{11}_{EC}$. The reduced state~\cite{Nielsen2010} of the electron is then $|\alpha|^2\dyad{0} + |\beta|^2 \dyad{1}$, so measuring in the standard basis yields the same statistics as measuring $\alpha\ket{0} + \beta\ket{1}$ in the same basis.

\paragraph{Readout noise}
Readout is modeled by performing a POVM measurement with the following Kraus operators (see~\cite{Nielsen2010} for definition)
\begin{equation}
M_0=\begin{pmatrix}\sqrt{f_0} & 0 \\ 0 & \sqrt{1-f_1}\end{pmatrix},\quad M_1=\begin{pmatrix}\sqrt{1-f_0} & 0 \\ 0 & \sqrt{f_1}\end{pmatrix}
\end{equation}
where $f_0$ ($f_1$) is the readout fidelity of the $\ket{0}$-state ($\ket{1}$) as given in table~\ref{tab:gate_noise}

% Readout is made noisy by flipping the measurement outcome with probability $p$, while leaving the post-measurement state untouched (note that this is purely classical noise). In all of these noise models, $p$ denotes the fidelity of the operation as found in the experimentalists' papers (see section above).

\subsection{Physical Entanglement Generation and Noise}
\label{sec:EG_noise}

\label{app:emit}
We here consider the single-click scheme of \Lab. 
To understand timing and quality, as well as parameter choices let us give a high-level overview of the single-click scheme: A microwave pulse is used to prepare the communication qubit in the state $\sqrt{\alpha} \ket{0} +  \sqrt{1-\alpha}\ket{1}$ (max. $5.5 \mu s$ for $A$ and $B$), where $\ket{0}$ is also called the bright state, and $\alpha$ the bright state population. 
A resonant laser pulse is
then used to cause emission of a photon, if the state was in the bright state, preparing the joint state of the communication qubit ($C$) and an emitted photon ($P$) in the state $\sqrt{\alpha}\ket{0}_C\ket{1}_P + \sqrt{1-\alpha}\ket{1}_C\ket{0}_P$, where $\ket{0}_P$ ($\ket{1}$) denotes the absence (presence) of a photon.
This process succeeds with probability $p_{\rm em} \approx 0.03$ without Purcell enhancement using optical cavities.
To ensure phase-stabilization only one laser may be used for both nodes, combined with local shutters to allow node control.
We remark that local control at the node is still desirable at a distance due to aligning with other operations such as performing gates.
to keep qubits stable.
The heralding station interferes both incoming photons on a beam splitter, thereby performing a probabilistic
entanglement swap. Intuitively, this can be understood as a measurement of the incoming qubits
in the Bell basis, where we can only obtain outcomes $\ket{\Psi^+}$, $\ket{\Psi^-}$ or ``other''. Since ``other'' is more than one possible state, we declare failure in this case.

Depending on the clicks observed in the detectors, we have projected the state of the communication qubits at $A$ and $B$ in the state $\ket{\Psi^+} = \frac{1}{\sqrt{2}}(\ket{0}_A\ket{1}_B+\ket{1}_A\ket{0}_B)$ (only left detector clicks), $\ket{\Psi^-} = \frac{1}{\sqrt{2}}(\ket{0}_A\ket{1}_B-\ket{1}_A\ket{0}_B)$ (only right detector clicks), or we declare failure (either none or both of the detectors click).
Success occurs with probability $p_{\rm succ} \approx 2\alpha p_\mathrm{det}$, where $p_\mathrm{det}$ is the probability of detecting an emitted photon.

During entanglement generation, a variety of noise processes occur:

\begin{enumerate}
    \item Dephasing of the nuclear spins (memory qubits) due to resetting the electron during entanglement generation attempts.
    \item Effective dephasing of the photon state due to uncertainty in the phase between the paths the photon travel to the beam-splitter.
    \item Effective dephasing of the photons state due to non-zero probability of emitting two photons.
    \item Effective amplitude damping of the photon state due to coherent emission, i.e. the photon is in a super-position of being emitted at different times.
    \item Effective amplitude damping due to collection losses from non-unity probability of emitting the photon in the zero-phonon line and non-unity collection efficiency into the fiber.
    \item Effective amplitude damping due to losses in fiber.
    \item Non-perfect beam-splitter measurement at the heralding stations due to photons not being perfectly indistinguishable.
    \item Errors in the classical outcome from the detectors due to non-unity detection efficiency and dark counts.
\end{enumerate}

% Currently in EasySquid, 1.-4. are performed by when the method \emph{QuantumProcessingDevice.photon\_emission} is called. \emph{QuantumProcessingDevice} uses an instance of \emph{NV\_PhotonEmissionNoiseModel} to perform the correct noise.
% Furthermore, 5. is performed by the channel object and 6. is performed by the \emph{NV\_MidPoint}.

\subsubsection{Dephasing mechanism on nuclear spins during entanglement attempts}

Between every entanglement attempt, the electron spin (communication qubit) needs to be reset.
The dominant source of noise on the nuclear spins (memory qubits) during the entanglement attempts is due to this re-initialization of the electron spin, as described in~\cite{Kalb2018}.
We model the noise on the nuclear spins as a fixed amount of dephasing noise
\begin{equation}
    \mathcal{D}_{p_d}(\rho_n) = (1-p_d)\rho_n + p_d Z\rho_n Z
\end{equation}
for every entanglement attempt.
The dephasing parameter depends on: the bright state population $\alpha$, the coupling strength $\Delta\omega$ and a decay constant $\tau_d$ as follows
\begin{equation}
    p_d = \frac{\alpha}{2} \left(1-\exp(-\Delta\omega^2\tau_d^2/2)\right),
\end{equation}
see equation (2) in~\cite{Kalb2018}.
If the length of the Bloch vector in the equatorial plane of the state in the nuclear spin is before the entanglement attempts $r_{XY}$, then after $N$ attempts the length will be
\begin{equation}
    (1-p_d)^Nr_{XY}.
\end{equation}

The bright state population can be chosen per experiment but the coupling strength and decay constant are fixed but vary between different nuclear spins. The decay constant can also vary by performing different microwave control pulses of the electron spin.
As an example of these parameters, for the nuclear spin $\mathrm{C}_1$ and the standard microwave control pulses, the coupling strength is $\Delta\omega = 2\pi\times 377$ kHz and the decay constant is $\tau_d = 82$ ns, see~\cite{Kalb2018}.
In the simulations we use these values for the coupling strength and decay constant.

\subsubsection{Phase uncertainty for photon paths}

There is uncertainty in the phase between the paths the photon travels from the nodes to the beam-splitter, due to for example uncontrolled stretching of the fiber.
If this phase difference is $\Delta\phi$ then the state after a successful measurement at the heralding station (conditioned on there being only one photon) is
\begin{equation}\label{eq:phase1}
    \ket{0}_{e_A}\ket{1}_{e_B}\pm\mathrm{e}^{\mathrm{i}\Delta\phi}\ket{1}_{e_A}\ket{0}_{e_B}.
\end{equation}
where $e_A$ is the electron spin at node $A$ and $e_B$ is the electron spin at node $B$.

We model this by performing dephasing noise to both qubits encoding the presence/absence of photon from $A$ and $B$.
As shown in~\cite{Rozpedek2018}, if we chose the dephasing parameter to be
\begin{equation}
    p_d = \left(1 - \frac{I_1(\sigma(\phi)^{-2})}{I_0(\sigma(\phi)^{-2})}\right)/2
\end{equation}
then the standard deviation of the phase $\phi$ in the state between the electron ($e_A$/$e_B$) and the photon ($p_A$/$p_B$)
\begin{equation}\label{eq:phase2}
    \ket{0}_{e_A}\ket{1}_{p_A}\pm\mathrm{e}^{\mathrm{i}\phi}\ket{1}_{e_A}\ket{0}_{p_A}
\end{equation}
is precisely $\sigma(\phi)$. To efficiently compute quotients of modified Bessel functions we implemented the algorithm described in~\cite{Amos1974}. Note that the variance of the phase in equation~\eqref{eq:phase1} is twice the variance of the phase in equation \eqref{eq:phase2}.
In experiments the standard deviation of the phase for the state between the electrons, i.e. \eqref{eq:phase1}, has in~\cite{Humphreys2018} shown to be $14.3^\circ$.
Thus, we set the standard deviation in equation~\eqref{eq:phase2} to be $14.3^\circ/\sqrt{2}$.

\subsubsection{Two-photon emission}
There is a probability that two photons are emitted from the electron during the entanglement generation attempt at a node.
For the physical setup we assume the probability of there being two photons emitted, given that at least one is emitted, to be approximately 4\%~\cite{Humphreys2018}.
As argued in~\cite{Humphreys2018}, the two-photon event can effectively be modeled as applying dephasing noise to the electron qubit which is part of the generated entanglement.

\subsubsection{Coherent emission, superposition of time-modes}

The emission of the photon is a coherent process and the photon is actually in a super-position of being emitted at different times.
As shown in~\cite{Rozpedek2018}, the effect of a finite detection window can be seen as effective amplitude damping noise to the qubit encoding the presence/absence of a photon.
The amplitude damping parameter is then given by
\begin{equation}
    p_a = \exp(-t_w/\tau_e),
\end{equation}
where $t_w$ is the detection time window and $\tau_e$ is the characteristic time of the NV emission (12 ns without cavity~\cite{Riedel2017} and 6.48 ns with cavity~\cite{Rozpedek2018}).

% There are in fact more complicated things happening during the emission of the photon which we currently do not model, as discussed with Arian.
% First thing we do not model is that during the pulse for emitting the photon there is a change that actually a second photon gets emitted, with probability of approximately 3\% as I understood it.
% The second thing we do not model is that during the first part of the emission process there is a lot of photons from the pulse and therefore more dark counts.

\subsubsection{Collection losses}

We model non-unity collection efficiencies by effective amplitude damping noise on the qubit encoding the presence/absence of a photon.
The amplitude damping parameter is given by
\begin{equation}
    p_a = (1 - p_{\mathrm{zero\_phonon}}\cdot p_{\mathrm{collection}}),
\end{equation}
where $p_\mathrm{zero\_phonon}$ is the probability of emitting a photon in the zero phonon line (3\% without cavity and 46\% with cavity~\cite{Riedel2017}) and $p_\mathrm{collection}$ is the probability of collection the photon into the fiber.
From~\cite{Humphreys2018} we know that the total detection efficiency of the system is $4\cdot10^{-4}$, which can be decomposed as
\begin{equation}\label{eq:total_det}
    p_{\mathrm{total}} = p_{\mathrm{zero\_phonon}}\cdot p_\mathrm{collection} \cdot p_\mathrm{transmission} \cdot p_\mathrm{detection},
\end{equation}
where $p_\mathrm{transmission}$ is the probability that the photon is not lost during transmission in the fiber and $p_\mathrm{detection}$ is the probability that the detector clicks, given that there was a photon.
Using equation~\eqref{eq:total_det} we find that $p_\mathrm{collection}=0.014$ given the numbers in~\cite{Humphreys2018}.
% \axel{Rewrite how we find the p collection}
% Currently, we set $p_\mathrm{collection}$ to be 2\%, the total detection probability of the setup is then close to the stated value of $4\cdot10^{-4}$ in~\cite{Humphreys2018}.
Frequency conversion succeeds with probability 30\%~\cite{Zaske2012}, so in this case we use $p_\mathrm{collection}=0.014\cdot0.3$.

\subsubsection{Transmission losses}
Since the qubit sent from the node to the heralding station is encoded in the presence/absence of a photon, the losses during transmission over fiber are modeled as amplitude damping noise.
We use an amplitude damping parameter $p_\mathrm{t\_loss}$ given as
\begin{equation}
    p_\mathrm{t\_loss} = 1 - 10^{-L\cdot \gamma / 10},
\end{equation}
where $L$ is the length of the fiber (in km) and $\gamma$ is assumed to be \SI{5}{\decibel/\kilo\meter} without frequency conversion and \SI{0.5}{\decibel/\kilo\meter} with frequency conversion.
% The parameter we use is $p_{\mathrm{t\_loss}}$ which is the probability of losing a photon per kilometer.
% Without frequency conversion $p_{\mathrm{t\_loss}}$ is $x$ (\SI{5}{\decibel}) and with frequency conversion $x$ (\SI{0.5}{\decibel}).

\subsubsection{Distinguishable photons}\label{sec:ind_photons}
Entanglement is generated between the electrons of the two nodes since the beam-splitter in the heralding station effectively deletes the information of which node a detected photon came from.
This information is only perfectly detected if the photons emitted from the nodes are completely indistinguishable.
In reality however, the photons properties (spectral, temporal etc.) can be slightly different and they are therefore not completely indistinguishable.
In section~\ref{sec:ind_photons} we derive effective measurement operators of a beam-splitter measurement, taking photon indistinguishability into account, which we make use of in our simulation.
For the physical setup we simulate, the overlap (visibility) of the photons coming from the nodes is approximately $0.9$~\cite{Humphreys2018}, i.e. $\abs{\mu}^2=0.9$ where $\mu$ is defined in equation~\eqref{eq:mu}.

\subsubsection{Detection losses and dark counts}
Detection losses and dark counts are modeled by probabilistically changing the ideal classical outcome from the detectors at the heralding station.
For each detector, if the ideal detector clicked the noisy detector also clicks with probability $p_\mathrm{detection}$ and otherwise not.
In the simulations we use $p_\mathrm{detection}=0.8$, as measured in~\cite{Hensen2015}.

If the ideal detector did not click the noisy detector does click with probability $p_\mathrm{dark}$.
The parameter used for the dark count is the dark count rate $\lambda_\mathrm{dark}=20$ per second~\cite{Humphreys2018}.
The dark counts follow a Poisson distribution and we have that
\begin{equation}
    p_\mathrm{dark} = 1 - \exp(- t_w \cdot \lambda_\mathrm{dark}),
\end{equation}
where $t_w$ is the detection time window.

%%% XX Heralding
\newcommand{\create}[2]{#1^\dagger(\omega_{#2})}
\newcommand{\annila}[2]{#1(\omega_{#2})}
\newcommand{\dirac}[2]{\delta(\omega_#1-\omega_#2)}
\newcommand{\intW}[1]{\int\!\!\mathrm{d}{\omega_{#1}}\;}
\newcommand{\expPW}[1]{\mathrm{e}^{\mathrm{i}\omega_{#1}\tau}}
\newcommand{\expMW}[1]{\mathrm{e}^{-\mathrm{i}\omega_{#1}\tau}}
\newcommand{\vacket}[1]{\ket{0}_{#1}} 
\newcommand{\vacbra}[1]{\bra{0}_{#1}} 
\newcommand{\vac}[1]{\ketbra{0}{0}_{#1}} 
\newcommand{\isom}{U_{lr\rightarrow cd}} 
\newcommand{\isomdag}{(U_{lr\rightarrow cd})^\dagger} 
\newcommand{\phiW}[1]{\phi(\omega_{#1})} 
\newcommand{\psiW}[1]{\psi(\omega_{#1})} 
\newcommand{\phiWconj}[1]{\phi^*(\omega_{#1})} 
\newcommand{\psiWconj}[1]{\psi^*(\omega_{#1})} 

\subsection{Heralding station}\label{app:heralding}

Let us now consider the measurement at the Heralding Station in more detail in order to understand its error models.

\subsubsection{Distinguishable photons}

We here describe how we model a beam-splitter measurement of two photons which are not perfectly indistinguishable.
This is relevant for many heralding entanglement generation schemes, since if photons are distinguishable the beam-splitter will not erase the information of where the photons came from.
Two perfectly indistinguishable photons incident on a beam-splitter will always go to the same output arm, as captured by the Hong-Ou-Mandel effect~\cite{hong}.
However, if the photons are distinguishable they do not necessarily go to the same output arm, which can be detected in experiment.
For a given setup, lets denote the probability that two incident photons on the beam-splitter go to different output arms as $\chi$.

We will in this section derive the effective POVM and Kraus operators correspond to detecting photons at the ends of the output arms of the beam-splitter in terms of $\chi$, under the assumptions described below and using ideas from the paper~\cite{branczyk} where $\chi$ is computed.

\subsubsection{Model}

Assume that there is a 50:50 beam-splitter with input arms $a$ and $b$ and output arms $c$ and $d$. At the end of the output arms there are photon detectors that can click.
We will assume that the detectors have a flat frequency response and at first that the detectors can count photons, i.e. there are different measurement outcomes for there being one or two photons incident on a detector.
However we will show below how one can easily consider detectors which cannot count photons from the analysis in this note.

\paragraph{Photons}

In many simulations we model the presence or absence of a photon as a two-level system, i.e. a qubit $\alpha\ket{0}+\beta\ket{1}$, where $\ket{0}$ means no photon and $\ket{1}$ one photon.
We would then describe the state before the beam-splitter as a state living in a 2-qubit Hilbert space spanned by the following four basis vectors:
\begin{equation}\label{eq:lr_basis}
    \ket{00}_{lr},\quad\ket{01}_{lr},\quad\ket{10}_{lr},\quad\ket{11}_{lr}
\end{equation}
describing 0 photons, photon on the right, photon on the left and two photons.
Here $l$ and $r$ corresponds to arm $a$ and $b$ of the beam-splitter, but we distinguish these since we will denote $a$ (and $b$) as the infinite dimensional Hilbert space describing the spectral property of the photon.
Note that we assume that there are never more than one photon per arm.

Describing the presence and absence of a photon as a qubit masks the fact that a photon can have many other degrees of freedom, such as polarization, spectral and temporal properties.
We will here focus on spectral and temporal properties and will therefore model a photon in arm $a$ with a spectral amplitude function $\phi$ as the state
\begin{equation}\label{eq:state1}
    \intW{}\phi(\omega)\create{a}{}\vacket{a},
\end{equation}
where $\create{a}{}$ is the creation operator of a photon in arm $a$ of frequency $\omega$ and $\vacket{a}$ is the vacuum and $\phi$ is normalized such that
\begin{equation}\label{eq:norm}
    \intW{}\abs{\phiW{}}^2 = 1.
\end{equation}
Furthermore, the state of arm $b$ will be described by a spectral amplitude function $\psi$ as
\begin{equation}
    \intW{}\psi(\omega)\create{b}{}\vacket{b}.
\end{equation}
Two photons arriving at the beam-splitter can have different spectral properties, captured by $\phi$ and $\psi$ being different.
We will also include a possible temporal shift $\tau$ between the arrival times of the two photons.
As described in equation (16) of~\cite{branczyk}, a temporal shift of a photon in arm $b$ induces the following action on the creation operators
\begin{equation}
    \create{b}{}\rightarrow\create{b}{}\expMW{}.
\end{equation}

\paragraph{Beam-splitter}
The 50:50 beam-splitter acts on the creation operators in the following way:
\begin{align}
    \create{a}{}\rightarrow\frac{1}{\sqrt{2}}(\create{c}{}+\create{d}{}) \\
    \create{b}{}\rightarrow\frac{1}{\sqrt{2}}(\create{c}{}-\create{d}{}).
\end{align}
Thus the state of a photon described as in equation~\eqref{eq:state1}, i.e. one photon in the input arm $a$,  will after the beam-splitter become
\begin{equation}\label{eq:cd_basis1}
    \ket{\phi}_{cd}=\frac{1}{\sqrt{2}}\intW{}\phi(\omega)(\create{c}{}+\create{d}{})\vacket{cd}.
\end{equation}
Furthermore, the three other cases of no photon, one photon in the input arm $b$ and one photon in each input arm becomes after the beam-splitter:
\begin{align}
    \ket{0}_{cd} & \label{eq:cd_basis2}\\
    \ket{\psi}_{cd}&=\frac{1}{\sqrt{2}}\intW{}\psi(\omega)\expMW{}(\create{c}{}-\create{d}{})\vacket{cd} \label{eq:cd_basis3}\\
    \ket{\phi,\psi}_{cd}&=\frac{1}{2}\intW{1}\intW{2}\phiW{1}\psiW{2}\expMW{1}(\create{c}{1}+\create{d}{1})(\create{c}{2}-\create{d}{2})\vacket{cd} \label{eq:cd_basis4}.
\end{align}
Where the states $\ket{0}_{cd}$, $\ket{\phi}_{cd}$, $\ket{\psi}_{cd}$ and $\ket{\phi,\psi}_{cd}$ should be thought of as the corresponding states to the states in equation~\eqref{eq:lr_basis}.
Below, we will in fact formally define an isometry between these two Hilbert spaces.

\paragraph{Detectors}
As mentioned we assume that the detectors have a flat frequency response.
The event that the detector in arm $c$ detected one photon can then be described by the projector
\begin{equation}\label{eq:P10}
    P_{1,0}=\intW{}\create{c}{}\vac{cd}\annila{c}{}
\end{equation}
Since we assume that there is maximally one photon arriving at each input arm of the beam-splitter the only other possible measurement outcomes are described by the following projectors:
\begin{align}
    P_{0,0} &= \vac{cd} \label{eq:P00}\\
    P_{0,1} &= \intW{}\create{d}{}\vac{cd}\annila{d}{} \label{eq:P01}\\
    P_{1,1} &= P_{1,0}\otimes P_{0,1} = \intW{1}\intW{2}\create{c}{1}\create{d}{2}\vac{cd}\annila{c}{1}{}\annila{d}{2} \label{eq:P11}\\
    P_{2,0} &= \frac{1}{2}\intW{1}\intW{2}\create{c}{1}\create{c}{2}\vac{cd}\annila{c}{1}\annila{c}{2} \label{eq:P20}\\
    P_{0,2} &= \frac{1}{2}\intW{1}\intW{2}\create{d}{1}\create{d}{2}\vac{cd}\annila{d}{1}\annila{d}{2} \label{eq:P02}
\end{align}
where $P_{0,0}$ corresponds to no photon, $P_{0,1}$ one photon in arm $d$, $P_{1,1}$ one photon in each arm, $P_{2,0}$ two photons in arm $c$ and $P_{0,2}$ two photons in arm $d$.
Note that the factors of $\frac{1}{2}$ are needed for $P_{20}$ such that $P_{20}^2=P_{20}$ and similarly with $P_{02}$.

\paragraph{Deriving effective POVM on presence/absence description}
The goal of this note is to derive the effective POVM on the Hilbert space $lr$, spanned by vectors in equation~\eqref{eq:lr_basis}, induced by the projective measurements in equations~\eqref{eq:P10}-\eqref{eq:P02}
on the infinite-dimensional Hilbert space $cd$.
To do this we will first define an isometry $\isom$ from the Hilbert space $lr$ to $cd$, using the states in equation~\eqref{eq:lr_basis} and equations~\eqref{eq:cd_basis1}-\eqref{eq:cd_basis4}.
This isometry will have the following action on the basis states of $lr$:
\begin{align}
    \ket{00}_{lr} &\rightarrow \ket{0}_{cd} \\
    \ket{01}_{lr} &\rightarrow \ket{\psi}_{cd} \\
    \ket{10}_{lr} &\rightarrow \ket{\phi}_{cd} \\
    \ket{11}_{lr} &\rightarrow \ket{\phi,\psi}_{cd}
\end{align}
and will therefore be given as
\begin{equation}
    \isom = \ket{0}_{cd}\bra{00}_{lr} + \ket{\psi}_{cd}\bra{01}_{lr} + \ket{\phi}_{cd}\bra{10}_{lr} + \ket{\phi,\psi}_{cd}\bra{11}_{lr}.
\end{equation}
One can easily check that the states $\ket{0}_{cd}$, $\ket{\phi}_{cd}$, $\ket{\psi}_{cd}$ and $\ket{\phi,\psi}_{cd}$ are mutually orthogonal and that $\isom$ is therefore indeed an isometry, i.e.
\begin{equation}
    \isomdag\isom = \mathbbm{1}_{lr}.
\end{equation}
Lets assume that $\ket{\Phi}_{lr}$ is a state in $lr$ and we wish to compute the probability of receiving a measurement outcome corresponding to the projector $P\in\{P_{00},P_{10},P_{01},P_{11},P_{20},P_{02}\}$ for the state $\isom\ket{\Phi}_{lr}$.
Using Born's rule we find that this probability is given as
\begin{equation}
    \bra{\Phi}_{lr}\isomdag P\isom\ket{\Phi}_{lr} = \tr[\isomdag P\isom\ketbra{\Phi}_{lr}].
\end{equation}
From the above equation we find that the effective POVM on $lr$ is given as
\begin{equation}\label{eq:povm_elem}
    \{\isomdag P\isom\;:\;P\in\{P_{00},P_{10},P_{01},P_{11},P_{20},P_{02}\}\}.
\end{equation}
whose elements we will denote as $M_{00}$, $M_{10}$, $M_{01}$, $M_{11}$, $M_{20}$ and $M_{02}$.
In section~\ref{sec:eff_povm} we compute what these POVM-elements are and find a choice of Kraus operators in section~\ref{sec:eff_krauss} for both the case then the detector can count photons and when it cannot.

\paragraph{Effective POVMs}\label{sec:eff_povm}
Here we compute the POVM-elements in equation~\eqref{eq:povm_elem} one-by-one.

\noindent\underline{\textbf{$M_{11}$:}}\newline
Let's start with $M_{11}$ since this will allow us to relate these POVM-elements to $\chi$, i.e. the probability that both detectors click, given that there were one photon in each input arm.
The operator $P_{11}$ only has non-zero overlap with the term $\ket{\phi,\psi}_{cd}\bra{11}_{lr}$ of $\isom$ and is therefore given as
\begin{equation}
    M_{11} = \isomdag P_{11}\isom = \ket{11}_{lr}\bra{\phi,\psi}_{cd}P_{11}\ket{\phi,\psi}_{cd}\bra{11}_{lr}.
\end{equation}
Lets evaluate the factor $\bra{\phi,\psi}_{cd}P_{11}\ket{\phi,\psi}_{cd}$.
Using equation~\eqref{eq:cd_basis4} and equation~\eqref{eq:P11} we find that the above expression evaluates to
\begin{align}
    \bra{\phi,\psi}_{cd}P_{11}\ket{\phi,\psi}_{cd} &= \frac{1}{2}\intW{1}\intW{2}\phiWconj{1}\psiWconj{2}\expPW{2}\vacbra{cd}(\annila{c}{1}+\annila{d}{1})(\annila{c}{2}-\annila{d}{2}) \nonumber\\
           &\quad\times \intW{3}\intW{4}\create{c}{3}\create{d}{4}\vac{cd}\annila{c}{3}{}\annila{d}{4} \nonumber\\
           &\quad\times \frac{1}{2}\intW{5}\intW{6}(\create{c}{5}+\create{d}{5})(\create{c}{6}-\create{d}{6})\vacket{cd}\phiW{5}\psiW{6}\expMW{6}\\
           &= \frac{1}{4}\intW{1}\intW{2}\intW{3}\intW{4}\intW{5}\intW{6}\phiWconj{1}\psiWconj{2}\phiW{5}\psiW{6}\expPW{2}\expMW{6}\Big(\nonumber\\
           &\quad+ \dirac{2}{3}\dirac{3}{6}\dirac{1}{4}\dirac{4}{5} \nonumber\\
           &\quad- \dirac{2}{3}\dirac{3}{5}\dirac{1}{4}\dirac{4}{6} \nonumber\\
           &\quad- \dirac{1}{3}\dirac{3}{6}\dirac{2}{4}\dirac{4}{5} \nonumber\\
           &\quad+ \dirac{1}{3}\dirac{3}{5}\dirac{2}{4}\dirac{4}{6}\Big)\label{eq:expr1}
\end{align}
where we used the fact that $\vacbra{cd}\annila{c}{1}\create{c}{2}\vacket{cd}=\dirac{1}{2}$ and similarly for arm $d$.
Using that
\begin{equation}
    \intW{2}f(\omega_2)\dirac{1}{2}=f(\omega_1)
\end{equation}
we find that equation~\eqref{eq:expr1} evaluates to
\begin{equation}
    \frac{1}{2}\intW{1}\intW{2}(\phiWconj{1}\psiWconj{2}\phiW{1}\psiW{2}\expPW{2}\expMW{2}-\phiWconj{1}\psiWconj{2}\phiW{2}\psiW{1}\expPW{2}\expMW{1}
\end{equation}
Finally using equation~\eqref{eq:norm} we find that $M_{11}$ evaluates to
\begin{equation}\label{eq:M11}
    M_{11}=\frac{1}{2}(1-\abs{\mu}^2)\ketbra{11}_{lr}
\end{equation}
where
\begin{equation}\label{eq:mu}
    \mu = \intW{}\phiWconj{}\psiW{}\expMW{}.
\end{equation}
From equation~\eqref{eq:M11} we can relate $\abs{\mu}$ to $\chi$ as
\begin{equation}
    \chi = \frac{1}{2}(1-\abs{\mu}^2).
\end{equation}

\noindent\underline{\textbf{$M_{20}$:}}\newline
The operator $P_{20}$ only has non-zero overlap with the term $\ket{\phi,\psi}_{cd}\bra{11}_{lr}$ of $\isom$ and is therefore given as
\begin{equation}
    M_{11} = \isomdag P_{20}\isom = \ket{11}_{lr}\bra{\phi,\psi}_{cd}P_{20}\ket{\phi,\psi}_{cd}\bra{11}_{lr}.
\end{equation}
Lets evaluate the factor $\bra{\phi,\psi}_{cd}P_{20}\ket{\phi,\psi}_{cd}$.
Using equation~\eqref{eq:cd_basis4} and equation~\eqref{eq:P20} we find that the above expression evaluates to
\begin{align}
    \bra{\phi,\psi}_{cd}P_{20}\ket{\phi,\psi}_{cd} &= \frac{1}{2}\intW{1}\intW{2}\phiWconj{1}\psiWconj{2}\expPW{2}\vacbra{cd}(\annila{c}{1}+\annila{d}{1})(\annila{c}{2}-\annila{d}{2}) \nonumber\\
           &\quad\times \frac{1}{2}\intW{3}\intW{4}\create{c}{3}\create{c}{4}\vac{cd}\annila{c}{3}{}\annila{c}{4} \nonumber\\
           &\quad\times \frac{1}{2}\intW{5}\intW{6}(\create{c}{5}+\create{d}{5})(\create{c}{6}-\create{d}{6})\vacket{cd}\phiW{5}\psiW{6}\expMW{6}\\
           &= \frac{1}{8}\intW{1}\intW{2}\intW{3}\intW{4}\intW{5}\intW{6}\phiWconj{1}\psiWconj{2}\phiW{5}\psiW{6}\expPW{2}\expMW{6}\nonumber\\
           &\quad\times \Big( \dirac{1}{4}\dirac{2}{3} +\dirac{1}{3}\dirac{2}{4}\Big)\nonumber\\
           &\quad\times \Big( \dirac{3}{6}\dirac{4}{5} +\dirac{3}{5}\dirac{4}{6}\Big)\label{eq:expr2}
\end{align}
where we used the fact that $\vacbra{cd}\annila{c}{1}\annila{c}{2}\create{c}{3}\create{c}{4}\vacket{cd}=\dirac{1}{3}\dirac{2}{4}+\dirac{2}{3}\dirac{1}{4}$.
Then similarly to $M_{11}$ we find that equation~\eqref{eq:expr2} evaluates to
\begin{equation}
    \bra{\phi,\psi}_{cd}P_{20}\ket{\phi,\psi}_{cd} = \frac{1}{4}(1+\abs{\mu}^2)
\end{equation}
and we thus find $M_{20}$ to be
\begin{equation}
    M_{20} = \frac{1}{4}(1+\abs{\mu}^2)\ketbra{11}_{lr}.
\end{equation}

\noindent\underline{\textbf{$M_{20}$:}}\newline
Similarly to $M_{20}$ we find that $M_{02}$ evaluates to
\begin{equation}
    M_{02} = \frac{1}{2}(1+\abs{\mu}^2)\ketbra{11}_{lr}.
\end{equation}

\noindent\underline{\textbf{$M_{10}$:}}\newline
The operator $P_{10}$ only has non-zero overlap with the terms $\ket{\phi}_{cd}\bra{10}_{lr}$ and $\ket{\psi}_{cd}\bra{01}_{lr}$ of $\isom$ and is therefore given as
\begin{equation}
    M_{10} = \isomdag P_{10}\isom = \Big(\ket{10}_{lr}\bra{\phi}_{cd}+\ket{01}_{lr}\bra{\psi}_{cd}\Big)P_{10}\Big(\ket{\phi}_{cd}\bra{10}_{lr}+\ket{\psi}_{cd}\bra{01}_{lr}\Big).
\end{equation}
Lets evaluate the factors $\bra{\phi}_{cd}P_{10}\ket{\phi}_{cd}$, $\bra{\psi}_{cd}P_{10}\ket{\psi}_{cd}$, $\bra{\phi}_{cd}P_{10}\ket{\psi}_{cd}$ and $\bra{\psi}_{cd}P_{10}\ket{\phi}_{cd}$ one-by-one.
First we have that:
\begin{align}
    \bra{\phi}_{cd}P_{10}\ket{\phi}_{cd} &= \frac{1}{\sqrt{2}}\intW{1}\phiWconj{1}\vacbra{cd}(\annila{c}{1}+\annila{d}{1}) \nonumber\\
           &\quad\times \intW{2}\create{c}{2}\vac{cd}\annila{c}{2}{} \nonumber\\
           &\quad\times \frac{1}{\sqrt{2}}\intW{3}(\create{c}{3}+\create{d}{3})\vacket{cd}\phiW{3}\nonumber\\
           &= \frac{1}{2}\intW{1}\intW{2}\intW{3}\phiWconj{1}\phiW{3}\dirac{1}{2}\dirac{2}{3}\nonumber\\
           &= \frac{1}{2}\intW{}\abs{\phiW{}}^2\nonumber\\
           &= \frac{1}{2}.
\end{align}
and similarly that
\begin{equation}
    \bra{\psi}_{cd}P_{10}\ket{\psi}_{cd} = \frac{1}{2}.
\end{equation}
Furthermore, we find that
\begin{align}
    \bra{\phi}_{cd}P_{10}\ket{\psi}_{cd} &= \frac{1}{\sqrt{2}}\intW{1}\phiWconj{1}\vacbra{cd}(\annila{c}{1}+\annila{d}{1}) \nonumber\\
           &\quad\times \intW{2}\create{c}{2}\vac{cd}\annila{c}{2}{} \nonumber\\
           &\quad\times \frac{1}{\sqrt{2}}\intW{3}(\create{c}{3}-\create{d}{3})\vacket{cd}\psiW{3}\expMW{3}\nonumber\\
           &= \frac{1}{2}\intW{1}\intW{2}\intW{3}\phiWconj{1}\psiW{3}\expMW{3}\dirac{1}{2}\dirac{2}{3}\nonumber\\
           &= \frac{1}{2}\intW{}\phiWconj{}\psiW{}\expMW{}\nonumber\\
           &= \frac{1}{2}\mu.
\end{align}
where $\mu$ is defined in equation~\eqref{eq:mu}.
One easily then finds that
\begin{equation}
    \bra{\psi}_{cd}P_{10}\ket{\phi}_{cd}=(\bra{\phi}_{cd}P_{10}\ket{\psi}_{cd})^*=\frac{1}{2}\mu^*.
\end{equation}
Combining the above results, we find that $M_{10}$ is given as
\begin{equation}
    M_{10} = \frac{1}{2}\Big(\ketbra{10}_{lr}+\ketbra{01}_{lr}+\mu\ketbra{10}{01}_{lr}+\mu^*\ketbra{01}{10}\Big)
\end{equation}

\noindent\underline{\textbf{$M_{10}$:}}\newline
Similarly to $M_{10}$ one finds that $M_{01}$ evaluates to
\begin{equation}
    M_{01} = \frac{1}{2}\Big(\ketbra{10}_{lr}+\ketbra{01}_{lr}-\mu\ketbra{10}{01}_{lr}-\mu^*\ketbra{01}{10}\Big)
\end{equation}

\noindent\underline{\textbf{$M_{00}$:}}\newline
Its easy to see that
\begin{equation}
    M_{00}=\ketbra{00}_{lr}.
\end{equation}

\paragraph{POVM for photon-counter detectors}
To summarize we found that the POVM-elements are given as
\begin{align}
    M_{00} &= \begin{pmatrix} 1 & 0 & 0 & 0 \\ 0 & 0 & 0 & 0 \\ 0 & 0 & 0 & 0 \\ 0 & 0 & 0 & 0 \end{pmatrix} \label{eq:povm1}\\
    M_{10} &= \frac{1}{2}\begin{pmatrix} 0 & 0 & 0 & 0 \\ 0 & 1 & \mu & 0 \\ 0 & \mu^* & 1 & 0 \\ 0 & 0 & 0 & 0 \end{pmatrix} \label{eq:povm2}\\
    M_{01} &= \frac{1}{2}\begin{pmatrix} 0 & 0 & 0 & 0 \\ 0 & 1 & -\mu & 0 \\ 0 & -\mu^* & 1 & 0 \\ 0 & 0 & 0 & 0 \end{pmatrix} \label{eq:povm3}\\
    M_{11} &= \frac{1}{2}\begin{pmatrix} 0 & 0 & 0 & 0 \\ 0 & 0 & 0 & 0 \\ 0 & 0 & 0 & 0 \\ 0 & 0 & 0 & 1-\abs{\mu}^2 \end{pmatrix} \label{eq:povm4}\\
    M_{20} &= \frac{1}{4}\begin{pmatrix} 0 & 0 & 0 & 0 \\ 0 & 0 & 0 & 0 \\ 0 & 0 & 0 & 0 \\ 0 & 0 & 0 & 1+\abs{\mu}^2 \end{pmatrix} \label{eq:povm5}\\
    M_{02} &= \frac{1}{4}\begin{pmatrix} 0 & 0 & 0 & 0 \\ 0 & 0 & 0 & 0 \\ 0 & 0 & 0 & 0 \\ 0 & 0 & 0 & 1+\abs{\mu}^2 \end{pmatrix}\label{eq:povm6}
\end{align}
where the rows and columns of the above matrices are ordered as $\ketbra{00}_{lr},\ketbra{10}_{lr},\ketbra{01}_{lr},\ketbra{11}_{lr}$ and $\mu$ is given as
\begin{equation}
    \mu = \intW{}\phiWconj{}\psiW{}\expMW{}.
\end{equation}
and is related by to the probability that both detectors click, given that there were one photon in each input arm $\chi$ as
\begin{equation}
    \chi = \frac{1}{2}(1-\abs{\mu}^2).
\end{equation}

\paragraph{POVM for non-photon-counter detectors}
If the detectors used cannot distinguish between one and two photons we can simply add the POVM elements $M_{10}$ and $M_{20}$ to get a new POVM given as
\begin{align}
    \tilde{M}_{00} &= \begin{pmatrix} 1 & 0 & 0 & 0 \\ 0 & 0 & 0 & 0 \\ 0 & 0 & 0 & 0 \\ 0 & 0 & 0 & 0 \end{pmatrix} \label{eq:povm7}\\
    \tilde{M}_{10} &= \frac{1}{2}\begin{pmatrix} 0 & 0 & 0 & 0 \\ 0 & 1 & \mu & 0 \\ 0 & \mu^* & 1 & 0 \\ 0 & 0 & 0 & (1+\abs{\mu}^2)/2 \end{pmatrix} \label{eq:povm8}\\
    \tilde{M}_{01} &= \frac{1}{2}\begin{pmatrix} 0 & 0 & 0 & 0 \\ 0 & 1 & -\mu & 0 \\ 0 & -\mu^* & 1 & 0 \\ 0 & 0 & 0 & (1+\abs{\mu}^2)/2 \end{pmatrix} \label{eq:povm9}\\
    \tilde{M}_{11} &= \frac{1}{2}\begin{pmatrix} 0 & 0 & 0 & 0 \\ 0 & 0 & 0 & 0 \\ 0 & 0 & 0 & 0 \\ 0 & 0 & 0 & 1-\abs{\mu}^2 \end{pmatrix}\label{eq:povm10}
\end{align}

\subsubsection{Effective Kraus operators}\label{sec:eff_krauss}
Given the POVMs in equation~\eqref{eq:povm1}-\eqref{eq:povm6} and equation~\eqref{eq:povm7}-\eqref{eq:povm10} 
one can choose corresponding Kraus operators for these measurements by taking the matrix square root of the corresponding POVM-elements.
Assuming that $\mu$ is real one finds a set of Kraus operators of the POVM $\{\tilde{M}_{00},\tilde{M}_{10},\tilde{M}_{01},\tilde{M}_{11}\}$ to be
\begin{align}
    \tilde{E}_{00} &= \begin{pmatrix} 1 & 0 & 0 & 0 \\ 0 & 0 & 0 & 0 \\ 0 & 0 & 0 & 0 \\ 0 & 0 & 0 & 0 \end{pmatrix} \label{eq:krauss1}\\
    \tilde{E}_{10} &= \frac{1}{2}\begin{pmatrix} 0 & 0 & 0 & 0 \\ 0 & (\sqrt{1+\mu}+\sqrt{1-\mu})/\sqrt{2} & (\sqrt{1+\mu}-\sqrt{1-\mu})/\sqrt{2} & 0 \\ 0 & (\sqrt{1+\mu}-\sqrt{1-\mu})/\sqrt{2} & (\sqrt{1+\mu}+\sqrt{1-\mu})/\sqrt{2} & 0 \\ 0 & 0 & 0 & \sqrt{1+\abs{\mu}^2} \end{pmatrix} \label{eq:eq:krauss2}\\
    \tilde{E}_{01} &= \frac{1}{2}\begin{pmatrix} 0 & 0 & 0 & 0 \\ 0 & (\sqrt{1+\mu}+\sqrt{1-\mu})/\sqrt{2} & (\sqrt{1-\mu}-\sqrt{1+\mu})/\sqrt{2} & 0 \\ 0 & (\sqrt{1-\mu}-\sqrt{1+\mu})/\sqrt{2} & (\sqrt{1+\mu}+\sqrt{1-\mu})/\sqrt{2} & 0 \\ 0 & 0 & 0 & \sqrt{1+\abs{\mu}^2} \end{pmatrix} \label{eq:eq:krauss3}\\
    \tilde{E}_{11} &= \frac{1}{\sqrt{2}}\begin{pmatrix} 0 & 0 & 0 & 0 \\ 0 & 0 & 0 & 0 \\ 0 & 0 & 0 & 0 \\ 0 & 0 & 0 & \sqrt{1-\abs{\mu}^2} \end{pmatrix}\label{eq:eq:krauss4}.
\end{align}

\subsection{Classical communication}
\subsubsection{Optical Link Error Model}
\label{app:classical}
\label{sec:optical_link_error_model}
\label{app:losses}

We claimed that we highly inflated losses in the simulation to stress test our protocol. We now consider more realistic values for such errors by considering a realistic \emph{packet-level} error model for the non-quantum optical link. For this we have assumed that two quantum internet end nodes are connected by a legacy 1000BASE-ZX single-mode \SI[mode=text]{1550}{\nano\meter} wavelength Gigabit Ethernet link. The reason for choosing 1000BASE-ZX interface is (i) its achievable long-distance transmission at least up to \SI[mode=text]{70}{\kilo\meter} with no dependency on optical repeaters and (ii) decades of its successful deployment within magnitude of networks worldwide. 

To be conservative, our optical Gigabit Ethernet model assumes a typical worst-case optical link budget (\SI[mode=text]{0.5}{\decibel/\kilo\meter} attenuation\footnote{Fibers measured for \Qlink\ have been found to have this loss level.}, \SI[mode=text]{0.7}{\decibel}/connector loss, \SI{0.1}{\decibel}/splice/(joint) loss, and \SI[mode=text]{3}{\decibel} safety margin)~\cite{cisco_linkloss_2005}. We also assume a typical worst-case \SI{-1}{\dBm} optical transmission power and \SI[mode=text]{-24}{\decibel} receiver sensitivity of a 1000BASE-ZX small form-factor hot pluggable transceiver, see e.g.~\cite{globix_spfzx_2013}. For a maximum realism of link error over an optical link we model a \emph{IEEE 802.3 frame errors}, instead of modeling individual bit errors of every message sent across the network. The latter would require a software implementation of a complete modulation and coding layer of IEEE 802.3 which is beyond the scope of this work. Using measurement trace-driven packet-level Gigabit Ethernet frame error data from~\cite[Table 6.1]{james_phdthesis_2005} we have mapped the received SNR per transmitter/receiver distance to the respective frame error probability, which was then applied to every classical message sent over an optical link between quantum end nodes. SNR values that were not represented in the measurements of~\cite{james_phdthesis_2005} have been linearly interpolated. We have not distinguished between the lengths of each classical message as the model of~\cite{james_phdthesis_2005} has aggregated over all messages captured over a measured campus Ethernet link (cf.~\cite[Fig. 6.1]{james_phdthesis_2005}). We note that our modeling approach is equivalent to the frame error models applied in e.g. NS-3~\cite{ns3_wifi_error} for WiFi frame errors.

For two example long-distance Quantum Internet typologies (node-to-node distance of \SI{15}{\kilo\meter} and \SI{20}{\kilo\meter}, respectively) we have ended up in a perfect frame error probability, with the assumption that amount of splices is zero\footnote{Which is consistent with the measurements, e.g. in~\cite[Section 4]{corndorf_ccr_2004}.} (we only start to observe frame errors only at transmitter/receiver distance exceeding 40\,km for the above model variables, with a very narrow transition error between no frame error rate and disconnected interface, i.e. frame error rate of one). Even when we increase the number of splices to an exaggerated level, say 30 splices for a \SI{15}{\kilo\meter} interface (with 0.3\,dB loss/splice), we still observe a very low frame error probability of 4$\times$10$^{\text{-8}}$. Therefore, to test the effect of frame errors on the non-quantum optical link on the Quantum Internet protocol stack---in the cases of extreme frame loss---we have increased the value of frame error to 10$^{\text{-4}}$ (and tested frame error rate to up to 10$^{\text{-10}}$---an error rate level of a \SI{20}{\kilo\meter} link with 21 splices---in steps of 10$^{\text{-1}}$). If our protocol would work in such a high (but unrealistic) condition then it would also work on a realistic low-error optical link.

\subsubsection{Optical Link CRC Error Model}
\label{sec:Optical_link_CRC_error_model}

Additionally, we have investigated a non-zero probability of CRC not being able to detect a frame error. Assuming the same optical link type (e.g. 1000BASE-ZX) we have used a model of~\cite{fujiwara_tcom_1989} to calculate the respective probability of not detecting a CRC frame error within a IEEE 802.3 frame. For this we have mapped the transmitter/receiver distance to the respective SNR (the same way as described in Section~\ref{app:classical}). Then we mapped the SNR to the respective BER using~\cite[Table 6.3]{james_phdthesis_2005} (performing the same process of interpolating SNR between the points not measured by~\cite{james_phdthesis_2005} as for the optical link error model, see again Section~\ref{sec:optical_link_error_model}) and then using~\cite[Fig. 1]{fujiwara_tcom_1989} mapped this resulting SNR to the respective probability of undetected error. We have assumed a worst case scenario of the longest IEEE 802.3 frame (i.e. $n=12144$ bits, that is a maximum MTU). Again, for any of the two Quantum Internet lengths mentioned above, we do not find any CRC errors. At the highly-spliced case, considered in Section~\ref{app:classical}, we obtain an \emph{extremely low} CRC error rate of 1.4$\times$10$^{\text{-23}}$. Therefore such errors were decided to be ignored in our implementation. Another reason for not considering these errors: it would require a full implementation of en- and decoding of classical frames which outside the scope of this work.

\section{Protocols}\label{app:protocol}
Here we give details of the implementation of the physical and link layer protocols in our simulations.  Python implementation is available upon request.
\subsection{Distributed Queue Protocol}
\label{sec:distributed_queue_protocol}

In order to track the individual applications that the entangled qubits belong to, the EGP makes use of a \emph{distributed queue} which shares request information between peers. Management of the distributed queue is performed by the Distributed Queue Protocol (DQP). In addition to storing the parameters supplied with a CREATE request generated by the layers above the link layer\footnote{Refer to Section~\ref{sec:designLinkLayer} of the main paper for the details of the CREATE request.}, DQP will keep additional information about each entanglement request including its \emph{create\_time}, \emph{min\_time} at which the request may be executed and \emph{MHP timeout cycle} by which the entanglement request will time out. We proceed with the introduction of the DQP by describing the structure of priority queues, followed by the queue establishment process, DQP message sequence diagram and DQP associated messages. 

\subsubsection{Priority Queues}
\label{sec:priority_queues}

Priorities are necessary to fulfill the use case requirements outlined in section \ref{sec:designArch}. This is accomplished by adding requests to different types of queues $\mathcal{Q} = \{Q_1,\ldots,Q_L\}$, where $L$ is the total number of queues in the distributed queue. Each queue can contain a maximum of $x$ \emph{items} simultaneously (in other words $x$ is the maximum size of each individual queue), where an item is an individual entanglement request with its associated metadata, e.g. create\_time, min\_time, MHP timeout cycle. Each CREATE request is assigned a queue number by the scheduler (see Section~\ref{sec:egp_scheduler} below), and receives an absolute queue ID which is a tuple $(j, i_j)$ where $j$ indicates the designated queue $Q_j$ (or, more abstractly, the \emph{queue ID} of the entanglement request) and $i_j$ is a unique ID within $Q_j$. Equivalently, for a finite number of queues we will denote $(j, i_j)$ as the \emph{absolute queue ID} or $a_{ID}$, and use $(j,i_j) \in \mathcal{Q}$ to indicate the ID of the request.

The queue ID must obey the following properties:

\begin{itemize}
\item \emph{Total order}: Items on each queue follow a total order of items waiting in the queue determined by $i_j$.
\item \emph{Arrival time}: ID of an entanglement (CREATE) request is a function of its arrival time. Let $t_1$ and $t_2$ denote the create\_time of entanglement requests 1 and 2, respectively. Then, let $i_1$ and $i_2$ denote their respective queue ID's. If both requests are added to the same queue $Q_j$, and $t_1 < t_2$, then $(i_2 - i_1) \mod x > 0$. That is, if a CREATE request arrives earlier, it will also receive a lower queue ID.
\end{itemize}

We will now outline the distributed queue establishment within DQP. For simplicity of exposition, we now assume there is only one queue, i.e., $L=1$.

\subsubsection{DQP Queue Establishment}
\label{sec:distributed_queue_establishment}

The core objective of DQP is to obtain shared queues at both nodes, i.e. the items and the order of the elements in the queues are agreed upon. That is, both controllable end nodes $A$ and $B$ hold local queues $\mathcal{Q}^A = \{Q_1^A,\ldots,Q_L^A\}$ and $\mathcal{Q}^B = \{Q_1^B,\ldots,Q_L^B\}$ respectively, which are synchronized using the DQP. CREATE request additions to the queue $Q_j$ can be made by either $A$ or $B$ invoking the DQP by the function $\texttt{ADD}(j, c_r)$, where $c_r$ is the entanglement request by CREATE message. $\texttt{ADD}$ returns a tuple $(i_j, R)$ where $R$ indicated success or failure. Failure can occur if 

\begin{itemize}
\item no acknowledgments are received within a certain time frame, i.e. a timeout occurs,
\item the remote node rejects addition to the queue, or 
\item the queue is full.
\end{itemize}

Success means that the request to create entanglement is placed into $\mathcal{Q}^A$ and $\mathcal{Q}^B$ such that the following properties are satisfied:

\begin{itemize}
\item \emph{Equal queue number}: If a request is added by $A$ as $(j,i_j) \in \mathcal{Q}^A$, then it will (eventually) be added at $B$ with the same absolute queue ID $(j,i_j) \in \mathcal{Q}^B$ (and vice versa);
\item \emph{Uniqueness of queue ID}: If a request is placed into the queue by either $A$ or $B$, then it is assigned a unique queue number. That is, if $(j,a) \in \mathcal{Q}^A$ and $(j,a') \in \mathcal{Q}^A$ reference two distinct CREATE requests, then $a \neq a'$;
\item \emph{Consistency}: If $(j,i_j) \in \mathcal{Q}^A$ and $(j,i_j) \in \mathcal{Q}^B$ then both absolute queue IDs refer to the same request\footnote{This is implied by the previous two conditions, but added for clarity.};
\item \emph{Fairness}: If $A$ (or $B$) is issuing requests continuously, then also the other node $B$ (or $A$) will get to add items to the queue after $A$ (or $B$) in a ``fair manner`` as determined by the window size, denoted as $W_A$ ($W_B$). More precisely, if $a_1,\ldots,a_N$ are CREATE requests submitted at $A$, and $b_1,\ldots,b_M$ are CREATE requests submitted at $B$ with $N > W_A$ and $M > W_B$---all assigned to the same queue $Q_j$ but not yet added---then the final ordering of the requests on the queue obeys $a_1,\ldots,a_m,b_1,\ldots,b_k,a_{m+1},\ldots$ with $m \leq W_A$ and $k \leq W_B$.
\end{itemize}

Recall that each request receives a minimum to be executed time---a time buffer before the request may begin processing which takes into account the processing time to add it into the queue (denoted by the min\_time)---which we will choose to be the expected propagation delay between $A$ and $B$. The purpose of this minimum time is to decrease the likelihood $A$ or $B$ wants to produce an entanglement before the other node is ready. If either $A$ or $B$ begins processing early, no penalty other than reduced performance due to increased decoherence of the quantum memory results. Refer to Section~\ref{sec:packet_formats_dqp} on how this minimum time is passed between nodes.

We recall that in the current implementation of quantum network we have two nodes only. This implies that the queue establishment can be realized by one node being the master controller of the queue marshaling access to the queue, and the other the slave controller. Extensions to multiple nodes are more complex, and a motivation to consider heralding station-centric protocols in the future versions of the protocol. Also, as we have two nodes only, there is no need for the introduction of leader election or a network discovery mechanism. We leave this as future work.

\subsubsection{DQP Sequence Diagrams}
\label{sec:sequence_diagram_dqp}

\begin{figure}
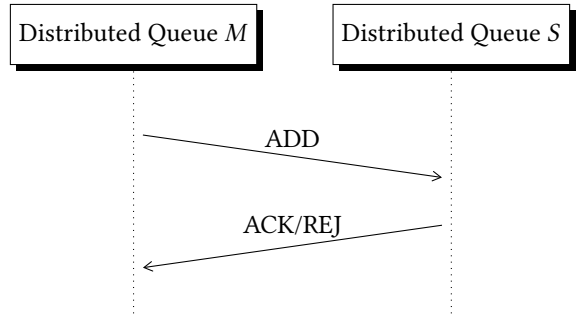

  \centering
  \begin{sequencediagram}
  \newinst[0]{a}{Distributed Queue $M$}
  \newinst[1]{b}{Distributed Queue $S$}
  \mess[1]{a}{ADD}{b}
  \mess[1]{b}{{ACK/REJ}}{a}
  \end{sequencediagram}

  \caption{DQP operation timeline. User $M$ (master) adds an item to the distributed queue by sending an ADD message to peer node $S$ (slave). $S$ either acknowledges or rejects the request using ACK and REJ messages, respectively. Note that this process is symmetric when $S$ attempts to add an item to the queue. For a definition of all messages refer to Figure~\ref{fig:dqp_add_packet}.}
  \label{fig:dqp_master_add}
\end{figure}

Figure~\ref{fig:dqp_master_add} shows a DQP sequence diagram of adding an item to the queue containing a request to the distributed queue. Specifically, an item is an entanglement create request with its associated properties that is passed inside an ADD message within its REQ field, refer to Figure~\ref{fig:dqp_add_packet} for details. 

Upon receiving an ADD message from master $M$, a slave $S$ may choose to acknowledge the item with an ACK message, should validation pass, or reject it with the REJ message for any of the previously mentioned reasons. In the case that master $M$ never receives an acknowledgment (ACK) or rejection (REJ) message after a timeout, the item will be removed from the queue and no processing will occur on the request. Loss of ADD, REJ, and ACK messages in the distributed queue protocol result in retransmissions of the original ADD to guarantee the receipt of rejection and acknowledgement messages.

\begin{figure}[h!]
	\centering
	\begin{bytefield}[boxformatting={\centering\small\sffamily}]{32}
		\bitheader{0-31} \\
		\bitbox{10}{OPT (reserved)} & 
		\bitbox{2}{FT} & 
		\bitbox{8}{CSEQ} &
		\bitbox{4}{QID} & 
		\bitbox{8}{QSEQ} & \\
		\wordbox{2}{Schedule Cycle} & \\
		\wordbox{2}{Timeout} & \\
		\bitbox{32}{Minimum Fidelity} & \\
		\bitbox{16}{Purpose ID} & 
		\bitbox{16}{Create ID} & \\ 
		\bitbox{16}{Number of Pairs} & 
		\bitbox{4}{Priority} &
		\bitbox{12}{(reserved)} & \\
		\bitbox{32}{Initial Virtual Finish} & \\
		\bitbox{32}{Estimated Cycles/Pair} & \\
		\bitbox{1}{\rotatebox{90}{\tiny STR}} &
		\bitbox{1}{\rotatebox{90}{\tiny ATM}} &
		\bitbox{1}{\rotatebox{90}{\tiny MD}} &
		\bitbox{1}{\rotatebox{90}{\tiny MR}} & 
		\bitbox{28}{(reserved)} 
	\end{bytefield}
	\caption{Packet format for \textsf{ADD}, \textsf{ACK}, and \textsf{REJ}. Explanation of the message fields---OPT: field reserved for future options, FT: frame type (\texttt{00}: ADD, \texttt{01}: ACK, \texttt{10}: REJ), CSEQ: the communication sequence number of the transmitted message; QID: the ID of the queue to add the request to; QSEQ: the sequence number within the specified queue to assign the request; Schedule Cycle: the first MHP cycle when the request may begin (equivalent to min\_time); Timeout: The MHP cycle when the request will time out; Initial Virtual Finish: Scheduling information for weighted fair queuing; STR: store flag, ATM: atomic flag, MD: measure directly flag, MR: master request flag.}
	\label{fig:dqp_add_packet}
\end{figure}

When the slave $S$ wishes to add an item to the queue, a message containing the request information and desired queue is included within the messages. Because the master controller has the final say on the state of the queue, a sequence number within the specified queue will be transmitted in return to the slave such that absolute queue IDs are consistent between the nodes.

\subsubsection{DQP Packet Formats}
\label{sec:packet_formats_dqp}

Figure~\ref{fig:dqp_add_packet} presents the packet format for messages exchanged in the DQP. \textsf{Schedule Cycle} and \textsf{Timeout Cycle} of 64\,bits is governed by the maximum number of MHP cycles in the scheduler. \textsf{Purpose ID} of 16 bits enables pointing to $2^{16}$ different applications and the total number of uniquely addressed applications and follows from the number chosen for IPv4. \textsf{Create ID} defines the identifier of locally created request. \textsf{Number of pairs} enables to request up to 2$^\text{16}$ pairs. \textsf{Priority} field of 4 bits is used as we enable 16 local queues composing the distributed queue and each one represents a priority lane. \textsf{Initial Virtual Finish} is used for weighted fair queuing. 

\subsection{Midpoint Heralding Protocol}
\label{sec:midpoint_herading_protocol}

The purpose of MHP is to create entanglement using a midpoint heralding protocol. The operation of the MHP is defined by Protocol~\ref{protocol:mhp}.

\newcounter{enumTemp2}
\label{protocol:mhp}
\begin{Lprotocol}{MHP for use with the Node-Centric EGP}
\sbline

\textit{Definition of functions and variables.}

\begin{itemize}
%\item \textsf{seq$_\text{MHP}$}: the sequence number sent in messages by the midpoint. Increments when heralding is successful at the midpoint;
\item POLLEGP: process to poll for entanglement parameters from EGP; it returns: 
\begin{itemize}
\item \textsf{flag}: true/false indicating whether entanglement should be attempted or not;
\item \textsf{PSEQ}: The pulse sequence identifier that should be issued to the hardware to initialize communication qubit and produce spin-photon entanglement.  May also instruct the hardware to store the spin state within a storage qubit.
\item \textsf{a$_\text{ID}$}: Absolute queue ID, i.e. $(j, i_j)$, of the request entanglement is being attempted for (\textsf{a$_\text{ID,A}$} and \textsf{a$_\text{ID,B}$} for nodes $A$ and $B$, respectively);
\item \textsf{params}: parameters to use for the entanglement attempt such as bright state population $\alpha$;
\end{itemize}
\item \textsf{mhp$_\text{err}$}: error in MHP reported to EGP through REPLY message (REPLY$_\text{A}$ and REPLY$_\text{B}$ sent to nodes $A$ and $B$, respectively), which can take the following values:
\begin{itemize}
\item GEN\_FAIL: general failure that occurs locally at the MHP (failed qubit initialization; other errors). Note: this error message is passed to EGP locally and not included in the REPLY message (see Figure~\ref{fig:mhp_reply_packet}); 
\item QUEUE\_MISMATCH: an error sent by the midpoint when \textsf{a$_\text{ID}$} included in frame from $A$ does not match \textsf{a$_\text{ID}$} included in frame from $B$;
\item TIME\_MISMATCH: when messages from $A$ and $B$ does not arrive at midpoint within the same time interval;
\item NO\_MESSAGE\_OTHER: when the midpoint receives a message from only one of $A$ or $B$;
\end{itemize}
\item GEN: the frame sent by $A$ and $B$ to the midpoint requesting entanglement. The contents include:
\begin{itemize}
%\item \textsf{produce}: a message type that specifies a request for the midpoint to produce entanglement;
\item \textsf{a$_\text{ID}$}:  same as \textsf{a$_\text{ID}$} above;
\end{itemize}
\item REPLY$_\text{A}$ and REPLY$_\text{B}$: the REPLY frames sent to $A$ and $B$ respectively. The contents include:
\begin{itemize}
\item \textsf{outcome}: the outcome of the attempted entanglement at the midpoint, also encodes the error that occurred for the attempt at entanglement (see errors listed above); stored locally at mid-point.
\item \textsf{seq$_\text{MHP}$}: the sequence number from the MHP;
\item \textsf{a$_\text{ID,receiver}$}: The absolute queue ID that was submitted by the node receiving the REPLY.
 \item \textsf{a$_\text{ID,peer}$}: The absolute queue ID that was submitted by peer node.
\end{itemize}
\end{itemize}

\textit{Initialization.} Initialize sequence numbers (set initial \textsf{seq$_\text{MHP}=0$} at $H$). Start timer using a global synchronized clock. 

\sbline
\textit{The protocol, executed at each time step:}
\begin{enumerate}
 \item \textbf{Executed at Node $A$ or Node $B$:}
 \begin{enumerate}
  \item Poll EGP, i.e. POLLEGP= (\textsf{flag}, \textsf{PSEQ}, \textsf{a$_\text{ID}$}, \textsf{params})
  \item If \textsf{flag}=\text{true}, i.e., we want to make entanglement:
  \begin{enumerate}
   \item Issue $PSEQ$ to hardware to initialize communication qubit and produce spin-photon entanglement. $PSEQ$ may also instruct the hardware to store spin state in a storage qubit. If any failures occur, send \textsf{mhp$_\text{err}$}$=\text{GEN\_FAIL}$ back to the EGP and skip to next time step.
   \item Use GEN = (\textsf{a$_\text{ID}$}) and transmit to midpoint upon photon emission.
  \end{enumerate}
 \end{enumerate}
\end{enumerate}
\setcounter{enumTemp2}{\value{enumi}}
\end{Lprotocol}
\addtocounter{Lprotocol}{-1}
\begin{Lprotocol}{(cont.) MHP for use with the Node-Centric EGP}
\begin{enumerate}
 \setcounter{enumi}{\value{enumTemp2}}
 \item \textbf{Heralding station $H$:}
 \begin{enumerate}
  \item Perform the following upon receipt of GEN messages:
  \begin{enumerate}
  \item If messages from $A$ and $B$ do not arrive within the same time interval, let \textsf{mhp$_\text{err}$} = TIME\_MISMATCH and send $\text{REPLY$_\text{A}$}=$(\textsf{mhp$_\text{err}$}, \textsf{seq$_\text{MHP}$}, \textsf{a$_\text{ID,A}$}, \textsf{$_\text{ID,B}$}) to $A$ and $\text{REPLY$_\text{B}$}=$(\textsf{mhp$_\text{err}$}, \textsf{seq$_\text{MHP}$}, \textsf{a$_\text{ID,B}$, \textsf{a$_\text{ID,A}$}}) to $B$.
  \item If \textsf{a$_\text{ID,A}$} $\neq$ \textsf{a$_\text{ID,B}$}, then set \textsf{mhp$_\text{err}$} = QUEUE\_MISMATCH and send $\text{REPLY$_\text{A}$}=$(\textsf{mhp$_\text{err}$}, \textsf{seq$_\text{MHP}$}, \textsf{a$_\text{ID,A}$},\textsf{a$_\text{ID,B}$}) to $A$ and $\text{REPLY$_\text{B}$}=$(\textsf{mhp$_\text{err}$}, \textsf{seq$_\text{MHP}$}, \textsf{a$_\text{ID,B}$},\textsf{a$_\text{ID,A}$}) to $B$.
  \item If GEN arrives only from A, set \textsf{mhp$_\text{err}$} = NO\_MESSAGE\_OTHER and send $\text{REPLY}=(\textsf{mhp$_\text{err}$}, \textsf{seq$_\text{MHP}$}, \textsf{a$_\text{ID,A}$}, \textsf{a$_\text{ID,B}$}\textsf{=null})$ to A, where \textsf{a$_\text{ID,B}$}=null indicates leaving the field as the zero string.  Perform vice versa if GEN arrives only from $B$.

%     \setcounter{enumTemp2}{\value{enumiii}}
%     \end{enumerate}
%  \end{enumerate}
%\end{enumerate}
%\end{Lprotocol}
%\addtocounter{Lprotocol}{-1}
%\begin{Lprotocol}{(cont.) MHP for use with the Node-Centric EGP}
%\begin{enumerate}
%   \item[] {}
%   \begin{enumerate}
%   \item[] {}
%   \begin{enumerate}
%   \setcounter{enumiii}{\value{enumTemp2}}

  \item If no errors occurred then execute quantum swap. Inspect detection result within corresponding time window with $r \in \{0,1,2\}$ where 0 denotes failure and 1 and 2 denote the creation of states one and two respectively. If $r \in \{1,2\}$, a unique and increasing sequence number \textsf{seq$_\text{MHP}$} is chosen by the heralding station (incrementing a counter) to be sent to both $A$ and $B$. Midpoint sends $\text{REPLY}=(\textsf{outcome}, \textsf{seq$_\text{MHP}$}, \textsf{a$_\text{ID}$}, \textsf{a$_\text{ID}$})$ to $A$ and $B$. 
  \end{enumerate}
 \end{enumerate}
 \item \textbf{Executed at Node $A$ or Node $B$}(here \textsf{a$_\text{ID,local}$=a$_\text{ID,A}$}, \textsf{a$_\text{ID,peer}$=a$_\text{ID,B}$} in $A$ and vice versa in $B$)\textbf{:}
 \begin{enumerate}
  \item If $\text{REPLY}=(\textsf{outcome}, \textsf{seq$_{\text{MHP}}$}, \textsf{a$_{\text{ID,local}}$}, \textsf{a$_\text{ID,peer}$})$ returns from midpoint: 
  \begin{enumerate}
  \item Set RESULT=$(\textsf{outcome}, \textsf{seq$_{\text{MHP}}$}, \textsf{a$_\text{ID,local}$}, \textsf{err=000}, \textsf{a$_\text{ID,peer}$})$ and pass to EGP.
  \end{enumerate}
  \item Else if \text{REPLY}=(\textsf{mhp$_\text{err}$}, \textsf{seq$_{\text{MHP}}$}, \textsf{a$_\text{ID,local}$}, \textsf{a$_\text{ID,peer}$}) returns from the midpoint:
  \begin{enumerate}
	\item  Set RESULT=(\textsf{outcome=0}, \textsf{seq$_\text{MHP}$}, \textsf{a$_\text{ID,local}$}, \textsf{mhp$_\text{err}$}, \textsf{a$_\text{ID,peer}$}) and pass to EGP.
  \end{enumerate}
 \end{enumerate}
\end{enumerate}
\end{Lprotocol}

\subsubsection{MHP Sequence Diagrams}
\label{sec:mhp_sequence_diagram}

\begin{figure}[h!]
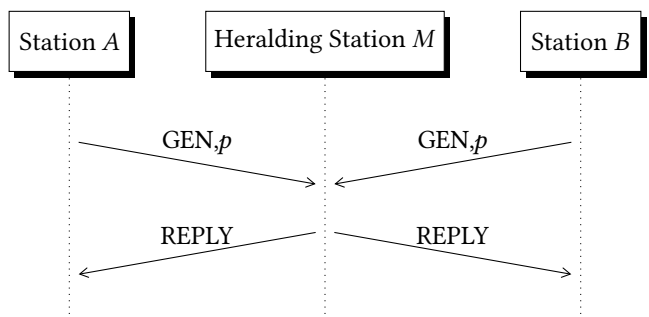

	\centering
	\begin{sequencediagram}
		\newinst[0]{a}{Station $A$}
		\newinst[1]{mid}{Heralding Station $M$}
		\newinst[1]{b}{Station $B$}
		
		\mess[1]{a}{{GEN,$p$}}{mid}
		\prelevel
		\prelevel
		\mess[1]{b}{{GEN,$p$}}{mid}
		
		\mess[1]{mid}{{REPLY}}{a}
		\prelevel
		\prelevel
		\mess[1]{mid}{{REPLY}}{b}
		
	\end{sequencediagram}
	\caption{Timeline of the MHP message exchange with a successful reply by the heralding station; $p$ is a photon associated with the GEN message. For a definition of GEN and REPLY message refer to Figure~\ref{fig:mhp_gen_packet} and Figure~\ref{fig:mhp_reply_packet}, respectively. }
	\label{fig:successful_mhp_timeline}
\end{figure}

\begin{figure}[h!]
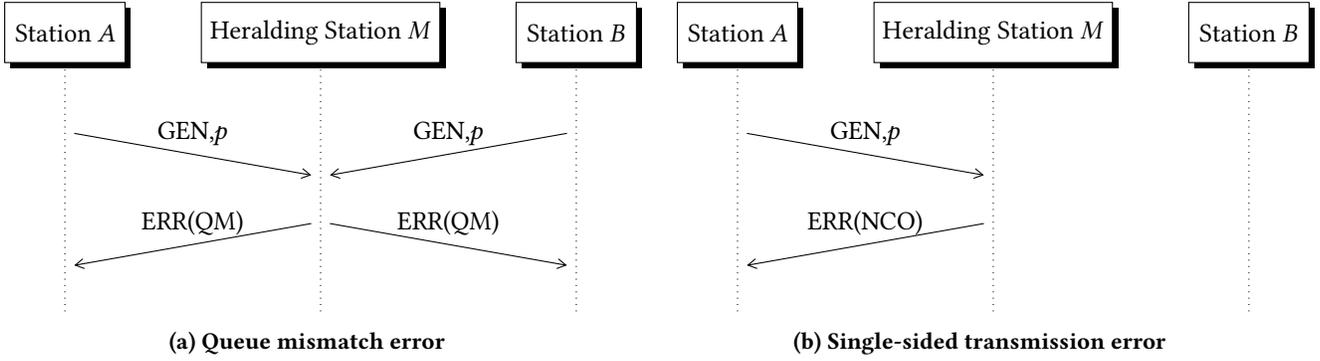

	\centering
\subcaptionbox{Queue mismatch error\label{fig:err_aid_timeline}}{%
	\begin{sequencediagram}
		\newinst[0]{a}{Station $A$}
		\newinst[1]{mid}{Heralding Station $M$}
		\newinst[1]{b}{Station $B$}
		\mess[1]{a}{{GEN,$p$}}{mid}
		\prelevel
		\prelevel
		\mess[1]{b}{{GEN,$p$}}{mid}
		\mess[1]{mid}{{ERR(QM)}}{a}
		\prelevel
		\prelevel
		\mess[1]{mid}{{ERR(QM)}}{b}
\end{sequencediagram}}
\hfill
\subcaptionbox{Single-sided transmission error\label{fig:err_nco_timeline}}{%
\begin{sequencediagram}
	\newinst[0]{a}{Station $A$}
	\newinst[1]{mid}{Heralding Station $M$}
	\newinst[1]{b}{Station $B$}
	\mess[1]{a}{{GEN,$p$}}{mid}
	\mess[1]{mid}{{ERR(NCO)}}{a}
\end{sequencediagram}}
\hfill
\caption{Timeline of two types of errors within MHP. For a definition of GEN and REPLY message refer to Figure~\ref{fig:mhp_gen_packet} and Figure~\ref{fig:mhp_reply_packet}, respectively. QM and NCO refer to specific fields of the REPLY message (i.e. \textsf{OT} field), i.e. QUEUE\_MISMATCH and NO\_MESSAGE\_OTHER, respectively; both error types are explained in Protocol~\ref{protocol:mhp}.}
\label{fig:mhp_error_diagrams}
\end{figure}

The MHP sequence diagram is defined by two cases: the successful ---see Figure~\ref{fig:successful_mhp_timeline}, and unsuccessful one---see Figure~\ref{fig:mhp_error_diagrams}. Specifically, there are two failure scenarios that may occur in the MHP protocol: queue mismatch error (Figure~\ref{fig:err_aid_timeline})---where the message consistency check fails at the midpoint---and single-sided transmission error (Figure~\ref{fig:err_nco_timeline}).

\subsubsection{MHP Packet Formats}
\label{sec:mhp_packet_format}

MHP relies on the exchange of the packets listed in the MHP sequence diagrams, see Figure~\ref{fig:successful_mhp_timeline} and Figure~\ref{fig:mhp_error_diagrams}: GEN and REPLY. 

GEN packet (Figure~\ref{fig:mhp_gen_packet}) is used by the midpoint to determine whether the nodes are consistent in their local information regarding their knowledge of the attempt at entanglement.

REPLY packet (Figure~\ref{fig:mhp_reply_packet}) is sent by the midpoint in the case of no error. It will include the senders' submitted absolute queue ID (i.e. \textsf{QID} and \textsf{QSEQ}) and additionally pass on the submitted queue ID of the peer node (i.e. \textsf{QIDP} and \textsf{QSEQP}). The sequence number, \textsf{SEQ}, denotes the number of successful heralded entanglement generations that have occurred at the midpoint heralding station and allows the end nodes to keep track of the number of entangled pairs that have been generated.  \textsf{OT} encodes the heralding signal from the midpoint upon successful operation and encodes errors in case of failures.

\begin{figure}[h!]
\centering
%  \begin{verbatim}
%+---------------+---------+----------+
%| Header (1-12) | QID (4) | QSEQ (8) |
%+---------------+---------+----------+
%  \end{verbatim}

\begin{bytefield}[boxformatting={\centering\small\sffamily}]{32}
	\bitheader{0-31} \\
	\bitbox{12}{Header} & 
	\bitbox{4}{QID} &
	\bitbox{8}{QSEQ} &
	\bitbox{8}{(reserved)}
\end{bytefield}

\caption{GEN packet format (used in MHP) sent by end stations to heralding station (midpoint). The pair (QID, QSEQ) represents the absolute queue ID, in other words they map to $(j, i_j)$---see Section~\ref{sec:priority_queues}.}
\label{fig:mhp_gen_packet}
\end{figure}

\begin{figure}[h!]
\centering
%\begin{verbatim}
%+---------------+-------------+----------+
%| Header (1-12) | Outcome (2) | SEQ (16) | ->
%+---------------+-------------+----------+
%+---------+----------+--------------+
%| QID (4) | QSEQ (8) | QID Peer (4) | ->
%+---------+----------+--------------+
%+--------------+
%| QSEQ Peer(8) |
%+--------------+
%\end{verbatim}

\begin{bytefield}[boxformatting={\centering\small\sffamily}]{32}
	\bitheader{0-31} \\
	\bitbox{12}{Header} & 
	\bitbox{4}{OT} &
	\bitbox{16}{SEQ} \\
	\bitbox{4}{QID} &
	\bitbox{8}{QSEQ} &
    	\bitbox{4}{QIDP} & 
    	\bitbox{8}{QSEQP} &
	\bitbox{8}{(reserved)}
\end{bytefield}

\caption{REPLY/ERR packet format (used in MHP) for replies by midpoint with no error. OT: outcome/error reported by midpoint; SEQ: sequence number; QIDP: QID Peer; QSEQP: QSEQ Peer; QID and QSEQ are defined the same as in for GEN message---see Figure~\ref{fig:mhp_gen_packet}. Error codes include (\texttt{001}: QUEUE\_MISMATCH: \texttt{010}: TIME\_MISMATCH, \texttt{100}: NO\_MESSAGE\_OTHER, (refer to Protocol~\ref{protocol:mhp} for the above error description).}
\label{fig:mhp_reply_packet}
\end{figure}

\subsection{Entanglement Generation Protocol}
\label{sec:entanglement_generation_protocol}

The role of the Entanglement Generation Protocol (EGP) is to produce the required entanglement between two end nodes or otherwise declare failure.

\subsubsection{Entanglement Generation Scheduler}
\label{sec:egp_scheduler}

We now proceed with the description of the scheduler---refer to Protocol~\ref{protocol:egp} for details. The EGP scheduler fulfills the following arbitrage functions, where we remark that for CREATE requests that demand multiple EPR pairs, only one request is added to the queue, and hence NEXT (function to select the next request from the local set of queues, see below) will return multiple pairs to be produced for the same request when called successively. 

\begin{itemize}
	\item GET\_QUEUE(\textsf{creq}): Once a request has been submitted, GET\_QUEUE deterministically chooses which queue $Q_j$ to assign the CREATE request \textsf{creq} to. This may depend on the details of the request, such as for example $t_{\max}$, or $F_{\min}$ as well as the purpose ID and priority.
	\item NEXT: Selects the next request from the local set of queues $\mathcal{Q}$ to serve, if any. Specifically, NEXT will determine:
		\begin{itemize}
			\item \emph{Flag}, set to \texttt{true} when a request is ready to be served;
			\item \emph{Absolute queue ID} (and corresponding request details) of request to be served;
			\item \emph{Parameters to use in the MHP} depending on the number of type of outstanding requests;
			\item \emph{Communication and storage qubits}, determined in cooperation with QMM.
		\end{itemize}
\end{itemize}

\newcounter{enumTemp}
\label{protocol:egp}
\begin{Lprotocol}{EGP - Node A (B analogous exchanging A and B)}
\tiny
\sbline
\textit{Definition of functions and variables.}
\begin{itemize}
\item Node ID: the ID of the peer we want to create entanglement with;
\item  $n$: the number of entangled pairs we wish to create;
\item  $F_{\min}$: the minimum acceptable fidelity required for the generated pairs;
\item  $t_{\max}$: the maximum amount of time the higher layer is willing to wait for the entanglement to be created
\item  Purpose ID: port/application ID the requested this entanglement---used for forwarding OK messages to the appropriate application;
\item  \textsf{priority}: The priority of the request;
\item  \textsf{seq$_\text{expected}$}: The expected sequence number from the midpoint. Initially set to 1;
\item  $j=$GET\_QUEUE($c_r$): The call to the scheduler to obtain the queue ID of $Q_j$ where the request should be placed in the distributed queue.
\item $(i_j,ok)$=ADD($j$, \textsf{creq}): The call to the distributed queue to add the create request \textsf{$c_r$} to $Q_j$. $i_j$ is the unique ID of the request within $Q_j$ and $ok$ is a status code of performing the ADD to the distributed queue. Can take the values \texttt{success} (item added), \texttt{timeout} (communication timeout with peer while adding), or \texttt{reject} (the peer rejected adding the item to the queue);
\item ERR\_NOTIME: Error issued to higher layers by the EGP upon receiving OK=\texttt{timeout} from adding item to queue;
\item  ERR\_REJECT: Error issued to higher layers by the EGP upon receiving OK=\texttt{reject} from adding item to queue;
\item  Trigger pair: equivalent to the POLLEGP call within the MHP protocol outline; %, should be made consistent?
\item  $(flag, (j, i_j), PSEQ, params)=$NEXT(): The call to the scheduler to obtain information for the next entanglement generation attempt where flag=\texttt{True/False} indicates whether entanglement should be attempted (same in MHP outline), $(j,i_j)$ is the absolute queue ID of the create request being served, PSEQ is a pulse sequence identifier encoding the communication and storage qubit information to use for entanglement attempts, and \textsf{params} encodes the parameters to use for entanglement attempts (same as params in MHP outline);
\item \textsf{proto\_err}: Status of the attempt at entanglement, encodes the \textsf{mhp$_\text{err}$} from the MHP outline if an error occurred, 0 if no errors happened;
\item \textsf{create\_time}: timestamp of when the entanglement was generated;
\item  $F_\text{est}$: the goodness passed in the frame, the estimate of the fidelity of the entangled qubits;
\item  \textsf{logical\_id}: The storage qubit ID where the entangled qubit is stored, for use by higher layers;
\item $tGoodness$ - A timestamp of when $F_\text{est}$ was record.
\item $tCreate$ - A timestamp of when the entanglement was created.
\item  $k$: the number of pairs left to generate for the request.
\end{itemize}

\begin{enumerate}
 \item \textbf{Adding to Queue:}
 \begin{enumerate}
  \item Ask scheduler which queue this request should be added to: $j =$ GET\_QUEUE(\textsf{$c_r$}).
  \item Try to add request to the queue using the DQP: $(i_j, ok) =$ ADD($j$, \textsf{$c_r$}).
  \item If $ok$ = \texttt{timeout} error, issue ERR\_NOTIME and stop.
  \item If $ok$ = \texttt{reject} error, issue ERR\_REJECTED and stop.
  \item Otherwise the request has been added to the Distributed Queue.
 \end{enumerate}
 \item \textbf{Trigger pair (polled by MHP):}
 \begin{enumerate}
  \item Ask the scheduler whether entanglement should be made and for which request: NEXT = (\textsf{flag}, ($j$,$i_j$) $\equiv$ \textsf{req}, \textsf{param}, \textsf{PSEQ}). For this end, the scheduler will employ its priority policy, as well as perform flow control depending on whether $B$ is likely to produce entanglement.
  \item If there is a generation waiting to be satisfied:
  \begin{enumerate}
   \item Construct response for MHP POLLEGP()=(\textsf{flag}, ($j$,$i_j$), \textsf{PSEQ}, \textsf{params}).
   \item Provide the response to the MHP.
  \end{enumerate}
  \item Otherwise if there are no generations to perform provide POLLEGP()=(\textsf{flag=False}, \textsf{a$_\text{ID,local}$=null}, \textsf{PSEQ=null}, \textsf{params=null}).
 \end{enumerate}

  \setcounter{enumTemp}{\value{enumi}}
\end{enumerate}
\end{Lprotocol}

\addtocounter{Lprotocol}{-1}
\begin{Lprotocol}{(cont.) EGP - Node $A$ ($B$ analogous exchanging $A$ and $B$)}
\begin{enumerate}
   \setcounter{enumi}{\theenumTemp}

 \item \textbf{Handle reply (message from MHP):}
 \begin{enumerate}
  \item Retrieve message from MHP including: result of generation $r \in \{0,1,2\}$, \textsf{seq$_\text{MHP}$}, absolute queue id $(j,i_j)$, and protocol error flag \textsf{proto\_err}.
  \item If the specified absolute queue ID is not found locally then the request may have timed out or expired:
  \begin{enumerate}
   \item Free the reserved communication/storage qubit in the Quantum Memory Manager.
   \item Update \textsf{seq$_\text{expected}$} with \textsf{seq$_\text{MHP}$}+1 and stop handling reply.
  \end{enumerate}
  \item Otherwise there is an absolute queue ID included in the response that is associated with an active request:
  \begin{enumerate}
   \item If \textsf{proto\_{err}} is not OK, a non quantum error occurred and no entanglement was produced. Update expected sequence number with \textsf{seq$_\text{MHP}$} and stop handling reply.
   \item If $r = 0$ we failed to produce entanglement, stop handling reply.
   \item Process \textsf{seq$_\text{MHP}$}:
   \begin{enumerate}
    \item If the \textsf{seq$_\text{MHP}$} is larger than the expected sequence number, (partially) ERR\_EXPIRE the request to higher layer and send EXPIRE message to peer. Stop handling reply.
    \item Else. if \textsf{seq$_\text{MHP}$} is smaller than the expected sequence number, ignore reply. Stop handling reply.
    \item Else, update next expected \textsf{seq$_\text{MHP}$} sequence number (increment current one modulo $2^{16}$).
   \end{enumerate}
   \item A pair is established. If $r=2$ and we are the origin of this request, apply correction information to transform state $|\Psi^- \rangle$ to state $|\Psi^+ \rangle$, if we are the peer then we instruct the scheduler to suspend subsequent generation attempts until we believe the request originator has completed correction.
   \item Look up queue item $(j,i_j)$:
   \begin{enumerate}
	\item If create\_time + $t_{\max} >$ current time or the request is not stored locally anymore, issue ERR\_TIMEOUT to higher layer and remove item from queue. Stop handling reply.
	\item Get fidelity estimate $F_{\rm est}$ from Fidelity Estimation Unit.
	\item Issue OK with Entanglement ID: ($A$, $B$, \textsf{seq$_\text{MHP}$}), \textsf{logical\_id}, Goodness $F_{\rm est}$, and $t_{\rm Goodness} = t_{\rm Create} = now$.
	\item If $k = 1$, i.e. this was the last pair to be produced for this request, remove item from queue.
	\item If $k > 2$, decrement $k$ on queue item.
   \end{enumerate}
  \end{enumerate}
 \end{enumerate}
\end{enumerate}
\end{Lprotocol}

\subsubsection{EGP Sequence Diagrams}
\label{sec:egp_sequence_diagram}

We proceed with the introduction of all message passing sequence diagrams for the EGP.

\begin{figure}[h!]
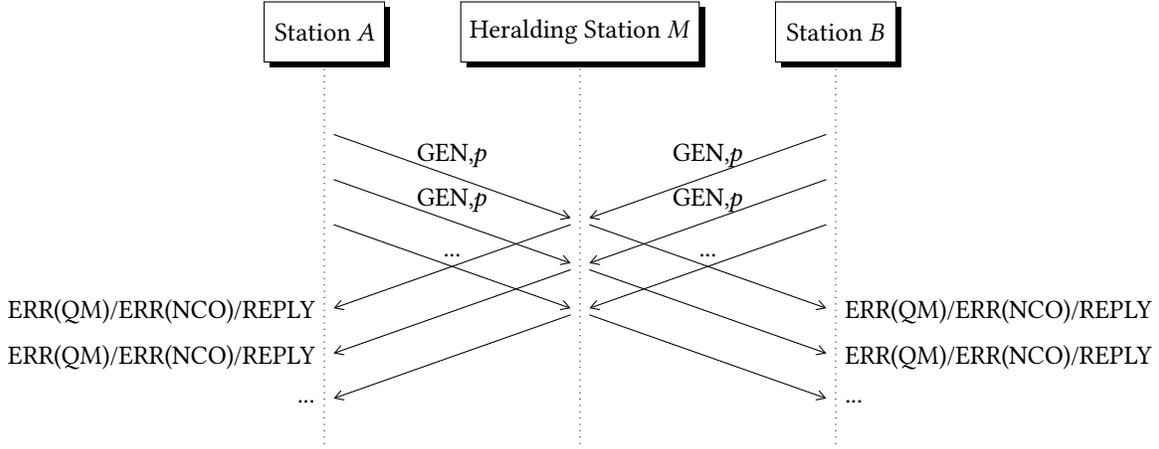

  \centering
  \begin{sequencediagram}
  \newinst[0]{a}{Station $A$}
  \newinst[1]{mid}{Heralding Station $M$}
  \newinst[1]{b}{Station $B$}
  
  \mess[2]{a}{{GEN,$p$}}{mid}
  \prelevel
  \prelevel
  \prelevel
  \mess[2]{b}{{GEN,$p$}}{mid}
  \prelevel
  \mess[2]{mid}{{}}{a}
  \node[anchor=east] (err) at (mess to) {ERR(QM)/ERR(NCO)/REPLY};
  \prelevel
  \prelevel
  \prelevel
  \mess[2]{mid}{{}}{b}
  \node[anchor=west] (err) at (mess to) {ERR(QM)/ERR(NCO)/REPLY};
  \prelevel
  \prelevel
  \prelevel
  \prelevel
  \mess[2]{a}{{GEN,$p$}}{mid}
  \prelevel
  \prelevel
  \prelevel
  \mess[2]{b}{{GEN,$p$}}{mid}
  \prelevel
  \mess[2]{mid}{{}}{a}
  \node[anchor=east] (err) at (mess to) {ERR(QM)/ERR(NCO)/REPLY};
  \prelevel
  \prelevel
  \prelevel
  \mess[2]{mid}{{}}{b}
  \node[anchor=west] (err) at (mess to) {ERR(QM)/ERR(NCO)/REPLY};
  \prelevel
  \prelevel
  \prelevel
  \prelevel
  \mess[2]{a}{{...}}{mid}
  \prelevel
  \prelevel
  \prelevel
  \mess[2]{b}{{...}}{mid}
  \prelevel
  \mess[2]{mid}{{}}{a}
   \node[anchor=east] (err) at (mess to) {...};
  \prelevel
  \prelevel
  \prelevel
  \mess[2]{mid}{{}}{b}
  \node[anchor=west] (err) at (mess to) {...};
  
  \end{sequencediagram}
  \caption{Timeline of multiplexing photon emission in the MHP: multiple GEN,$p$ messages are sent one by one for which any reply messages (ERR(QM), ERR(NCO) or REPLY) is possible. Compare this with the sequence diagram presented in Figure~\ref{fig:mhp_error_diagrams} and Figure~\ref{fig:successful_mhp_timeline} for the MHP.}
  \label{fig:successful_mhp_multiplexing}
\end{figure}

Figure~\ref{fig:successful_mhp_multiplexing} presents a sequence diagram for the EGP operation when performing emission multiplexing. In some cases (such as the CM use case) it is not necessary to wait for the REPLY message from the midpoint before attempting entanglement generation again if one simply desires to generate correlated bit streams.

\begin{figure}[h!]
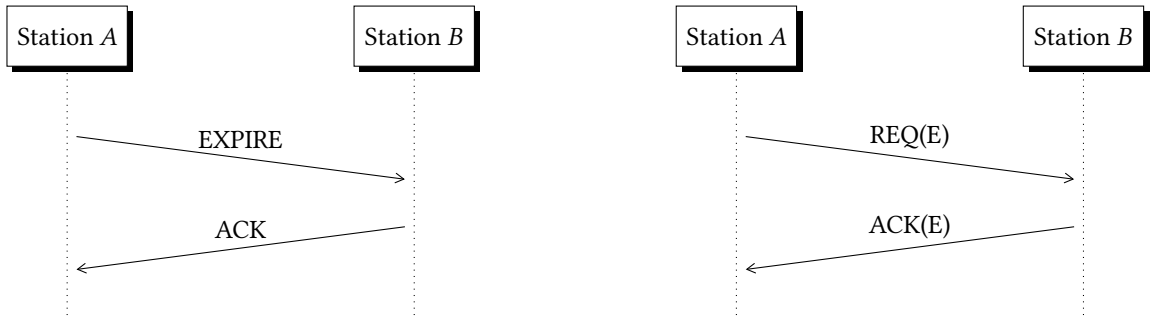

  \centering
  \begin{subfigure}{.5\textwidth}
  \centering
  \begin{sequencediagram}
  \newinst[1]{a}{Station $A$}
  \newinst[3]{b}{Station $B$}
  
  \mess[1]{a}{{EXPIRE}}{b}
  \mess[1]{b}{{ACK}}{a}
  
  \end{sequencediagram}
  \label{fig:egp_exp}
\end{subfigure}%
\begin{subfigure}{.5\textwidth}
  \centering
  \begin{sequencediagram}
  \newinst[1]{a}{Station $A$}
  \newinst[3]{b}{Station $B$}
  
  \mess[1]{a}{{REQ(E)}}{b}
  \mess[1]{b}{{ACK(E)}}{a}
  
  \end{sequencediagram}
  \label{fig:egp_mem_req}
  \end{subfigure}
  \caption{(Left) Timeline of request expiration within EGP. Definitions of EXPIRE and ACK message are given in Figure~\ref{fig:egp_expire_packet} and Figure~\ref{fig:egp_expire_ack_packet}, respectively. (Right) Timeline of memory advertisement requests within EGP. Definition of  REQ(E) and ACK(E) message is given in Figure~\ref{fig:egp_mem_packet}.}
\label{fig:egp_messages}
\end{figure}

Figure~\ref{fig:egp_messages} shows a sequence diagram detailing the message flow should station A obtain information that it is behind as well as a timeline of the message exchange when EGP processes at both nodes exchange available quantum memory information. Sharing this information allows both nodes to know whether there are available resources in order to proceed with satisfying an entanglement request. In the absence of resources of either peer there is no use in photon emission. Simply, both nodes must be able to emit photons for the protocol to 
operate properly.

 Imperfect message transmission may cause any of the GEN, REPLY, EXPIRE, REQ(E), and ACK messages to become lost or corrupted in transit between nodes. Depending on which messages are lost in the protocol, different actions are taken to prevent deadlock. $A$ lost EXPIRE or corresponding ACK results in a retransmission of the EXPIRE to ensure that OK messages are properly revoked at the peer node. Loss of a REQ(E) message or its corresponding ACK results in a retransmission of the REQ(E) to make sure that both nodes have up-to-date information of available resources.

Losing a GEN message is handled by the midpoint heralding station when only one GEN message arrives from an end node. In this case the REPLY message containing NO\_CLASSICAL\_OTHER (see Protocol~\ref{protocol:mhp}) is issued to alert the nodes of the failure. In this case no attempt at entanglement is made and the sequence number at the midpoint remains the same.

Losing a REPLY message from the midpoint that contains an outcome of 0 has no impact on $A$ or $B$ as \textsf{seq$_\text{expected}$} is only updated when a successful attempt at entanglement occurs.  When a REPLY message containing an outcome of 1 or 2 (for successful entanglement) is lost, the end node(s) that lost the message will continue attempting entanglement generation in subsequent polls by the MHP as there are outstanding pairs to be generated for the request. Upon successful receipt of any message from the midpoint (REPLY), the included SEQ will be ahead of the receiving node(s) \textsf{seq$_\text{expected}$} and the loss will be detected.  The detecting node will then transmit an EXPIRE message to its peer containing the old \textsf{seq$_\text{expected}$} that did not agree with the SEQ received from the midpoint along with its new \textsf{seq$_\text{expected}$}, so that any OK messages containing the missing set of sequence numbers are revoked at the peer.

\subsubsection{EGP Packet Formats}
\label{sec:egp_packet_formats}

Finally, we present the definitions of all messages being used by the EGP. Figure~\ref{fig:egp_to_mhp_packet} describes the information passed from the EGP to the MHP during each periodic cycle, while Figure~\ref{fig:mhp_to_egp_packet} shows the packet format for replies from the physical layer MHP to the EGP. Figure~\ref{fig:egp_expire_packet} and Figure~\ref{fig:egp_expire_ack_packet}, define EXPIRE and ACK messages, respectively, exchanged in entanglement request expiration sequence diagram described in Figure~\ref{fig:egp_messages}. Figure~\ref{fig:egp_mem_packet} shows the REQ(E) and ACK(E) packet formats exchanged by memory advertisement requests made by the EGP sequence diagram described in Figure~\ref{fig:egp_messages}.Figure~\ref{fig:egp_ck_ok_packet} and Figure~\ref{fig:egp_md_ok_packet} present the format of OK messages passed from the EGP to higher layers, in case of \emph{create and keep request} and \emph{measure directly request}, respectively.

\begin{figure}[h!]
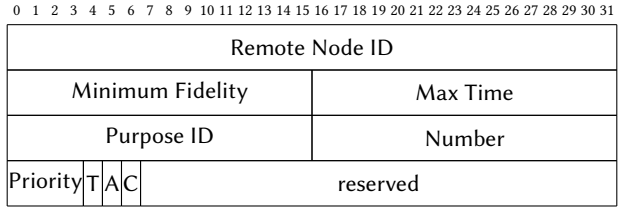

\centering
\begin{bytefield}[boxformatting={\centering\small\sffamily}]{32}
	\bitheader{0-31} \\
	\bitbox{32}{Remote Node ID} \\
	\bitbox{16}{Minimum Fidelity} &
	\bitbox{16}{Max Time} \\  % timestamp of entnglement creations
  	\bitbox{16}{Purpose ID} &
	\bitbox{16}{Number} \\
	\bitbox{4}{Priority} &
	\bitbox{1}{T} &
	\bitbox{1}{A} &
	\bitbox{1}{C} &
	\bitbox{25}{reserved}
\end{bytefield}
	\caption{Packet format for CREATE message to EGP.  Explanation of the message fields---Remote Node ID: Used if the node is directly connected to multiple nodes.  Indicates which node to generate entanglement with; Minimum Fidelity---The desired minimum fidelity, between 0 and 1, of the generated entangled pair; Max Time---The maximum time in seconds the higher layer is willing to wait for the request to be fulfilled; Purpose ID---Allows the higher layer to tag the request for a specific purpose; Number---The number of entangled pairs to generate; Priority---Can be used to indicate if this request is of high priority and should ideally be fulfilled early; T---the type of request, Either create and keep (K) or measure directly (M), where K stores the generated entanglement in memory and M measures the entanglement directly; A---atomic flag, indicates that the request should be satisfied as a whole, i.e. that all entangled pairs are available in memory at the same time; C---consecutive flag, indicates that the entangled pairs of the request should be generated close in time.}
	\label{fig:egp_create_packet}
\end{figure}

\begin{figure}[h!]
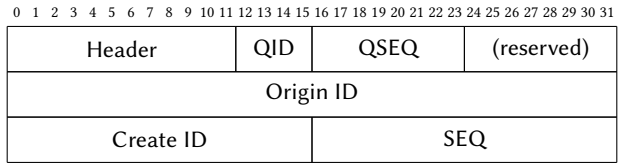

	\centering
	\begin{bytefield}[boxformatting={\centering\small\sffamily}]{32}
		\bitheader{0-31} \\
		\bitbox{12}{Header} & 
		\bitbox{4}{QID} & 
		\bitbox{8}{QSEQ} &
		\bitbox{8}{(reserved)} \\
		\bitbox{32}{Origin ID}\\ 
		\bitbox{16}{Create ID} &
		\bitbox{16}{SEQ}
	\end{bytefield}
	
	\caption{Packet format for \textsf{EXPIRE} message. Explanation of the message fields---Origin ID: ID of the node the request originated from; Create ID: creation ID of the request. SEQ: up-to-date expected MHP sequence number at the node the EXPIRE originates from. Recall that (QID, QSEQ) represent the absolute queue ID.}
	\label{fig:egp_expire_packet}
\end{figure}

\begin{figure}[h!]
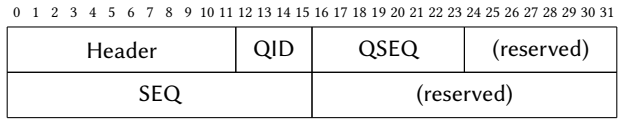

	\centering
	\begin{bytefield}[boxformatting={\centering\small\sffamily}]{32}
		\bitheader{0-31} \\
		\bitbox{12}{Header} &
		\bitbox{4}{QID} &
		\bitbox{8}{QSEQ} &
		\bitbox{8}{(reserved)} \\
		\bitbox{16}{SEQ} &
		\bitbox{16}{(reserved)}
	\end{bytefield}
	
	\caption{Packet format for \textsf{ACK} message. Recall that (\textsf{QID}, \textsf{QSEQ}) is the absolute queue ID and SEQ is the acknowledger's up-to-date expected MHP sequence number (the same as in the case of EXPIRE message, see Figure~\ref{fig:egp_expire_packet}).}
	\label{fig:egp_expire_ack_packet}
\end{figure}

\begin{figure}[h!]
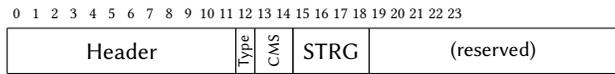

	\centering
	% \begin{verbatim}
	%+---------------+-----------+-------------+
	%| Header (1-12) | COMMS (2) | STORAGE (4) |
	%+---------------+-----------+-------------+
	% \end{verbatim}
	
	\begin{bytefield}[boxformatting={\centering\small\sffamily}]{24}
		\bitheader{0-23} \\
		\bitbox{12}{Header} &
		\bitbox{1}{\rotatebox{90}{\tiny Type}} &
		\bitbox{2}{\rotatebox{90}{\tiny CMS}} & % communcation sequence number encodes the no of comm qbits avaibale 
		\bitbox{4}{{STRG}} % encodes the numer of storage qubits available
		\bitbox{13}{\footnotesize (reserved)}
	\end{bytefield}
	\caption{REQ(E)/ACK(E) Packet format for EGP memory requests. Explanation of the message fields---Type: message type (\texttt{0}: REQ(E), \texttt{1}: ACK); CMS: the number of available communication qubits, STRG: the number of available storage qubits.}
	\label{fig:egp_mem_packet}
\end{figure}

\begin{figure}[h!]
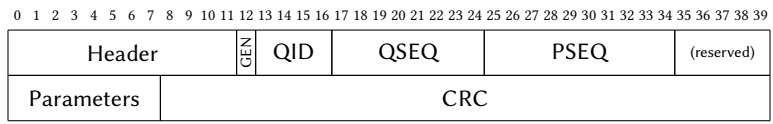

	\centering
	\begin{bytefield}[boxformatting={\centering\small\sffamily}]{40}
		\bitheader{0-39} \\
		\bitbox{12}{Header} & 
		\bitbox{1}{\rotatebox{90}{\tiny GEN}} & %emit a photon or not
		\bitbox{4}{QID} &
		\bitbox{8}{QSEQ} &
		\bitbox{10}{PSEQ} & %pulse sequence identifier
		\bitbox{5}{\tiny (reserved)} \\
		\bitbox{8}{Parameters} &
		\bitbox{32}{CRC}
	\end{bytefield}
	
	\caption{Packet format of POLLEGP messages sent from EGP to MHP. Explanation of the message fields---GEN: emit photon flag; PSEQ: pulse sequence identifier which instructs the underlying quantum communication device of the parameters to use when emitting the photon. See figure \ref{fig:egp_architecture}.}
	\label{fig:egp_to_mhp_packet}
\end{figure}

\begin{figure}[h!]
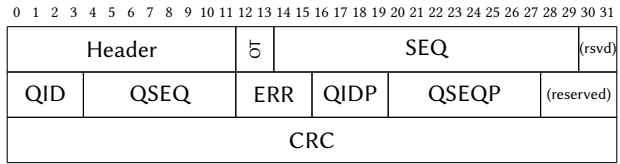

	\centering
	\begin{bytefield}[boxformatting={\centering\small\sffamily}]{32}
		\bitheader{0-31} \\
		\bitbox{12}{Header} & 
		\bitbox{2}{\rotatebox{90}{\tiny OT}} & % the same as OT 
		\bitbox{16}{SEQ} &
		\bitbox{2}{\tiny (rsvd)} \\
		\bitbox{4}{QID} &
		\bitbox{8}{QSEQ} &
		\bitbox{4}{ERR} &
		\bitbox{4}{QIDP} &
		\bitbox{8}{QSEQP} &
		\bitbox{4}{\tiny (reserved)} \\
		\bitbox{32}{CRC}
	\end{bytefield}
	
	\caption{Packet format of messages from MHP to EGP. Explanation of the message fields---OT: measurement outcome (the same field as in REPLY message define in Figure~\ref{fig:mhp_reply_packet}). Error codes in ERR field encode the errors described in the MHP protocol (refer to Protocol~\ref{protocol:mhp}). See figure \ref{fig:egp_architecture}.}
	\label{fig:mhp_to_egp_packet}
\end{figure}

\begin{figure}[h!]
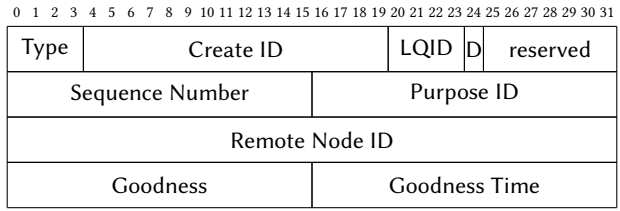

  \centering
%  \begin{verbatim}
%  +---------------+--------------+
%  | Header (1-12) | CreateID (2) | ->
%  +---------------+--------------+
%  +-------------+---------------+
%  | originID (4)| peerID (1-12) | ->
%  +-------------+---------------+
%  +---------------+------------+
%  | peerID (1-12) | MHPSeq (2) | ->
%  +---------------+------------+
%  +----------------+------------------+
%  | logical ID (2) | tGoodness (1-12) | ->
%  +----------------+------------------+
%  +-------------+-------------+
%  | tCreate (2) | Goodness (4)|
%  +-------------+-------------+
%  \end{verbatim}

\begin{bytefield}[boxformatting={\centering\small\sffamily}]{32}
	\bitheader{0-31} \\
	\bitbox{4}{Type} & 
	\bitbox{16}{Create ID} &
	\bitbox{4}{LQID} & % stirage qubit ID in the device
	\bitbox{1}{D} &
	\bitbox{7}{reserved} \\
	\bitbox{16}{Sequence Number} &
	\bitbox{16}{Purpose ID} \\
	\bitbox{32}{Remote Node ID} \\ 
	\bitbox{16}{Goodness} &
	\bitbox{16}{Goodness Time}\\ % timestamo of the fidelity of etanglement (estimation
\end{bytefield}

\caption{Packet format for \textsf{OK} message corresponding to a \textsf{create and keep} request. Explanation of the message fields---Type: Indicates that this is a create and keep OK; Create ID: The same as the Create ID returned to the requester; LQID: Logical Qubit ID where the entanglement is stored; D: Directionality flag indicating the source of the request; Sequence Number: A sequence number for identifying the entangled pair; Purpose ID: The purpose ID of the request; Remove Node ID: Used if connected to multiple nodes; Goodness: An estimate of the fidelity of the generated pair; Goodness Time: Time of the goodness estimate. See figure \ref{fig:egp_architecture}.}
\label{fig:egp_ck_ok_packet}
\end{figure}

\begin{figure}[h!]
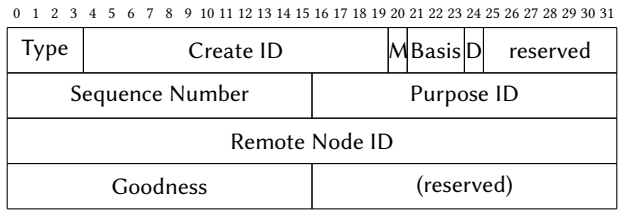

  \centering

\begin{bytefield}[boxformatting={\centering\small\sffamily}]{32}
	\bitheader{0-31} \\
	\bitbox{4}{Type} & 
	\bitbox{16}{Create ID} &
	\bitbox{1}{M} & % measurement outcome 
	\bitbox{3}{Basis} & % measurement basis for entanglement creation
	\bitbox{1}{D} &
	\bitbox{7}{reserved}\\
	\bitbox{16}{Sequence Number} &
	\bitbox{16}{Purpose ID} \\
	\bitbox{32}{Remote Node ID} \\
    	\bitbox{16}{Goodness} &
	\bitbox{16}{(reserved)}
\end{bytefield}

	\caption{Packet format for \textsf{OK} message corresponding to a \textsf{measure directly} request. Explanation of the message fields---Type: Indicates this is an OK for a measure direclty request; Create ID: The same Create ID given to the requester; M: Outcome of the measurement performed on the entangled pair; Basis: Which basis the entangled pair was measured in, used if the basis is random.  The remainder of the fields are explained in Figure \ref{fig:egp_ck_ok_packet}. See figure \ref{fig:egp_architecture}.}
\label{fig:egp_md_ok_packet}
\end{figure}

\begin{figure}[h!]
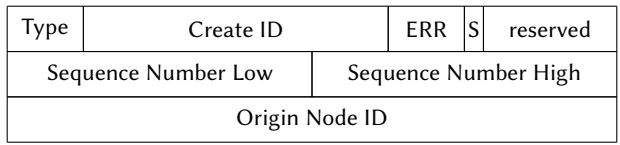

\centering
\begin{bytefield}[boxformatting={\centering\small\sffamily}]{32}
        \bitbox{4}{Type} &
	\bitbox{16}{Create ID} &
	\bitbox{4}{ERR} &
	\bitbox{1}{S} &
	\bitbox{7}{reserved} \\
	\bitbox{16}{Sequence Number Low} &
	\bitbox{16}{Sequence Number High} \\
	\bitbox{32}{Origin Node ID}
\end{bytefield}
	\caption{Packet format for \textsf{ERR} messages containing errors from EGP.  Explanation of the message fields---Type: Indicates this is an ERR message; ERR: The error that occurred in the EGP, may be any of the errors described in Section \ref{sec:entanglement_generation_protocol}; S: Used by ERR\_EXPIRE, to specify whether a range of sequence numbers should be expired (S=1) or all sequence numbers associated with the given Create ID and Origin Node (S=0); Sequence Number Low/High: Use together to specify a range of sequence numbers to expire.  The remaining fields are explained in figure \ref{fig:egp_ck_ok_packet}.}
	\label{fig:egp_err_packet}
\end{figure}

\end{document}